\documentclass[onecolumn,epjc3]{svjour3}

\usepackage[a4paper, total={6in, 9in}]{geometry}
\usepackage[utf8]{inputenc}
\usepackage{amsmath,amsthm}
\usepackage{slashed}
\usepackage{amssymb}
\usepackage{calrsfs}
\usepackage{xcolor}
\usepackage{graphicx}
\usepackage{breqn}
\usepackage{float}
\usepackage{subcaption} 
\usepackage{physics}
\usepackage{hyperref}\definecolor{myred1}{RGB}{255, 0, 0}

\usepackage{soul}            

\usepackage[normalem]{ulem}   

\usepackage{comment}          

\usepackage{color}

\newcommand{\beqn}{\begin{eqnarray}}
\newcommand{\Eqn}{\end{eqnarray}}
\newcommand{\be}{\begin{equation}}
\newcommand{\E}{\end{equation}}
\newcommand{\bear}{\begin{eqnarray}}
\newcommand{\Ear}{\end{eqnarray}}

\newcommand{\ba}{\begin{array}{c}}
\newcommand{\bat}{\begin{array}{cc}}
\newcommand{\ea}{\end{array}}
\newcommand{\bi}{\begin{itemize}}
\newcommand{\ei}{\end{itemize}}
\newcommand{\mA}{\mathcal{A}}

\newcommand{\mF}{\mathcal{F}}
\newcommand{\mG}{\mathcal{G}}

\newcommand{\mL}{\mathcal{L}}

\newcommand{\mO}{\mathcal{O}}

\newcommand{\lsim}{\stackrel{<}{_\sim}}
\newcommand{\gsim}{\stackrel{>}{_\sim}}

\renewcommand\[{\begin{dmath}}
\renewcommand\]{\end{dmath}}

\begin{document}

\title{On the relevance of fermion loops for $W^+W^-$ scattering at LHC }

\author{Carlos~Quezada-Calonge\thanksref{e1,addr1}
        \and
        Antonio~Dobado\thanksref{e2,addr1}  
        \and
        Juan~Jos\'e~Sanz-Cillero\thanksref{e3,addr1}  
}

\thankstext{e1}{e-mail: cquezada@ucm.es}
\thankstext{e2}{e-mail: dobado@fis.ucm.es}
\thankstext{e3}{e-mail: jjsanzcillero@ucm.es}

\institute{ Departamento de F\'\i sica Te\'orica and IPARCOS, Facultad de Ciencias F\'\i sicas, \\
Universidad Complutense de Madrid, 28040 Madrid, Spain \label{addr1}
}

\date{\today }

\maketitle

\begin{abstract}
        We study the one-loop corrections to Vector Boson Scattering (in particular $W^+W^-$ elastic scattering) within the framework of effective theories. 
        Re-scattering via intermediate electroweak would-be-Goldstone bosons dominate at high energies, as the corresponding loop diagrams with these intermediate bosons scale like $\mathcal{O}(s^2/v^4)$ in the chiral effective counting.  
    In the present article, we focus our attention on fermion-loop corrections which  scale like $\mathcal{O}(M_{\rm Fer}^2 s/v^4)$ in the Higgs Effective Field Theory (HEFT). Although this dependency is formally suppressed for $s\to\infty$ with respect to that from boson loops, the large top mass can lead to a numerical competition between fermion and boson loops at intermediate energies of the order of a few TeV.    
        For the study of these fermion effects we have calculated  the imaginary part induced by loops of top and bottom quarks in $W^+ W^-\to W^+W^-$ elastic scattering and compared it to the  loop contributions from purely bosonic loops. We have examined the dependence of both amplitudes on the effective couplings, allowing an $\mathcal{O}(10 \%)$ deviation from the SM. In some cases, boson loops dominate over top and bottom corrections, as expected. However, we find that there are regions in the space of effective parameters that yield a significant --and even dominant-- imaginary contribution from fermion loops. 
        In addition to our conclusions for the general HEFT, we also provide 
        analyses particularized to some benchmark points in the $SO(5)/SO(4)$ Minimal Composite Higgs Model. 
    
\end{abstract}

\section{Introduction}

The discovery of the Higgs boson in 2012  by CMS and ATLAS \cite{CMS:2012qbp,ATLAS:2012yve} has provided the last missing piece of the Standard Model (SM). Over the last decade,
in the absence of new direct signals which may suggest new physics (NP), much effort has been put into high precision tests of the SM through LHC data. The hope is that by observing small deviations we may be able to elucidate the underlying NP at higher energies.

Within this context, one of the main processes for this exploration is Vector Boson Scattering (VBS). Deviations from the SM arising from a strongly interacting electroweak symmetry breaking sector  (EWSBS)~\cite{Delgado:2013loa} are expected to enhance the scattering of the longitudinal components of $W$ and $Z$ bosons at high energies.   
In the absence of new states, the most general description of the NP is the so-called Higgs Effective Field Theory (HEFT), which is a sort of Higgs-equipped Electroweak Chiral Lagrangian (EChL)~\cite{Appelquist,EWChL-HEFT}. 
For VBS at a center-of-mass (CM) energy well over the $WW$ threshold ($\sqrt{s}\gg M_W$), an important tool is the Equivalence Theorem (ET)~\cite{ET}. 
This relates, up to $\mathcal{O}(M_W/ \sqrt{s})$ corrections, processes with longitudinal electroweak (EW) gauge bosons $W^\pm,Z$ and  amplitudes with EW would-be Goldstone-bosons (WBGB) $\omega^a$. By neglecting these $\mathcal{O}(M_W/ \sqrt{s})$ contributions, the so-called na\"ive Equivalence Theorem (nET), the calculation of the amplitudes gets highly simplified. For instance, in the case of this article, the more involved $W_L^+ W_L^-\to W_L^+ W_L^- $ computation would be traded for the simpler $\omega^+\omega^-\to \omega^+\omega^-$ calculation. Notice that in the nET we have replaced the external longitudinal gauge bosons with WBGB's, but all particles (gauge bosons and WBGB's) must be considered in the internal lines. However, WBGB's interact through derivative operators and formally dominate at high energies in strongly interacting models. For this reason, it often works in this framework  to consider only WBGB loops as a sensible first approach to the problem~\cite{DelgadoLopez:2016cty}.~\footnote{
 It is important to note that the full --generalized-- Equivalence Theorem also  provides the subdominant corrections~\cite{ET} and an exact relation can be established at the price, nonetheless, of making the computation 
 more involved.}

In HEFT, leading order (LO) contributions to the amplitude appear at tree-level and scale like $\mO(p^2/v^2)\sim \mO(s/v^2)$.  
At next-to-leading-order (NLO) in HEFT's chiral expansion, the amplitudes get $\mO(p^4/v^4)$ corrections, with $p$ representing soft scales of the low-energy effective theory (masses, CM energy, etc.). More precisely, WBGB loops are NLO in the chiral expansion and scale like $\mathcal{O}(s^2/v^4)$, whereas fermion loops show an $\mathcal{O}(M_{\rm Fer}^2 s/v^4)$ dependence. Hence, the latter are usually neglected: WBGB loops will produce stronger deviations from the SM as we increase the center-of-mass energy.

In the present article, we provide a systematical quantitative study of the importance of these fermion-loop contributions to the $W^+W^-$ scattering at the energies relevant at LHC within the context of the HEFT. 
During this analysis, we have realized that in some cases an accurate calculation of the boson loops requires going beyond the nET. At high energies, HEFT models with Higgs couplings very close to the SM ones have boson loop contributions which are identically zero in the zero mass limit, $M_{W,Z,h,{\rm Fer}}\to 0$.

Hence, in this situation, the deviations from the SM enter in numerical competition with the corrections to the nET. 
For this reason, in this article we go beyond the nET and perform the analysis of $W^+W^-$ scattering, rather than $\omega^+\omega^-$. 
Some preliminary results in the nET were provided in Ref.~\cite{Dobado:2021ozt}.

It is well known that  fermion-loops are proportional to the masses of the particles in the EW fermion doublet inside the loop and to their Higgs effective couplings. 
Experimentally, Higgs-fermion couplings are still allowed for deviations within a $\pm\mO( 10 \%)$ with respect to the SM values or larger~\cite{pdg}. 
We will focus on the heaviest quark doublet, given by the $(t,b)$ quarks,  but results can be extended to the remaining Standard Model EW doublets in a straightforward way.  Nevertheless, they will be numerically negligible because their masses are much smaller than the Higgs vacuum expectation value ($M_{\rm Fer}\ll v\approx 246$~GeV).

In this work we will focus on the imaginary part; since it first appears in the scattering amplitude at NLO in the low-energy chiral counting, this imaginary part is not masked by the purely real LO amplitude or the real tree-level corrections at NLO, determined by additional counter-terms. The quantity of interest in this article will be the ratio of fermion and boson loop contributions to the imaginary part of the scattering amplitude. More specifically, we will study the first two partial wave amplitudes (PWA), $J=0$ and $J=1$. For this we will make use of perturbative unitarity which connects the imaginary part of an intermediate {\it two-particle}-loop contribution with the amplitude of tree-level processes with the same {\it two-particle} as a final state. 
The calculation and study of the real parts of these one-loop amplitudes will be provided elsewhere~\cite{in-preparation}. 

The custodial limit (sometimes called isospin limit) also provides a convenient approximation to our calculation. By neglecting explicit custodial breaking terms in the HEFT Lagrangian, expressions are simplified and calculations become in general simpler. However, we have two sources of custodial symmetry breaking. In the first place, the components of the EW fermion doublets have very different masses ($M_t\neq M_b$). 
In addition, $g'\neq 0$ introduces a small custodial symmetry violation which leads, e.g., to the EW gauge boson mass difference ($M_W\neq M_Z$). 
In this article, we will always consider the physical top and bottom masses while the custodial breaking due to the $U(1)_Y$ coupling will be neglected in a first approximation to the problem ($(M_Z^2 -M_W^2)\ll M_W^2$). 
This $g'=0$ limit makes the analysis simpler and clearer as the number of intermediate channels is much smaller (photons decouple when $g'\to 0$). 
However, we will later complement this computation with the full calculation for $g'\neq 0$, finding similar results.

 In this article  we have concluded previous preliminary  studies~\cite{Dobado:2021ozt,Dobado:2020lil} by including all possible two-particle intermediate physical states for the elastic $W^+W^-$ scattering, 
 including all possible intermediate gauge boson polarizations, and with $M_t\neq M_b$ and $g'\neq 0$. Thus, the available two-particle absorptive cuts are $t\bar{t}$ and $b\bar{b}$ in the case of fermionic cuts, and $W^+W^-$, $ZZ$, $hh$, $Zh$, $\gamma\gamma$, $\gamma Z$ and $\gamma h$ for bosonic intermediate states.  


\section{Electroweak Chiral Lagrangian}

In this Section we present the relevant EW Chiral Lagrangian for the elastic $WW$ scattering analysis discussed in this article. However, in a first approximation, we approached the study by making use of the nET, where the longitudinal gauge bosons in the external legs of the amplitude are replaced by EW Goldstone bosons, which is accurate up to $M_{W,Z}/\sqrt{s}$ corrections. Though not stated in the theorem, it is also common in the literature to ignore gauge boson intermediate exchanges in this ET approximation, considering only scalar exchanges (Goldstone boson interactions carry additional derivatives in their interaction with respect to the gauge boson ones). Thus, individual scattering diagrams with Goldstone vertices grow like $E^2$, eventually violating the unitarity bound. 
Nevertheless, in the exact SM limit there is a fine cancellation between the various contributions to the total amplitude, which behaves like $E^0$; the unitarity bound is always preserved and the theory is renormalizable. In that SM limit, the contribution of the intermediate gauge boson exchanges is crucial.    
Hence, in BSM scenarios that are nonetheless close to the SM these contributions cannot be ignored. Moreover, for energies below the TeV, near the $WW$ production threshold, the corrections to the nET eventually become important. For these two reasons, we have also performed  the present analysis beyond the nET limit: in addition to the $\omega^+ \omega^- $ 
scattering in the nET, we have also computed the actual $W_L^+W_L^- $  
longitudinal gauge boson scattering. Although we will focus on the latter, we will briefly discuss the difference in the following subsection.

\subsection{\bf Effective Lagrangian in the equivalence theorem limit}

In this first approach we will just consider in our EFT description the scalar bosons and the fermions we are interested in. Since the fermion contributions will be proportional to the masses of the fermions in the weak doublets, we will only include the top and bottom quarks in the effective Lagrangian below. The remaining fermions nonetheless can also be incorporated into the theory in a straightforward way, if required. 

At leading order (LO), $\mathcal{O}(p^2)$, the relevant part of our effective Lagrangian is given by~\cite{Appelquist,EWChL-HEFT,deFlorian:2016spz,Pich:2018ltt,Herrero:1993nc,Pich:2016lew,Buchalla:2012qq}:
\begin{eqnarray}
\mL_{2} &=& \mL_S \, +\, \mL_{\rm kin-F}\, +\, \mL_{\rm Yuk}\, , 
\label{eq:HEFT-Lagr-ET}
\end{eqnarray}
where,
%
\begin{eqnarray}
\mathcal{L}_S  &=& 
%
\frac{v^4}{4} \mathcal{F}(h) {\rm Tr}\{ \partial_\mu U^\dagger \partial^\mu U\} 
+\frac{1}{2}  \partial_\mu h \partial^\mu h  - V(h) \, ,
\label{eq:S-Lagr-ET}
\\
\mL_{\rm kin-F} &=&  i \bar{t} \partial \hspace*{-0.17cm} \slash\,  t
\, + \,  i \bar{b} \partial \hspace*{-0.17cm} \slash\,  b \,,
\\
\mathcal{L}_{\rm Yuk} & =&-\mathcal{G}(h) \bigg [\sqrt{ 1-\frac{ \omega^2  }{v^2}}(M_t \bar{t} t+M_b \bar{b}b)
+i\frac{\omega ^0}{v}\left( M_t \bar{t}\gamma^5t-M_b\bar{b}\gamma^5b \right)  
\nonumber \\ 
&& \qquad \qquad + \, 
i\frac{ \sqrt{2} \omega ^+}{v}\left(M_b \bar{t} P_R b-M_t \bar{t} P_L b   \right) \, +\,  
   i\frac{ \sqrt{2} \omega ^-}{v}\left(M_t \bar{b} P_R t-M_b \bar{b} P_L t   \right) \bigg ] \, , 
\label{eq:Yuk-Lagr}
\end{eqnarray}
with $\mathcal{L}_{\rm Yuk}$ providing the Yukawa interactions between fermions and scalars~\footnote{ The Yukawa Lagrangian provided in Eq.~(\ref{eq:Yuk-Lagr}) is indeed the general chiral expression of the Yukawa interaction ${ \mL_{\rm Yuk}= - \mG(h) \overline{Q}_L \, U\, M_Q\, Q_R + \mbox{h.c.}}$ 
expressed in the spherical coordinate coset representation $U=\sqrt{1-\omega^2/v^2} + i\omega^a \sigma^a/v$ that we will be using throughout  the article, with $Q^T=(t\,,\, b)$ and $M_Q=$diag$(M_t,M_b)$.} 
($h$ is the Higgs, $\omega^a$ the WBGB 
fields with  $\omega^2 =\sum _j(\omega_j)^2$), 
%
%
%
$P_{R,L}=\frac{1}{2}(1\pm \gamma_5)$ are the chirality projectors and $v\simeq 246$~GeV. For the Goldstones in Eq.~(\ref{eq:S-Lagr-ET}) we are using in this article the coset representation $U=\sqrt{1-\omega^2/v^2} + i\omega^a \sigma^a/v$~\cite{Dobado:1997jx}. In front of these operators, symmetry invariance allows us to insert a general function of the Higgs field singlet $h$ with an analytical expansion of the form, 
\begin{equation}
    \mathcal{G}(h)=1+{c_1} \frac{h}{v}+ ..., \quad  \mathcal{F}(h)=1+2{a} \frac{h}{v}+{b} \frac{h^2}{v^2}+... \quad\text{and}\quad V(h)= \frac{M_h^2}{2}  h^2+ d_3 \frac{M_h^2}{2v}  h^3+ d_4 \frac{M_h^2}{8v^2}  h^3+...
\end{equation}
%
%
%
%
In the SM case, one has $a=b=c_1=d_3=d_4=1$,  
and zero for any higher powers of $h$. 
These couplings $(a,b,c_1,d_3)$ are the only relevant parameters in $\mF(h)$, $\mG(h)$ and $V(h)$ for the present $W^+W^-$ elastic scattering study.

As it was mentioned in the introduction, 
we first computed the boson loop contributions to $WW$ scattering in the context of the nET, neglecting diagrams with gauge bosons in the intermediate internal lines.  
For strongly interacting beyond-SM (BSM) scenarios with $a\neq 1$, these nET simplified calculations do reproduce well the behavior of $W_L W_L$ scattering.  
However, in the SM case, the nET does not recover the right prediction for $W_L W_L\to W_LW_L$ scattering if intermediate gauge boson exchanges are not taken into account and the $\omega\omega\to \omega\omega$ scattering fail to yield the precise prediction in this important case. Hence, we move beyond the nET and compute the loop contributions, including physical gauge bosons in the tree-level calculation of this amplitude.   
This full tree-level amplitude  $W_L W_L\to W_LW_L$ has already been used for the $g'=0$ case in Ref.~\cite{Dobado:2021ozt}.

\begin{figure}[!t]  
\begin{subfigure}{.5\textwidth}
  \centering
  \includegraphics[width=.9\linewidth]{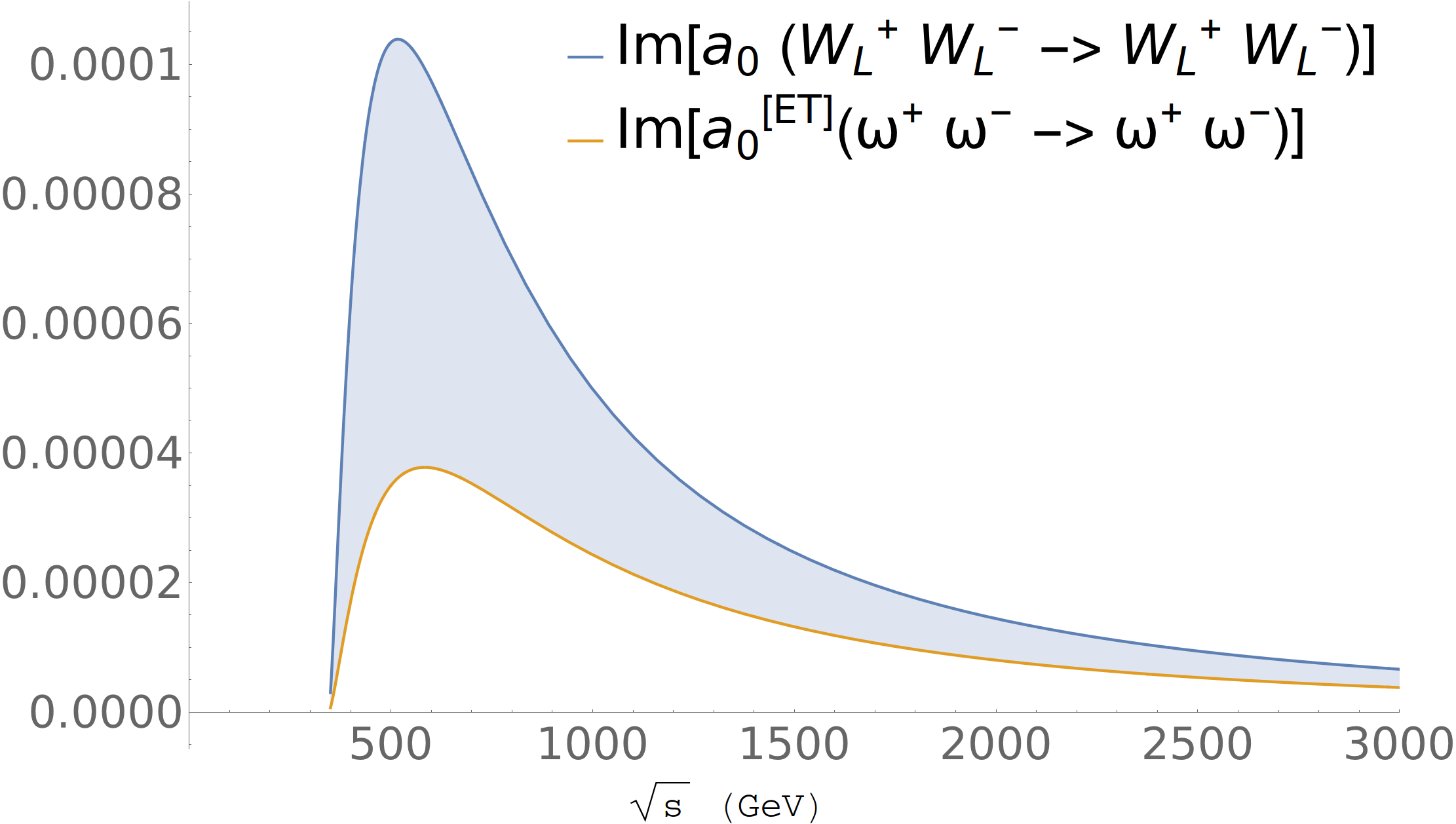}  
  \caption{}
  \label{fig:j0fullgaugevset}
\end{subfigure}
\begin{subfigure}{.5\textwidth}
  \centering
  \includegraphics[width=.9\linewidth]{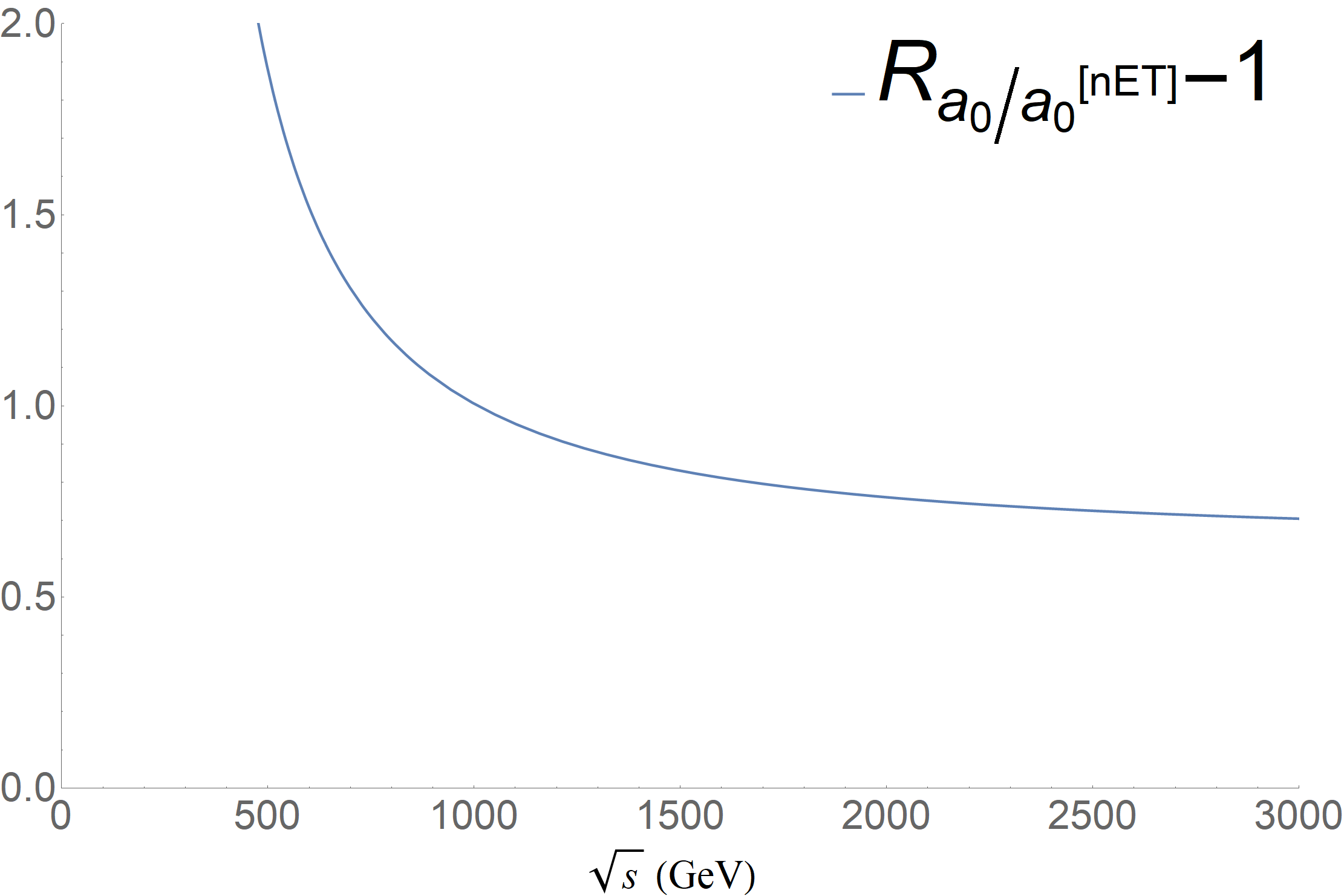}  
  \caption{}
  \label{fig:ratiofullgaugevset}
\end{subfigure}
\caption{\small Left: imaginary part of the SM top quark one-loop diagram in the $a_{J=0}$ partial wave amplitude (coming from tree-level $\mathcal{A}(W_L^+W_L^- \rightarrow t \bar{t})$ and its ET analogue partial wave $a_{J=0}^{\rm [ET]} $ (given by tree-level $\mathcal{A}(\omega^+ \omega^- \rightarrow t \bar{t})$ in the Feynman-t'Hooft gauge). Right: ratio ${ R_0=Im [a_0]/Im[a_0^{\rm [ET]}]  }$ of these same partial wave amplitudes.  
}
\label{fig:fullgaugevset}
\end{figure}

Regarding the fermion contribution via the tree-level scattering $\mA(W_L^+W_L^- \to f\bar{f})$, we have first reproduced the results in Ref.~\cite{Dawson:1990ux} for $\omega^+\omega^- \to f\bar{f}$.~\footnote{ Notice that the $\omega^+\omega^- \to f\bar{f}$ amplitude in Ref.~\cite{Dawson:1990ux} does not include gauge bosons in the internal lines. We have also removed these contributions in our calculation for the comparison with this work.}
Nonetheless, we find that the nET shows further complications in this case. 
The SM amplitude with same-sign fermion helicities ($++,\, --$) does not match its nET counterpart $\mA(\omega^+\omega^- \to f\bar{f})$ at high energies.
%
In the limit when $M_W/\sqrt{s}\to 0$, at fixed $s$ and $M_t$, we find that this amplitude coincides with the corresponding WBGB scattering. 
However, this is different to  the high-energy limit $s\to\infty$, at fixed $M_W$ and $M_t$, where we have found an important discrepancy between both amplitudes in the SM case. 
This difference can be observed directly in the first partial wave amplitude $a_{J=0}$. Fig.~\ref{fig:fullgaugevset} shows the SM $J=0$ PWA for $W_L^+W_L^-\to t\bar{t}$ and $\omega^+\omega^-\to t\bar{t}$, where one can see that the difference is significant even at high energies (of the order of $ 70\%$).  
%
Fortunately, it is possible to recover 
the full $W_L^+W_L^- \rightarrow f \bar{f}$ amplitude
if instead of applying the nET one employs the full generalized ET~\cite{ET,Dobado:1997jx}. 
Similar concerns about the nET were raised in previous works when dealing with WBGB amplitudes, effective Lagrangians and possible heavy scalars~\cite{espriu,Pal:1994jk,Grosse-Knetter:1994lkr}. The discussion of this topic is beyond the scope of this article and it is relegated to a future work~\cite{in-preparation}.

It is important to remark 
that this high-energy discrepancy only occurs in the SM  
due to a fine cancellation. 
For BSM theories with  $a c_1\neq 1$, the nET works well and the WBGB scattering amplitudes reproduce the longitudinal gauge boson scatterings at high energies. 
For instance, we find that in BSM scenarios $\mathcal{A}(W^+W^- \rightarrow t \bar{t})\approx \mA(\omega^+ \omega^- \rightarrow t \bar{t})\sim \sqrt{N_C }(1- a c_1)\sqrt{s} M_t/v^2 $ for $s\gg M_W^2,\, M_t^2$.  
However, the latter leading term is cancelled in the SM. The first non-vanishing contributions for both amplitudes differ at high energies by a term $\propto \sqrt{N_c} \frac{M_t M_W ^2}{\sqrt{s} v^2}$.

In summary, given all these considerations, we will always be working with the actual $W^+_{L}W^-_{L}$ scattering amplitudes, both for bosonic and fermionic intermediate absorptive cuts.

\subsection{\bf Effective Lagrangian beyond the equivalence theorem limit}

Ultimately, for a study beyond the ET, one must also add the EW gauge boson interactions to the EChL~\cite{Appelquist}. Thus, the relevant part of the  LO, $\mO(p^2)$, Lagrangian for the $WW$ study in this article is given by~\cite{Appelquist,EWChL-HEFT,deFlorian:2016spz,Pich:2018ltt,Herrero:1993nc,Pich:2016lew,Buchalla:2012qq},
\begin{eqnarray}
\mL_{2} &=& \mL_S\,  +\,  \mathcal{L}_{\rm Yuk} \,+\, \mL_{\rm kin-F} \, \, +\, \mathcal{L}_{\rm YM}\, \, , 
\label{eq:HEFT-Lagr}
\end{eqnarray}
with 
\begin{eqnarray}
\mathcal{L}_S  &=& \frac{v^4}{4} \mathcal{F}(h) {\rm Tr}\{( D_\mu U)^\dagger D^\mu U\} 
+\frac{1}{2}  \partial_\mu h \partial^\mu h  - V(h) \,,
\label{eq:S-Lagr}
\\
\mL_{\rm kin-F} &=&  i \bar{t}  D \hspace*{-0.2cm} \slash\, t
\, + \,  i \bar{b}  D \hspace*{-0.2cm} \slash\, b \,,
\end{eqnarray}
where the covariant derivatives in $\mL_{\rm kin-F}$ and $\mL_S$ now contain the couplings with the EW gauge bosons, $\mathcal{L}_{\rm YM}$ is the standard $SU(2)_L\times U(1)_Y$ Yang-Mills Lagrangian and $\mathcal{L}_{\rm Yuk}$ is the previous Yukawa Lagrangian in Eq.~(\ref{eq:Yuk-Lagr}).

\section{Loop corrections to elastic $W_L^+W_L^-$ scattering}

Starting from this Lagrangian, we have computed the fermion-loop contribution to the elastic $W^+_LW^-_L$ scattering amplitude for the ($t,b)$ quark doublet. At LO in the chiral expansion, $\mO(p^2)$, the amplitude $\mathcal{A}_2$ is purely real and it is given by tree-level diagrams made from $\mL_2$ vertices. Its first correction, $\mathcal{A}_4$, shows up at $\mO(p^4)$ in the chiral counting. It acquires a real tree-level contribution $\mathcal{A}_{4,\rm tree}$ from the corresponding effective couplings in the next-to-leading order (NLO) Lagrangian $\mL_4$ (namely $a_4$ and $a_5$). Likewise, one-loop diagrams made of $\mL_2$ vertices also yield a $\mathcal{A}_{4,1\ell}$ contribution to the $\mO(p^4)$ amplitude and  provide the first contribution to the imaginary part of the amplitude $\mathcal{A}$.

Up to the order studied in this work, $\mO(p^4)$, the real part of the amplitude is provided by the aforementioned three contributions, $\mbox{Re}\mathcal{A}=\mathcal{A}_2+\mathcal{A}_{4\rm tree}+ \mbox{Re}\mathcal{A}_{4,1\ell}$.  
This makes the study of the NLO one-loop corrections cumbersome. On the other hand, the imaginary part only receives contributions from one-loop diagrams up to this order, $\mbox{Im}\mathcal{A}=\mbox{Im}\mathcal{A}_{4,1\ell}$. This makes the study of the importance of fermion corrections much simpler and clearer, and it will be the procedure followed in this article.   
More specifically, we will be studying the imaginary part of the projected Partial Wave Amplitudes (PWA) $a_J(s)$, 
with 
\begin{equation} 
\mathcal{A}(s,t)=\sum_J 16\pi K (2J+1) P_J(\cos\theta) \, a_J(s)\, ,
\end{equation} 
with $K=1$ ($K=2$) for distinguishable (indistinguishable) final particles. In the physical energy region, Im~$a_J(s)$ will be provided by the one-loop absorptive cuts in the $s$-channel, which we will use to label the various contributions.

In scattering amplitudes with only bosons in the external legs, it is possible to clearly separate fermion and boson loops.  
We will measure the relevance of each of these two contributions. For this, we will use the following notation to refer the corresponding absorptive cuts:
\begin{eqnarray}
 {\rm Fer}_J&=&\mbox{Im}\, a_J|_{b\bar{b}      ,t\bar{t}}    \, ,\nonumber\\ 
 {\rm Bos}_J&=&\mbox{Im}\ a_J|_{\gamma \gamma, \gamma Z, \gamma h,W^+ W^-,ZZ, Zh, hh}      \, .
 \label{eqn:cuts}
\end{eqnarray}
Notice that the channels are arranged by increasing mass, as they will be presented later in the figures.
%
%
%
The absorptive cuts with intermediate longitudinal vector bosons $WW$ and $ZZ$, and $hh$  can be found in Refs.~\cite{espriu,wwzz,wwhh}, respectively. 
The rest are provided in the Appendix. In this work we have not only included the contribution from intermediate longitudinal modes but also the transverse ones.
Beyond the nET approximation 
there are also contributions from the intermediate channels $Zh$ that we did not include in a previous work \cite{Dobado:2021ozt,Dobado:2020lil}.

However, in the massless limit, all the mentioned one-loop corrections contain forward ($\cos\theta=1$) and/or backward ($\cos\theta=-1$) divergences. In Bos$_J$ these singularities arise in the limit $M_W,\, M_Z\to 0$ due to the exchange of $W,Z,\gamma$ gauge bosons in crossed channels (as the photon is massless, one always finds a forward divergence for the $W^+W^-$ intermediate cut). On the other hand, the amplitudes with intermediate $t\bar{t}$ and $b\bar{b}$ absorptive cuts have a forward divergence for $M_b\to 0$ and $M_t\to 0$, respectively.    
One can also identify a distinctive pattern for this large forward/backward contribution to the different partial waves: the singular behavior of the $ZZ$, $hh$, $\gamma\gamma$ and $\gamma h$ channels is only relevant for even $J$; the forward/backward divergences of the $Zh$ and $\gamma Z$ cuts arise just for odd $J$; finally, since in the massless limit, the $t\bar{t}$, $b\bar{b}$ and $W^+W^-$ channels only have forward divergences, they are relevant both for even and odd $J$ PWA.

In general, the non-zero mass of weak gauge bosons and fermions regulate the indicated divergences, except for one present in the $W^+W^-$ absorptive cut. For the latter, we encounter a divergent diagram arising from the exchange of a photon in the $t$-channel, making its PWA projection integral divergent at $\cos\theta\to 1$.  
To confront this issue, we will consider two strategies:  
\begin{enumerate}

     \item {\bf Assume $g'=0$ and integrate over the whole solid angle:}  
     in this scenario, $M_Z=M_W$  and thus, the photon decouples ($e=g'\cos\theta_W=0$) because the $W^+W^-$ cut  forward photon divergence is absent. 
     In addition, the   
     $\gamma\gamma$, $\gamma Z$ and $\gamma h$ channels  
     vanish, simplifying the analysis. 
     In this case we can perform the complete angular integration and project onto PWA. This would correspond to the custodial limit but for the fact that we keep $M_t \neq M_b$. \\
     
     \item {\bf Impose angular cuts:} 
      In order to deal with the divergence from the $W^+W^-$ channel, we perform the PWA integration within the angular limits $\abs{\cos{\theta}} \leq (\cos\theta)_{\rm max}$ for the intermediate particles (with, e.g., $\cos\theta_{\rm max}=0.9$).   
     This approach allows us to go beyond the $g'=0$ limit, incorporating all the aforementioned cuts in Eq.~(\ref{eqn:cuts}). 
     We will refer to these amplitudes $\widetilde{a}_J(s)$ as pseudo-PWA (p-PWA). Although the p-PWA are now finite and well-defined (even in the massless limit), we note that they lose many of the interesting PWA properties: the clear separation of the different angular momenta no longer holds and analogous PWA unitarity relations fail.
     
\end{enumerate}

%

Moving on, it is important to note which particular couplings enter in each PWA:
\begin{equation} 
\begin{split}
\mbox{\bf $J=0$:} \qquad 
&{\rm Fer}_0 \quad \longrightarrow\quad  a,c_1 ,\\
&{\rm Bos}_0 \hspace{-0.067cm} \quad \longrightarrow\quad a,b, d_3, \\
\mbox{\bf $J=1$:} \qquad 
&{\rm Fer}_1  \quad\longrightarrow\quad  \mbox{no \  dependence on } a,\, b,\, c_1 \quad =\quad \mbox{SM,} \\ 
& {\rm Bos}_1 \hspace{-0.067cm} \quad\longrightarrow\quad  a\, .
\end{split}
\end{equation}

The main goal of the present work is to point out that there are regions of the parameter phase-space where fermion-loops become as important as the bosonic ones and should not be neglected. 
To this goal we introduce the ratio:
\begin{equation}
    R_J=\frac{{\rm Fer}_J}{{\rm Bos}_J+{\rm Fer}_J} .
\end{equation}

Values of $R_J$ close to zero will indicate that we can safely drop fermion-loops, while deviations from this value will point out the relevance of fermions in $WW$ scattering. Although it is commonly assumed that fermion-loops are negligible in most of the parameter space, we will see that this is not true for some particular channels and in some regions of the effective couplings. 

In the following, we will focus on the contributions from fermion-loops to the first two partial waves $J=0,1$. By using perturbative unitarity, we can write down the fermionic contribution to the one-loop imaginary part of the partial waves in terms of the tree level amplitudes $\mathcal{A}(W^+W^-\to F\overline{F})\equiv Q^{\Delta \lambda,F}$ (one for the production of each intermediate fermion state $F\overline{F}$), with  $Q^{0,\, F}=\frac{1}{\sqrt{2}}(Q^{++, F}-Q^{--,F})=\sqrt{2} Q^{++,F} $, $ Q^{+-,F}$  and $Q^{-+,F}$. For $J=0$ only the $Q^{0,F}$ combination is necessary for the partial-wave projection $Q_J^{\Delta \lambda,F}$, while for $J=1$ three ($Q^{0,F}$, $Q^{ +-,F}$, $Q^{-+,F}$) enter in the projection:
%
%
%
\begin{eqnarray}
     \rm{Fer}_0&=& \mbox{Im}\  a_0(s)\bigg|_{t\bar{t},b\bar{b}}  \,=\, \displaystyle{\sum_{F=t,b}}   \beta_F \, \left|Q^{0,\, F}_{0}\right|^2 \,  \theta(s-4M_F^2)  \, ,
\\
    \rm{Fer}_1&=&  \mbox{Im} \ a_1(s)\bigg|_{t\bar{t},b\bar{b}}  \,=\, \displaystyle{\sum_{F=t,b}}  
    \beta_F\, \left(\left|Q^{0,F}_1\right|^2+\left|Q^{+-,F}_1\right|^2+\left|Q^{-+,F}_1\right|^2\right)  \theta(s-4M_F^2) \, , 
%
\end{eqnarray}
where $\beta_F=\sqrt{1-4M_F^2/s}$ and the partial-wave projections are defined as~\cite{DelgadoLopez:2016cty}, 
\begin{equation}
Q^{\Delta \lambda }_J \, =\, \frac{1}{64\pi^2 K}\sqrt{\frac{4\pi}{2J+1}} \int  Q^{\Delta \lambda}(s,\Omega)\, Y^*_{J,\Delta \lambda}(\Omega)\, d\Omega\,,  
\end{equation}
where $Y_{JM}(\Omega)$ are the spherical harmonics  and $\Delta \lambda$ is the helicity difference $\Delta \lambda=\lambda_1-\lambda_2$, with the super-index $F$ omitted for simplicity. 

Additionally, we can also calculate the relative cumulative PWA, which we will denote as:

\begin{equation*}
    \chi^J_i=\frac{\displaystyle{ \sum_{n=1}^{i} \mbox{Im}\ a_j\, \bigg|_{n}  }}{ \mbox{Im}\ a_J}, 
\end{equation*}
 where $\rm N_{ch}$ is the total number of absorptive channels, Im $ a_J= \displaystyle{ \sum_{n=1}^{\rm N_{ch}} \mbox{Im}\ a_J\bigg|_n }$ is the total imaginary part of the $a_J$ PWA, and $\mbox{Im}\ a_J\bigg|_n$ represents the absorptive contribution from channel $n$  --either bosonic or fermionic-- which are  arranged in increasing order of their mass threshold. The analytical expression of the tree-level amplitudes that provide the imaginary part of the one-loop diagrams is rather lengthy and has been relegated to \ref{section:scatterin_gamplitudes}. 
 All calculations have been performed within arbitrary renormalizable $R_\xi$ gauges with parameters $\xi_W$, $\xi_Z$ and $\xi_A$. We have checked that the full amplitudes are gauge independent, as expected.



\section{Importance of fermion-loops in the $g'=0$ limit}
\label{sec:gpr=0}

We will start our phenomenological study by considering the $g'=0$ limit. In the absence of fermion masses, this implies that custodial symmetry is preserved.  Actually, fermion masses are not the problem but rather the mass splitting of the fermion multiplets: custodial symmetry is restored in the limit  $g'=0$ and $M_t=M_b$ (and similarly for each quark and lepton doublet). 
This approximate custodial/isospin symmetry is very useful to simplify and classify the contribution from bosonic channels, as the weak bosons turn into a degenerate multiplet ($M_Z=M_W=gv/2$, at LO), the $W^3$--$B$ mixing vanishes ($\tan\theta_W=g'/g=0$, at LO) and amplitudes with photons become zero (since $e=g'\cos\theta_W=0$, at LO). 
Custodial symmetry breaking corrections are proportional to $\sin^2\theta_W \sim 0.2$, which makes the isospin limit scenario a suitable first approximation to the problem. In the following section, we will go beyond this limit and consider the numerical relevance of the $g'\neq 0$ corrections.

Nonetheless, for the purpose of the phenomenological analyses in this article, we will never consider the true isopin limit, which also requires $M_t=M_b$. While it has been used for some theoretical checks of the analytical expressions, the large experimental hierarchy $M_t\gg M_b$ is crucial for the numerical studies of the cross section and any comparison with the experiment.

For this work we will assume a   10\% deviation on the parameters of the Effective Lagrangian. 
The relevant effective couplings for the present one-loop computation are $a$ ($hWW$), $b$ ($hhWW$), $d_3$ ($hhh$) and $c_1$ ( $t \bar{t}h$), with their corresponding vertices within the brackets. While the experimental values of $a$ and $c_1$ fall within this range \cite{deBlas:2018tjm,ATLAS:2020jgy}, $b$  and $d_3$  present a much wider uncertainty~\cite{Aad:2020kub}. Since the aim of this study is to call attention on the often neglected fermion corrections which are proportional to $c_1$, we will not use its precise experimental range. A 10\% deviation from the SM already shows their relevance and the need to include them in future calculations. 

Concerning the center-of-mass energy we have considered the interval  $0.5$~TeV$\leq \sqrt{s}\leq 3$~TeV, which is the relevant one to look for NP at the LHC. We will use as inputs: $M_W$=80.38 GeV, $M_Z$=91.19 GeV, $M_H$=125.25 GeV, $v$=246.22 GeV, $M_t$=172.76 GeV and $M_b$= 4.18 GeV~\cite{pdg}. The value of the Weinberg angle is found in the standard way from $M_W$ and $M_Z$, $\cos^2 \theta_W = M_W^2/M_Z^2$ at LO.

\subsection{$J=0$  PWA: $R_0$}

\begin{figure}[!t]  
\begin{subfigure}{.5\textwidth}
  \centering
  \includegraphics[width=.9\linewidth]{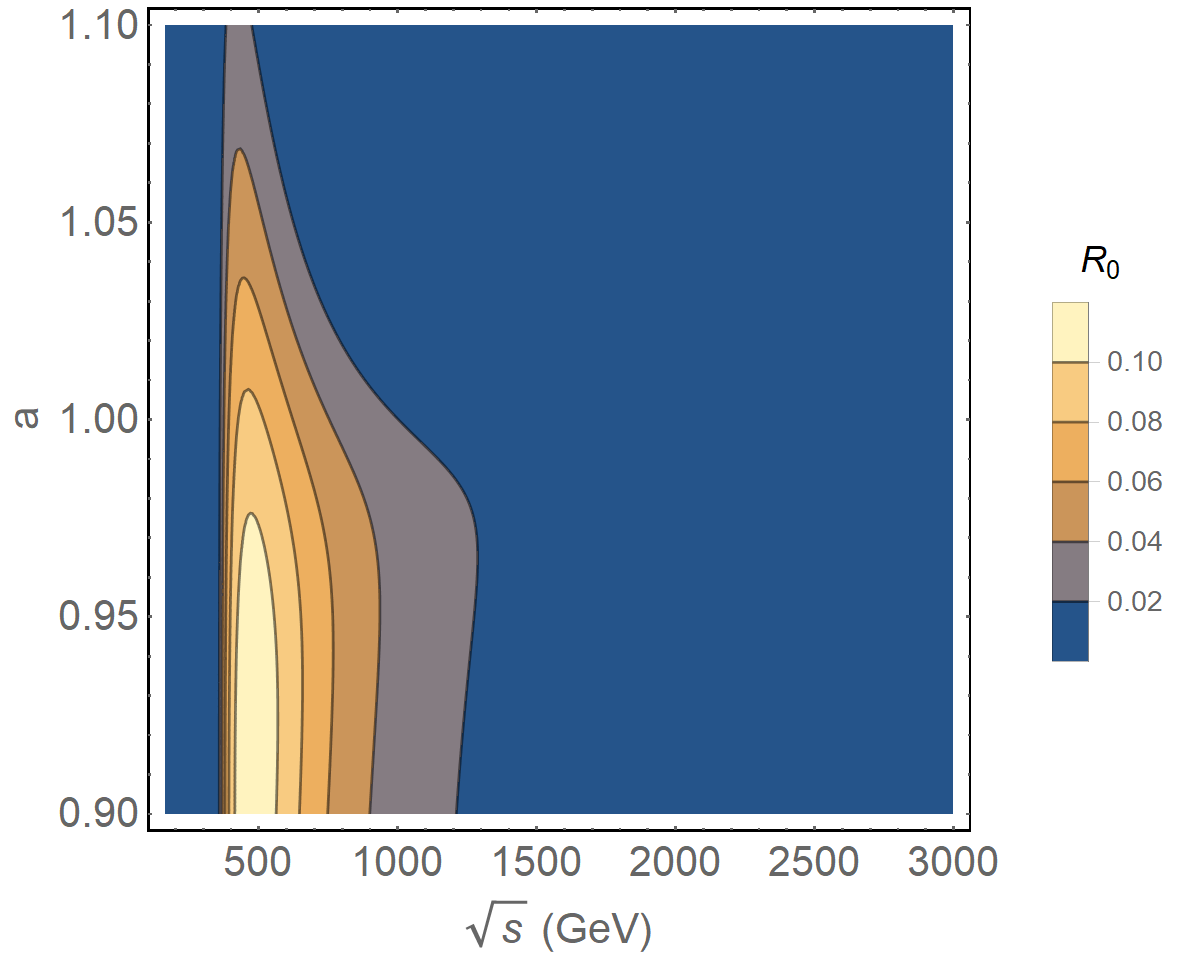}  
  \caption{$R_0$ dependence on $a$ for $b=c_1=d_3=1$.}
  \label{fig:r0_countour_a_isospin}
\end{subfigure}
\begin{subfigure}{.5\textwidth}
  \centering
  \includegraphics[width=.9\linewidth]{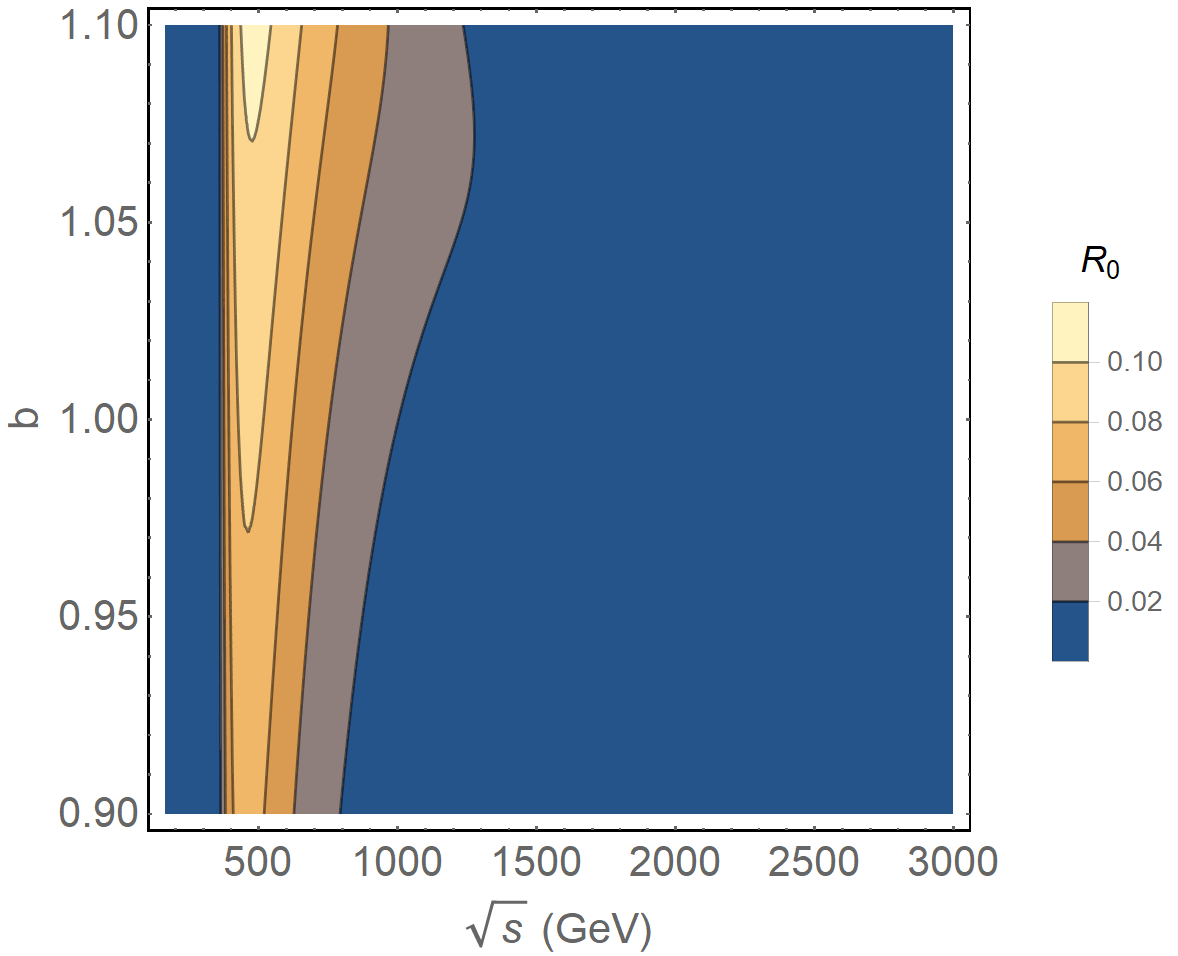}    
  \caption{$R_0$ dependence on $b$ for $a=c_1=d_3=1$.}
  \label{fig:r0_contour_b_isospin}
\end{subfigure}
\\[3ex]
\begin{subfigure}{.5\textwidth}
  \centering
  \includegraphics[width=.9\linewidth]{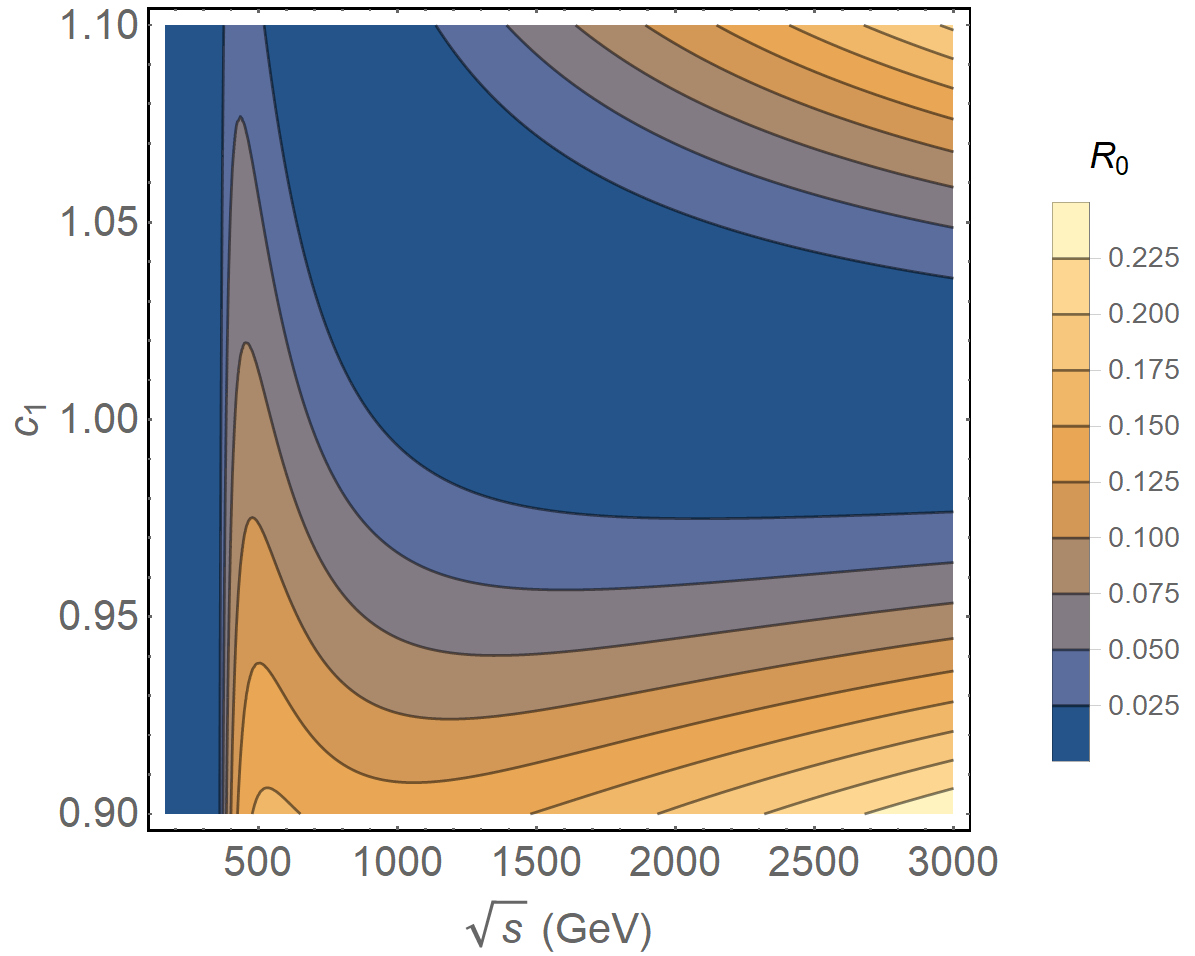}  
  \caption{$R_0$ dependence on $c_1$ for $a=b=d_3=1$.}
  \label{fig:r0_countour_c1_isospin}
\end{subfigure}
\begin{subfigure}{.5\textwidth}
  \centering
  \includegraphics[width=.9\linewidth]{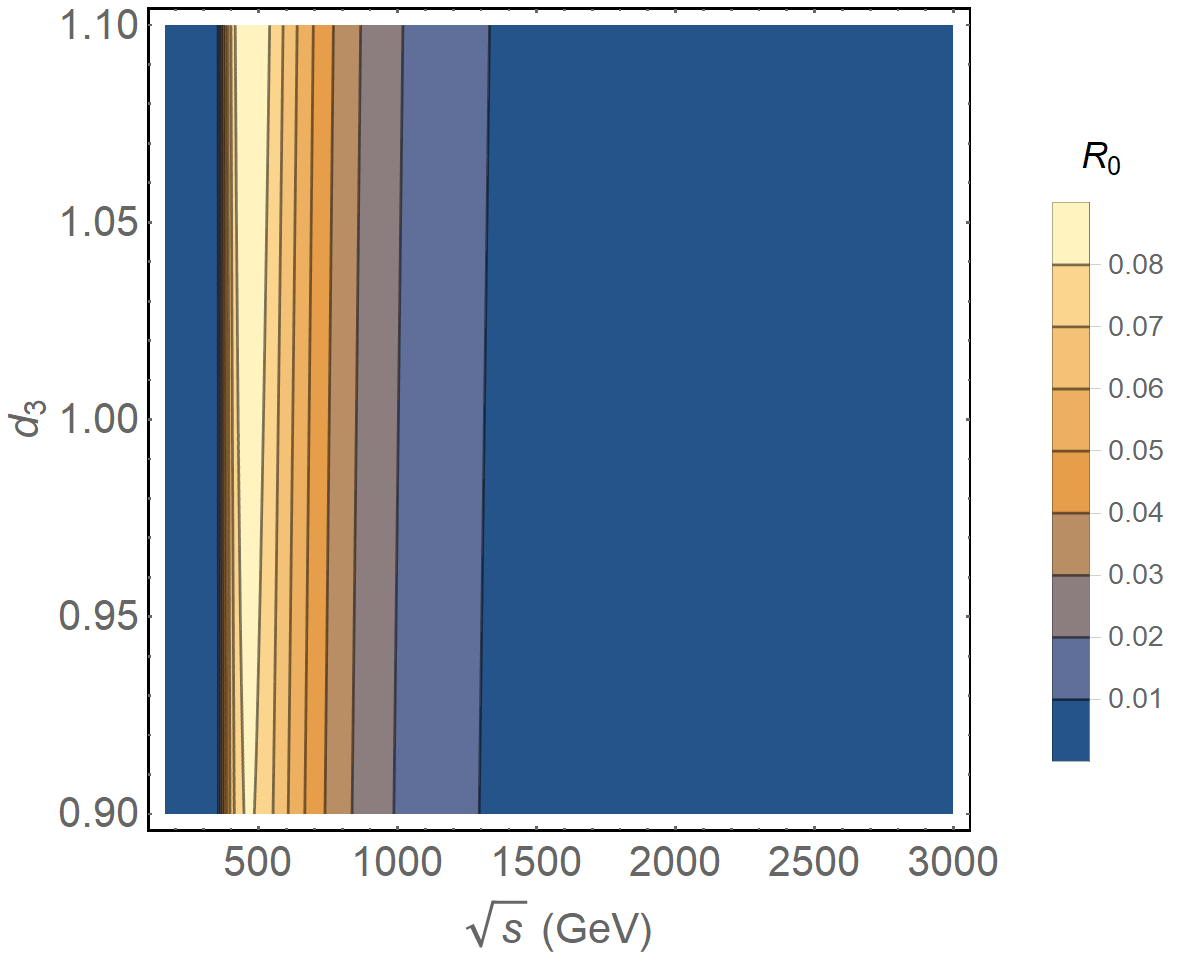}  
  \caption{$R_0$ dependence on $d_3$ for $a=b=c_1=1$.}
  \label{fig:r0_contour_d3_isospin}
\end{subfigure}
\caption{}
\label{fig:figr0contour}
\end{figure}

\begin{figure}[!t]    
\includegraphics[width=10cm]{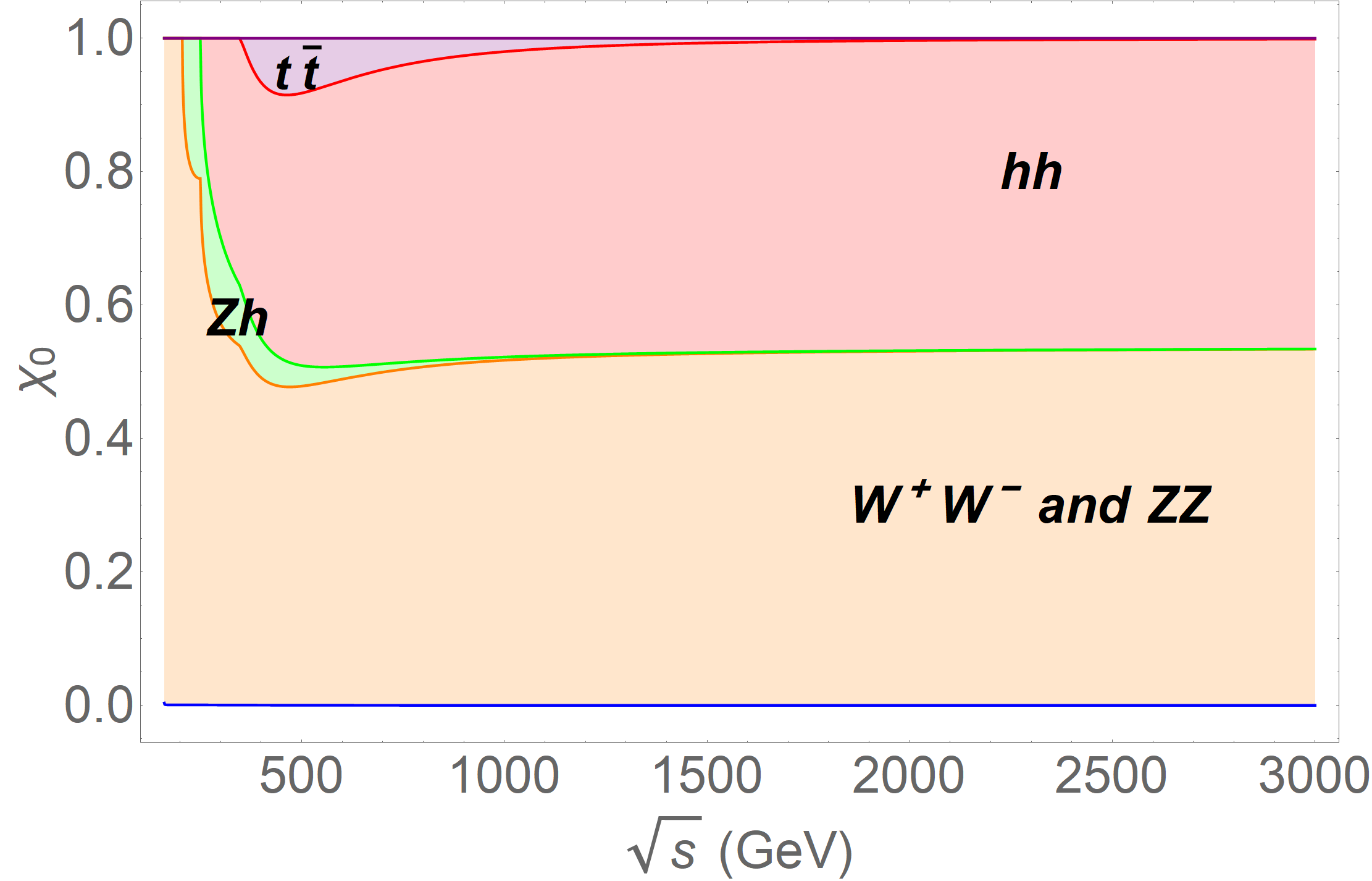}
\centering
\caption{\small Cumulative relative contribution of each channel to $J=0$ PWA in the SM.}
\label{fig:r0ratio_isospin_SM}
\end{figure}

In the  following plots, we have scanned the value of $R_0$ in the aforementioned region of the coupling space one parameter at a time, while keeping the others fixed to their SM values for reference.

As we see in Figs.~\ref{fig:r0_countour_a_isospin} and \ref{fig:r0_contour_b_isospin}, when we explore  $a$ and $b$, respectively, we find $\mathcal{O}(10 \%)$ corrections around $\sqrt{s}=500$~GeV. For $\sqrt{s}\gsim 1.5$~TeV bosons completely dominate, as expected. When it comes to the dependence on $c_1$, we can see in Fig.~\ref{fig:r0_countour_c1_isospin}, 22\% corrections at  $\sqrt{s} \sim$ 3 TeV when $c_1$ deviates from the SM. If we considered a broader phenomenological range for $c_1$ this correction would be even larger. For the case of $d_3$, we observe in Fig.~$\ref{fig:r0_contour_d3_isospin}$ fermion corrections of the order of 8\% around 500 GeV. Although in absolute terms $R_0$ barely changes with $d_3$,  it decreases when the center-of-mass-energy is increased. This lack of sensitivity is due to the fact that $d_3$ only enters in the $hh$ cut and via a non-derivative interaction.

\begin{figure}[!t]    
\centering
\begin{subfigure}{0.4 \columnwidth}
  \centering
  \includegraphics[width=1\linewidth]{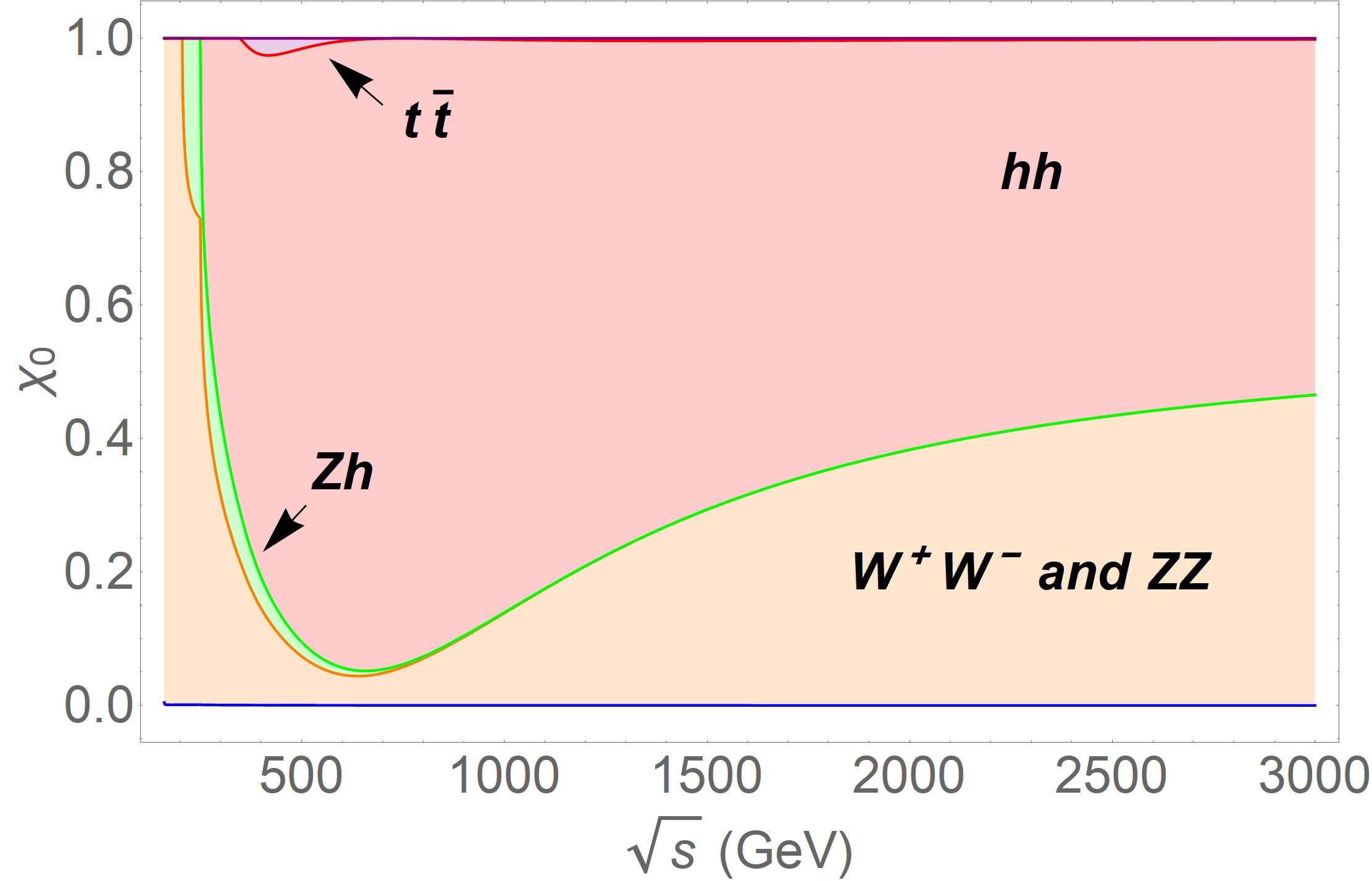}  
  \caption{$a=1.10$ and $b=c_1=d_3=1$.}
  \label{fig:r0_ratio_a_11}
\end{subfigure}
\hspace*{0.75cm}
\begin{subfigure}{0.4 \columnwidth}
  \centering
  \includegraphics[width=1 \linewidth]{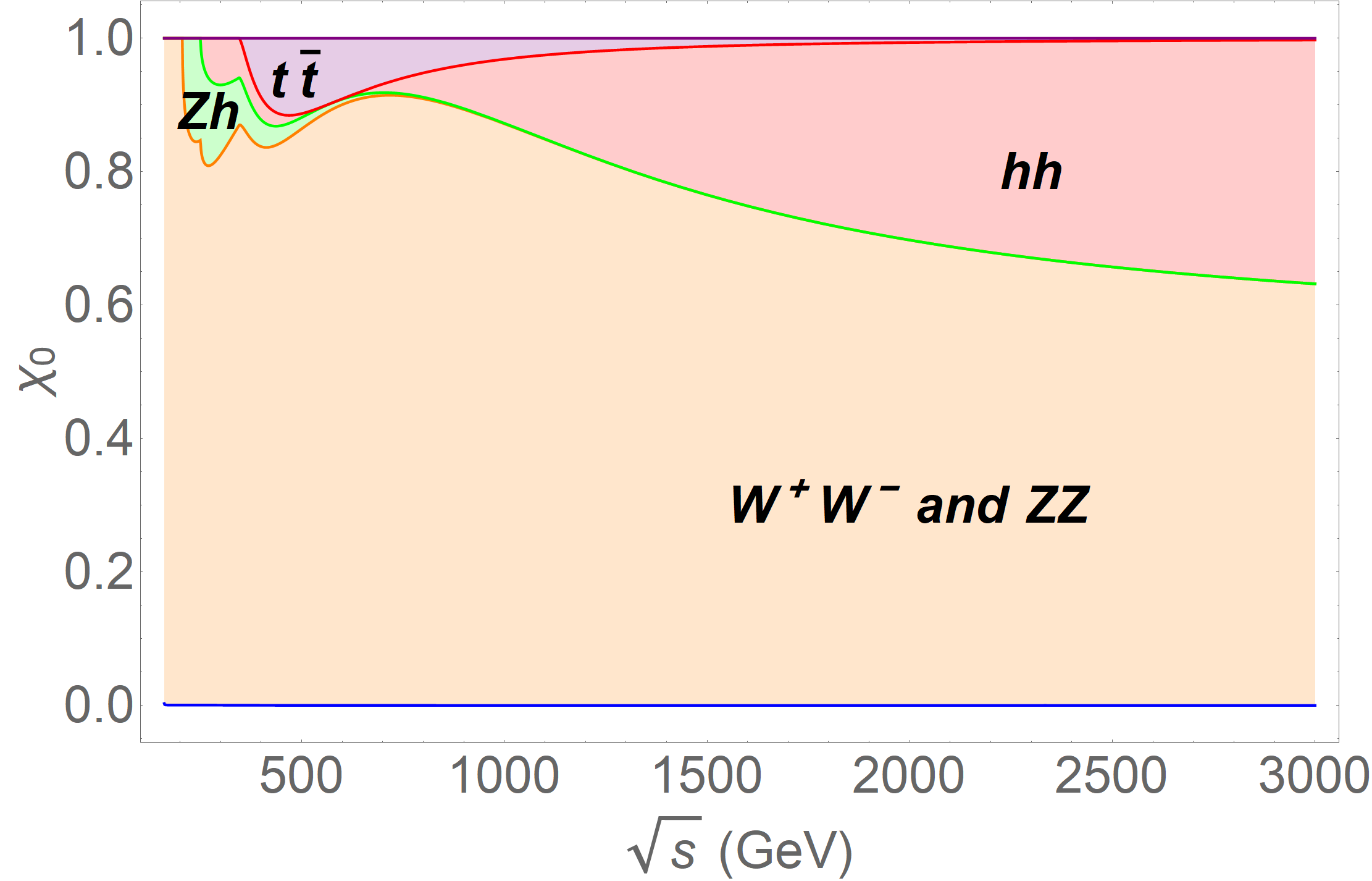}  
  \caption{ $a=0.90$ and $b=c_1=d_3=1$.}
  \label{fig:r0_ratio_a_09}
\end{subfigure}
%
\\[8pt]
\centering
\begin{subfigure}{0.4 \columnwidth}
  \centering
  \includegraphics[width=1\linewidth]{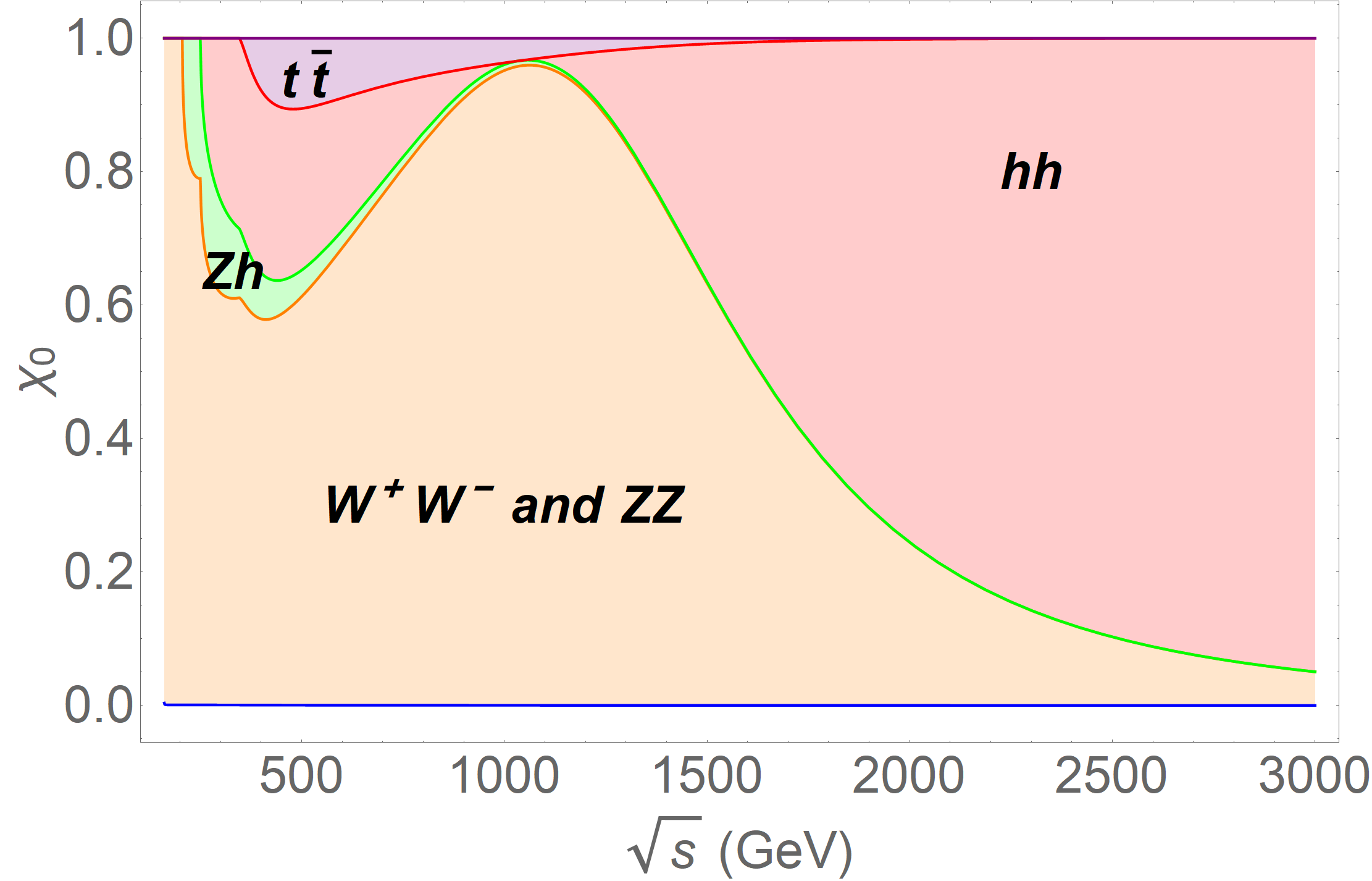}  
  \caption{ $b=1.10$ and $a=c_1=d_3=1$.}
  \label{fig:r0_ratio_b_11}
\end{subfigure}
\hspace*{0.75cm}
\begin{subfigure}{0.4 \columnwidth}
  \centering
  \includegraphics[width=1 \linewidth]{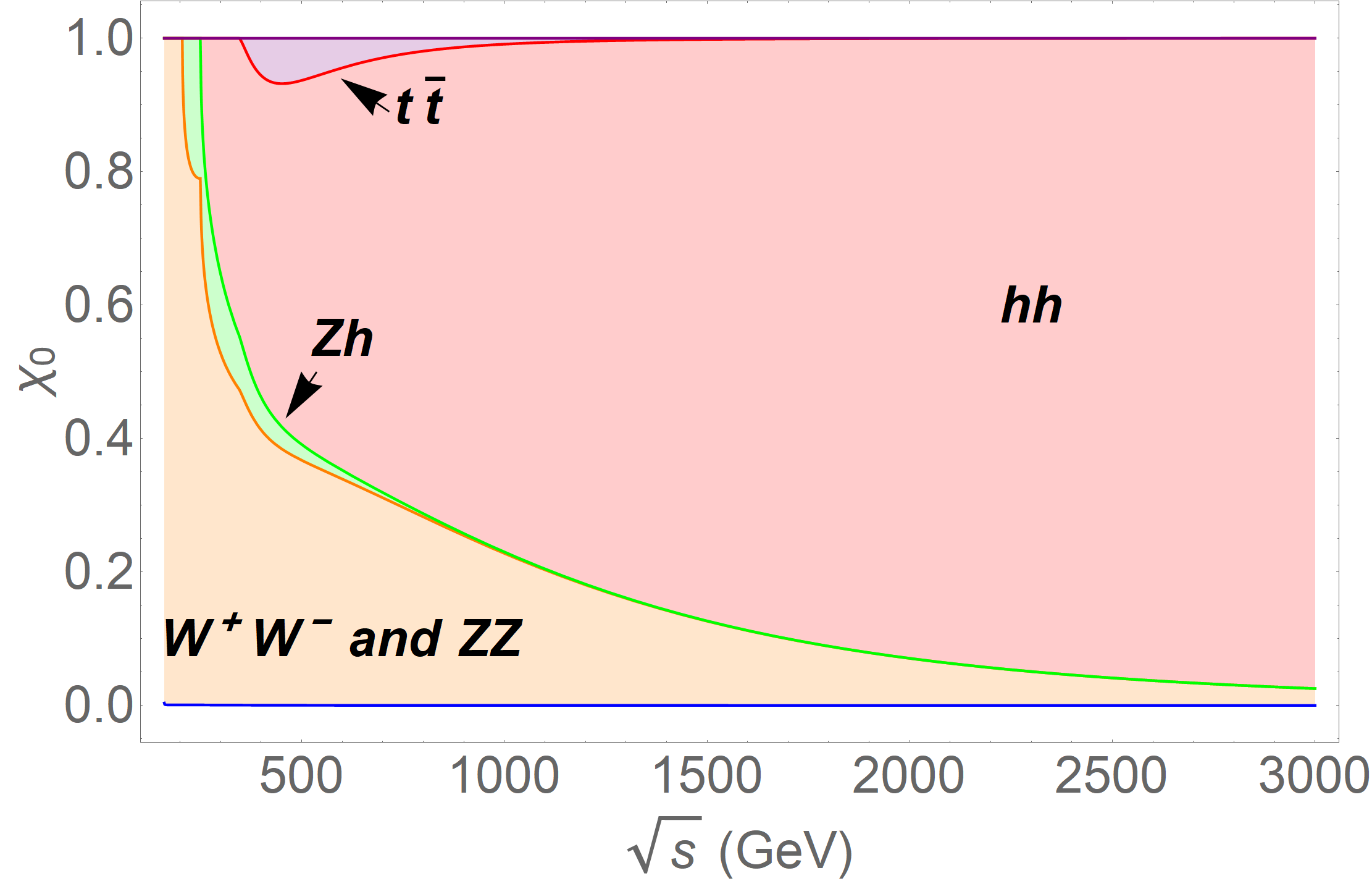}  
  \caption{$b=0.90$ and $a=c_1=d_3=1$.}
  \label{fig:r0_ratio_b_09}
\end{subfigure}
\\[8pt]
\centering
\begin{subfigure}{0.4 \columnwidth}
  \centering
  \includegraphics[width=\linewidth]{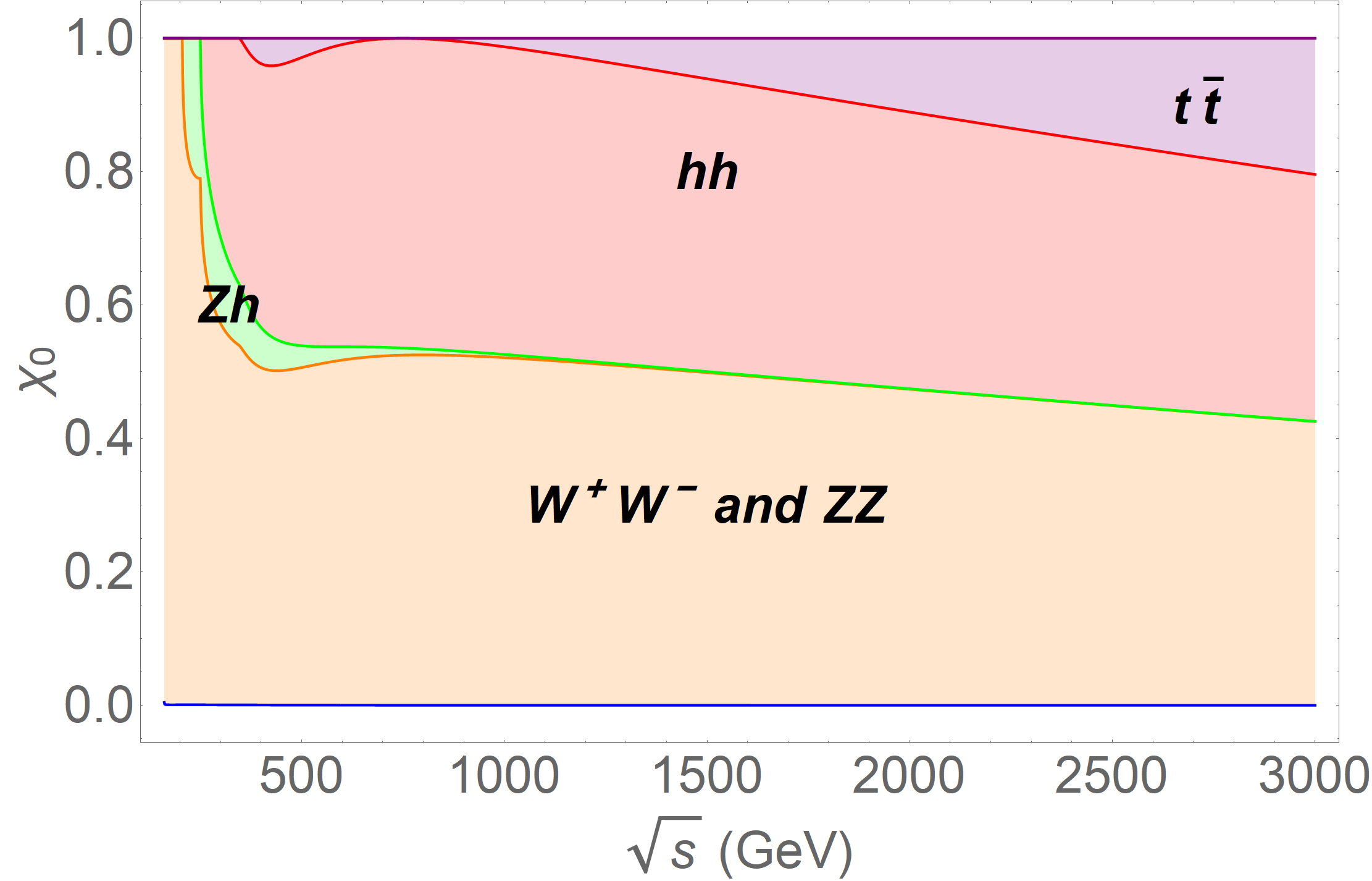}  
  \caption{ $c_1=1.10$ and $a=b=d_3=1$ . }
  \label{fig:r0_ratio_c1_11}
\end{subfigure}
\hspace*{0.75cm}
\begin{subfigure}{0.4 \columnwidth}
  \centering
  \includegraphics[width=\linewidth]{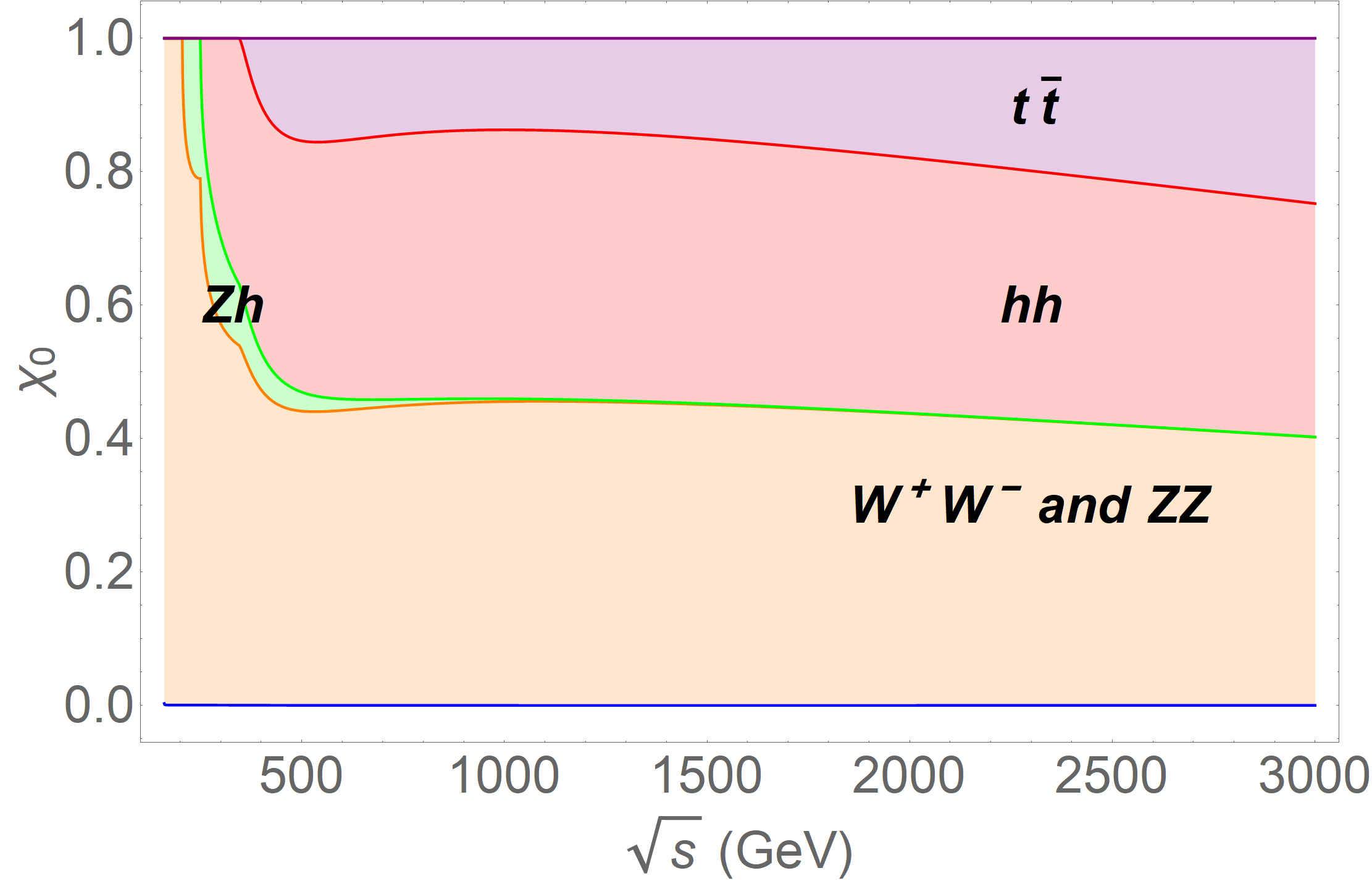}  
  \caption{ $c_1=0.90$ and  $a=b=d_3=1$ . }
  \label{fig:r0_ratio_c1_09}
\end{subfigure}
\caption{\small Cumulative relative contributions for each absorptive cut to the $J=0$ PWA for $a,b$ and $c_1$ at the borders of the considered parameter space. The $b\bar{b}$ contribution is numerically  negligible for this PWA.}
\label{fig:ratiosr0}
\end{figure}

From this analysis we extract that for $R_0$ the most relevant parameter is $c_1$. The further it is from its SM value, the larger the fermion contribution is, as expected from the analytical expression of the fermionic cuts.

It is also illustrative to show how each cut contributes to the total amplitude. These cumulative relative amplitude curves $\chi$ for the SM are shown in Fig. \ref{fig:r0ratio_isospin_SM}. Each curve contains the contribution of all intermediate cuts below the mentioned cut. They are ordered according to the value of the mass threshold of the intermediate state: the first cut is $b \bar{b}$, then $WW$ and $ZZ$ at the same energy ($g'=0$), $Zh$, $hh$ and finally $t \bar{t}$. Clearly, in the SM case, in Fig \ref{fig:r0ratio_isospin_SM}, top loop-corrections are only relevant around $\sqrt{s} \sim $500 GeV, reaching a maximum of $R_0 \sim$ 10\%. The $b\bar{b}$ cut is present (blue line at the bottom) but its contribution is absolutely negligible for $J=0$.

Now aware that $d_3$ is not relevant, we will explore the cumulative curves for BSM scenarios where $a$, $b$ and $c_1$ have been modified one at a time. As seen in Fig. \ref{fig:ratiosr0}, $WW$, $ZZ$ and $hh$ provide a large section of the total amplitude (Figs. \ref{fig:r0_ratio_a_11},  \ref{fig:r0_ratio_a_09},\ref{fig:r0_ratio_b_11} and \ref{fig:r0_ratio_b_09}), while $t \bar{t}$ is only important (corrections of order 22 \%) when $c_1$ takes the extreme values ($c_1=0.9$ or $c_1=1$) and the rest of the parameters are set to their SM values (Figs. \ref{fig:r0_ratio_c1_11} and \ref{fig:r0_ratio_c1_09}). The $Zh$ cut is relevant only below $\sqrt{s} \sim $~500 GeV and then rapidly becomes insignificant, as can be seen in Fig.~\ref{fig:ratiosr0}.

\begin{figure}[!t]    
\centering
\begin{subfigure}{.4 \columnwidth}
  \centering
  \includegraphics[width=1\linewidth]{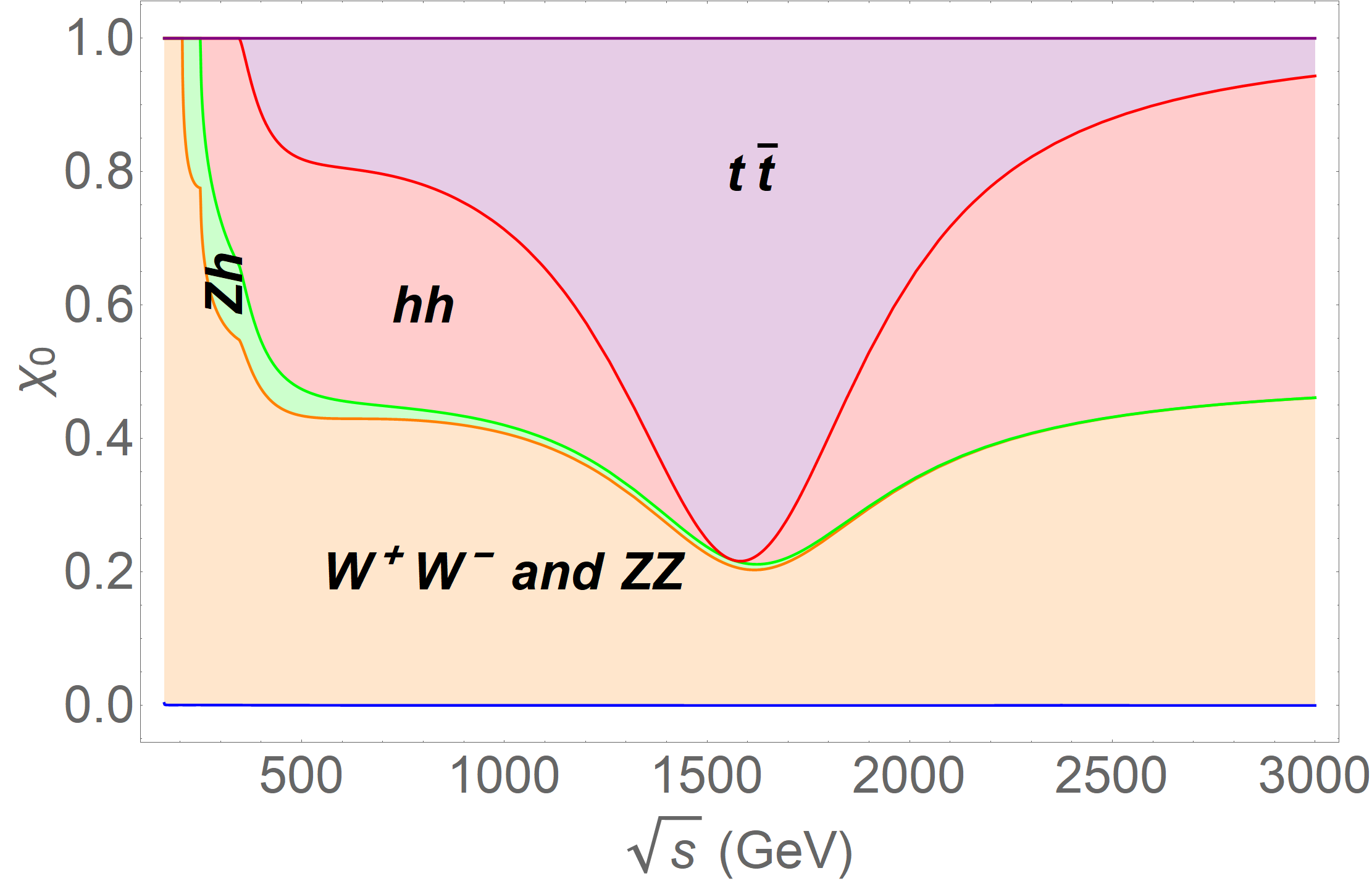}  
  \caption{$J=0$ PWA: largest fermion-loop contribution of 75\% found at 1.5 TeV for ${a=1.023}$, $b=1.100$, $c_1=0.900$ and ${d_3=1.100}$.  }
  \label{fig:ratior0_bestfit_1500gev}
\end{subfigure}
\hspace*{0.75cm}
\begin{subfigure}{.4 \columnwidth}
  \centering
  \includegraphics[width=1 \linewidth]{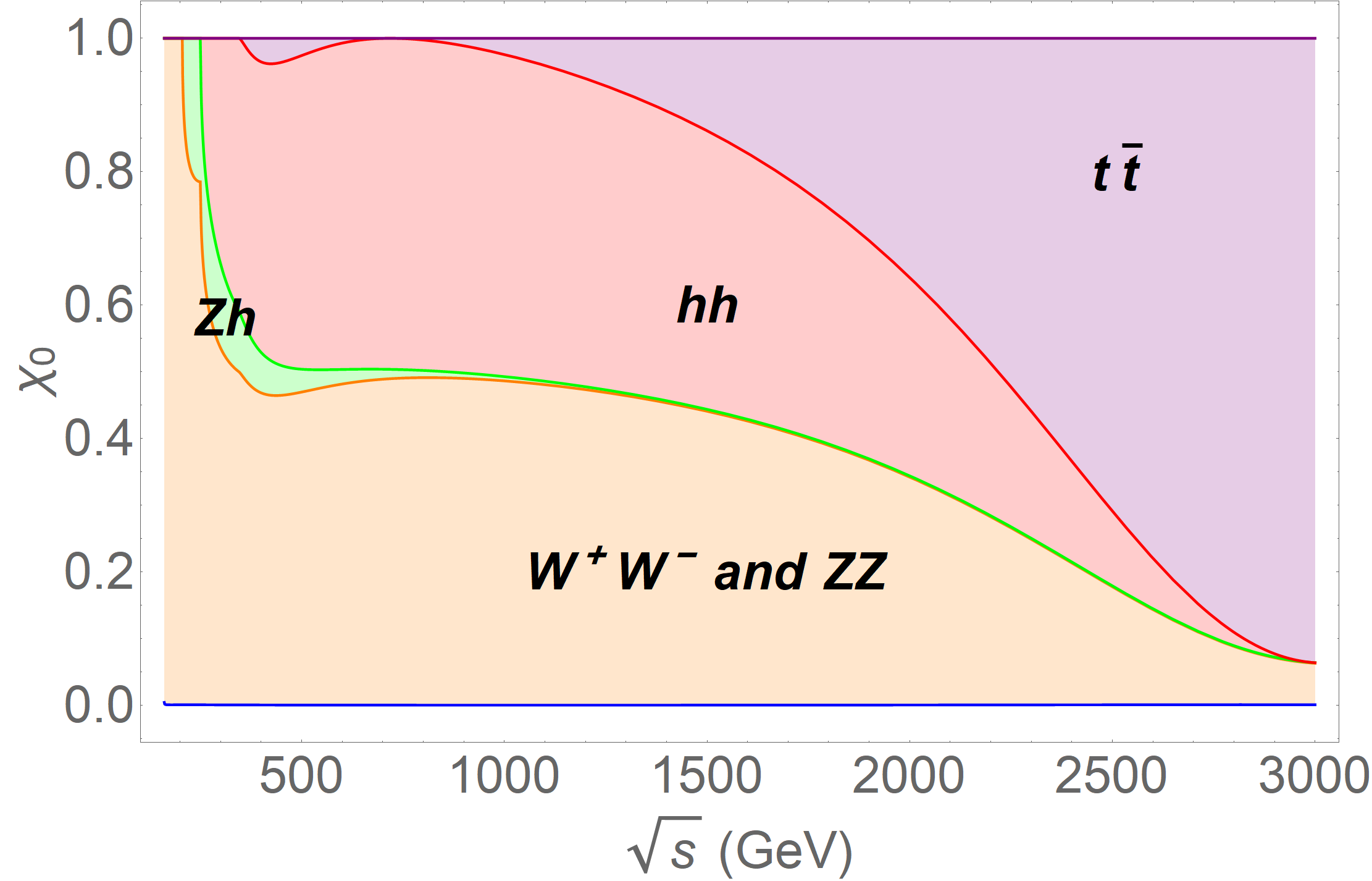}  
  \caption{$J=0$ PWA: largest fermion-loop contribution of 94\% found at 3 TeV for  $a=1.008$, $b=1.035$, $c_1=1.100$ and $d_3=0.900$.  }
  \label{fig:ratior0_bestfit_3000gev}
\end{subfigure}
\caption{}
\label{fig:ratiobestr0}
\end{figure}

These previous plots  give us a notion of the behavior of the amplitude at different energies and values of the HEFT parameters. Now, we will explore the whole possible range these parameters can take to find a set which maximizes $R_0$. We will do this  for two benchmark energies: 1.5 TeV and 3 TeV, a pair of typical energies at which NP is usually expected. We scanned the space of effective parameters $(a,b,c_1,d_3)$ within the aforementioned 10\% deviation from the SM and located the point that maximized $R_0$ at a given CM energy $\sqrt{s}$. We find that ${a=1.023}$, $b=1.100$, $c_1=0.900$ and ${d_3=1.100}$ give rise to a $R_0=75\%$ 
at 1.5 TeV, and $a=1.008$, $b=1.035$, $c_1=1.100$ and $d_3=0.900$ to a $R_0=94\%$ 
at 3 TeV. Some optimal couplings are found to lie on the boundaries of the considered parameter space due to the structure of their analytical expression in the amplitude. We have plotted the relative ratio for both of these configurations in Fig. \ref{fig:ratiobestr0}. As seen in these optimal cases, fermion-loop corrections provide most of the amplitude for $J=0$ .

\begin{figure}[!t]  
\begin{subfigure}{.5\textwidth}
  \centering
  \includegraphics[width=.9\linewidth]{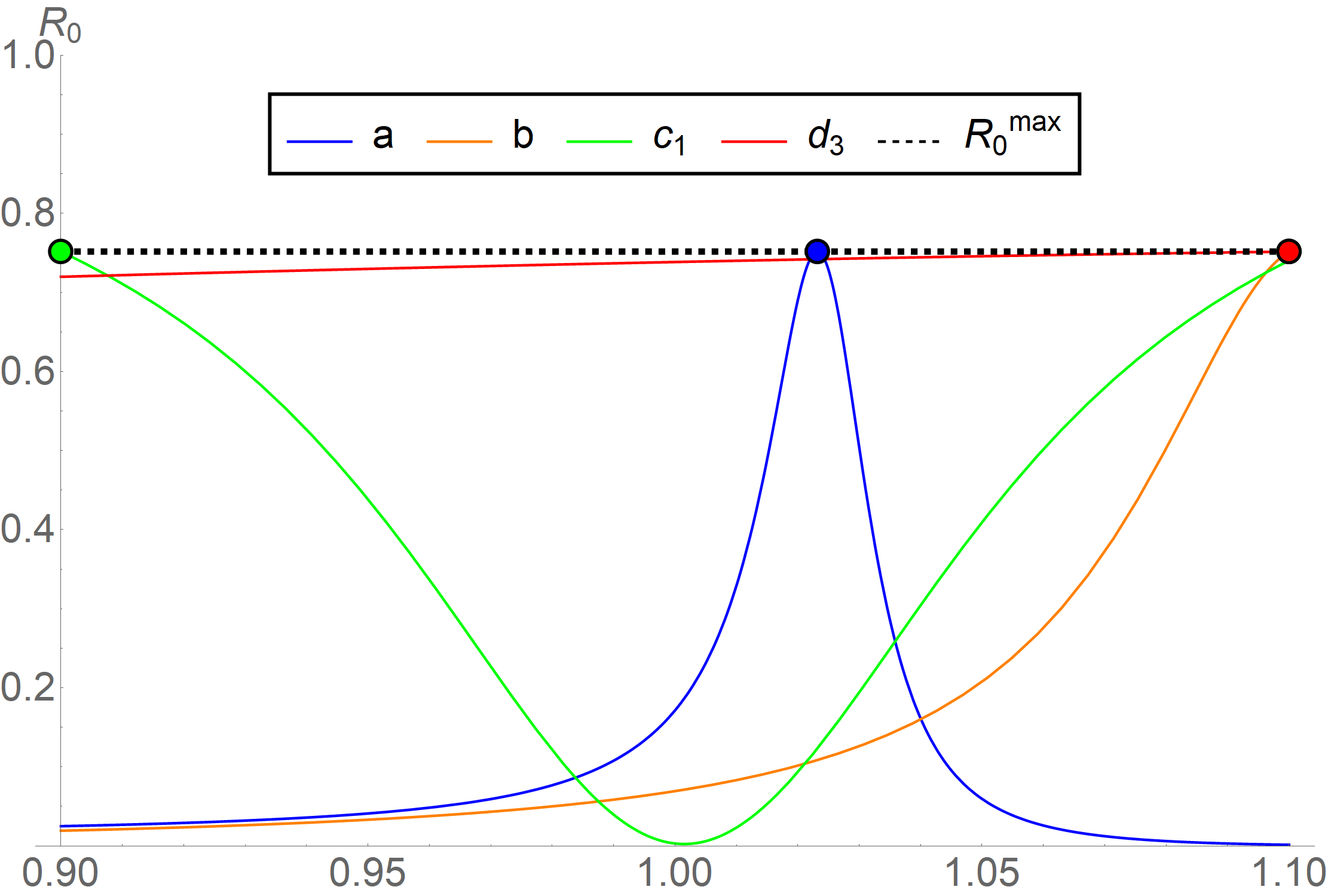}  
  \caption{}
  \label{fig:r0_best1500_sensitivity}
\end{subfigure}
\begin{subfigure}{.5\textwidth}
  \centering
  \includegraphics[width=.9\linewidth]{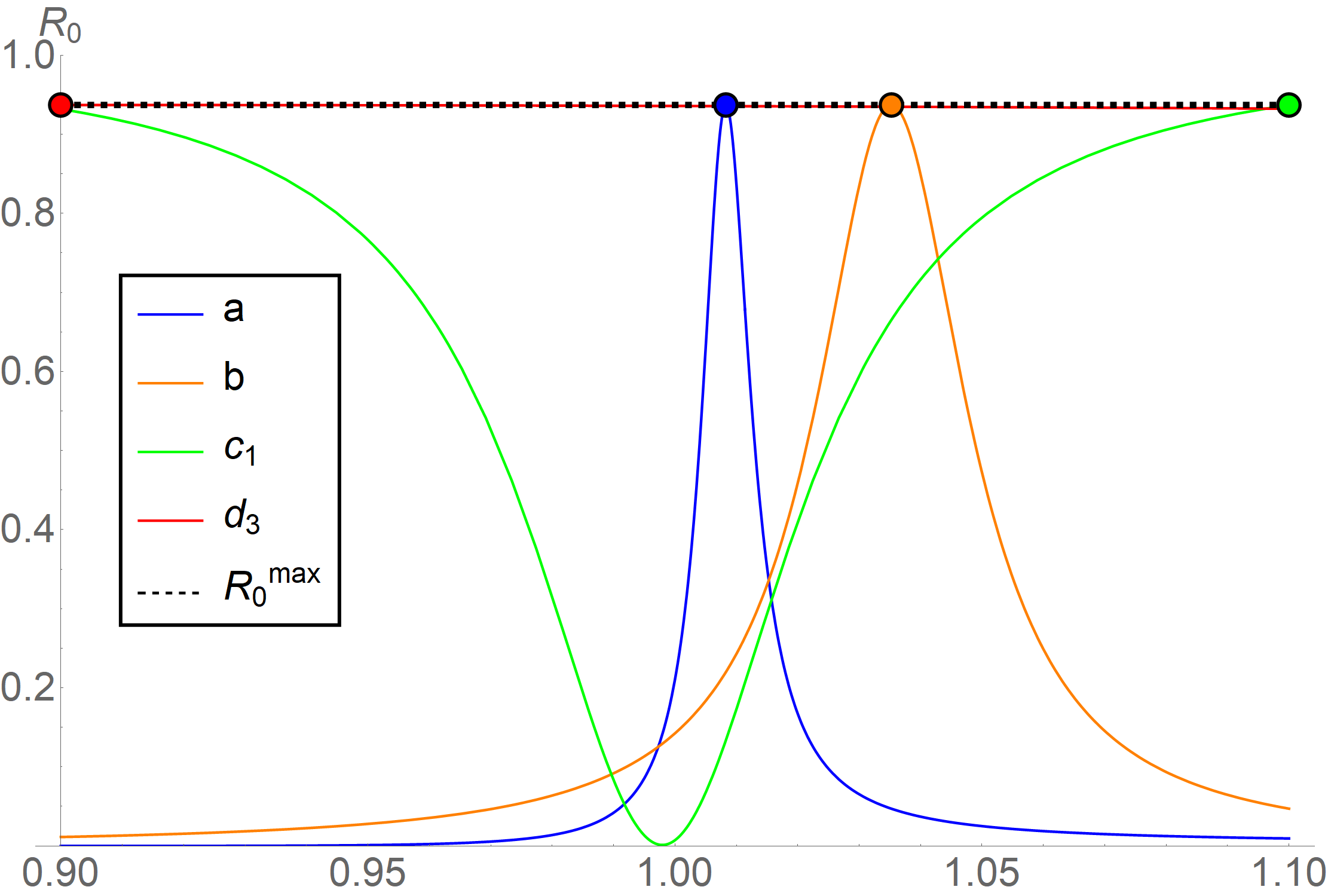}  
  \caption{}
  \label{fig:r0_best3000_sensitivity}
\end{subfigure}
\caption{\small Sensitivity of $R_0$ to each parameter when the rest are set to the highest correction value at $\sqrt{s}$=1.5 TeV (left) and $\sqrt{s}$=3 TeV (right).}
\label{fig:figr0sensitivity}
\end{figure}

To test the sensitivity of $R_0$ to these optimal points $(a, b, c_1,d_3)$, we have plotted $R_0$ varying one parameter at a time while keeping the others fixed to the values that maximize $R_0$. This is shown in Fig. \ref{fig:figr0sensitivity} for 3 TeV and 1.5 TeV, respectively. The full dots on each curve represent the optimal value of the parameter which maximizes the ratio. We can observe that the $R_0$ correction rapidly drops if we  change one of the values of  $a,b$ and $c_1$. Thus, a fine interplay is needed among the couplings  to produce these large fermion-loop corrections. Again, $R_0$ remains essentially constant with respect to $d_3$ variations.

In summary, it is possible to say that, in general, the assumption of neglecting the imaginary part of top-quark loop corrections for the $J=0$ channel is not well sustained since we have found many sets of values of the HEFT parameters which yield meaningful contributions. Moreover, in some cases they even dominate the total amplitude.

\subsection{$J=1$  PWA: $R_1$}

\begin{figure}[!t]    
\includegraphics[width=10cm]{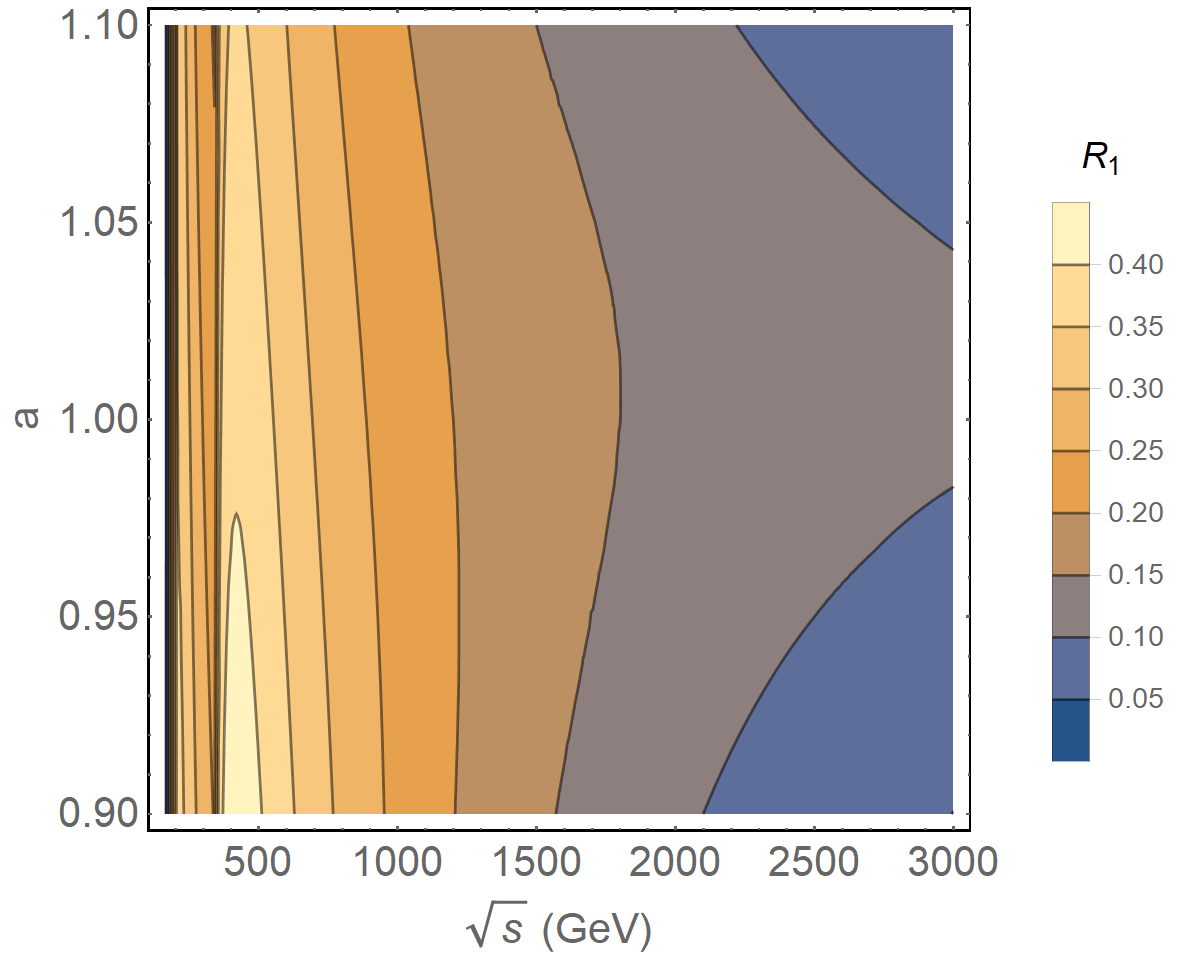}
\centering
\caption{\small $R_1$ dependence on the $a$ parameter.}
\label{fig:r1_countour_a_isospin}
\end{figure}

Now we consider the $J=1$ PWA. The only diagrams from fermion loops which yield a non-zero contribution to $J=1$ are the boxes. However, they do not contain any NP parameter (i.e., deviations from SM).   Hence, Fer$_1$ does not depend on the $c_1$ parameter and is fully determined by the SM gauge-fermion interactions. On the other hand, the bosonic part Bos$_1$ depends only on $a$ through the $WW, ZZ$ and $Zh$ intermediate channels (the isoscalar $hh$ channel does not contribute to $J=1$). As can be seen in Fig.~\ref{fig:r1_countour_a_isospin}, we find a wide range of corrections for low energies (30-40\% at 0.5 TeV for $0.9\lsim a\lsim 1.1$) and for high energies (10-15\% at 3 TeV in the whole range of $a$).

\begin{figure}[!t]    
\includegraphics[width=10cm]{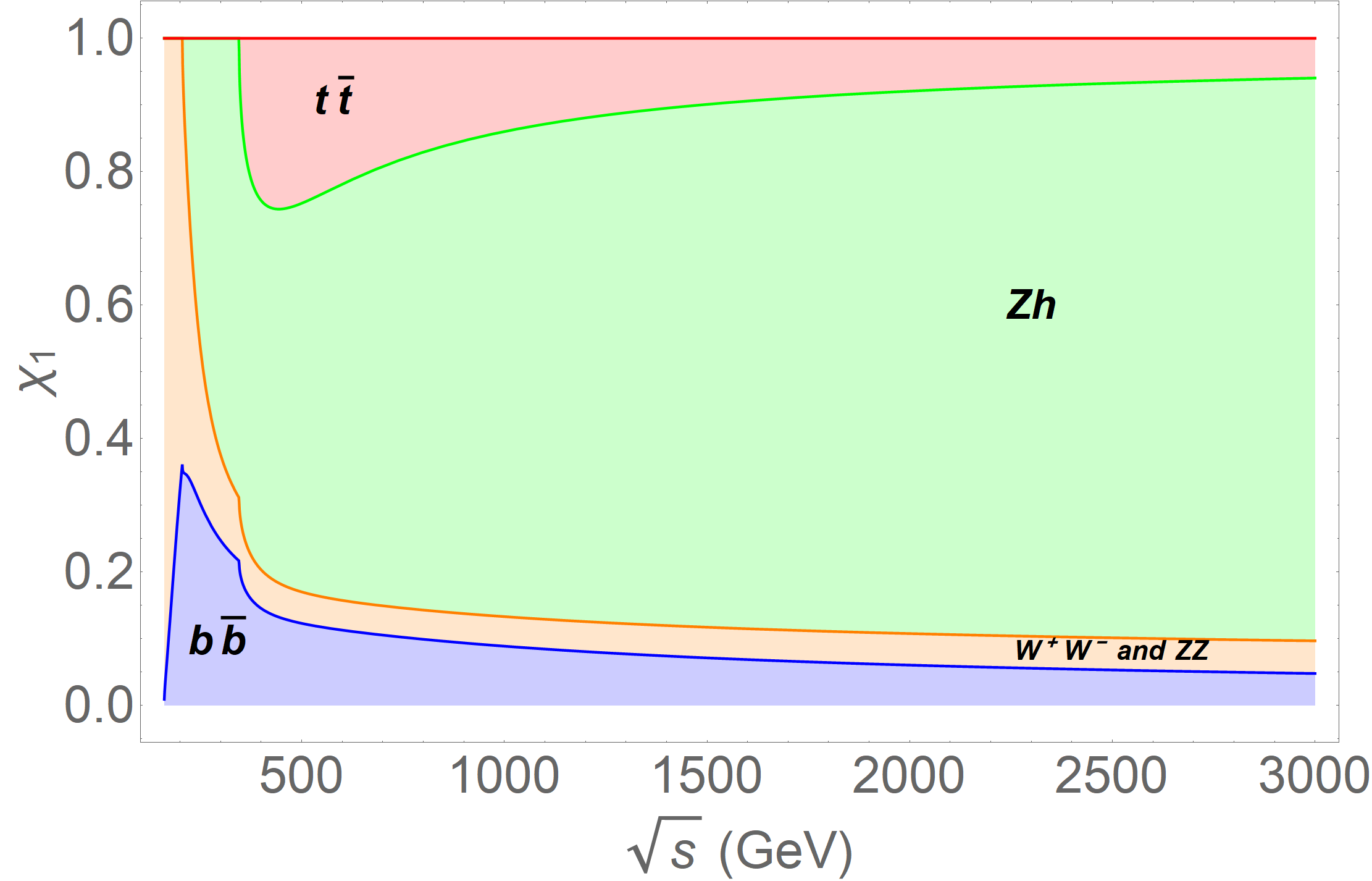}
\centering
\caption{\small Cumulative relative contribution of each channel to $J=1$ PWA in the SM.}
\label{fig:r1ratio_isospin_SM}
\end{figure}

\begin{figure}[!t]    
\centering
\begin{subfigure}{.4 \columnwidth}
  \centering
  \includegraphics[width=1\linewidth]{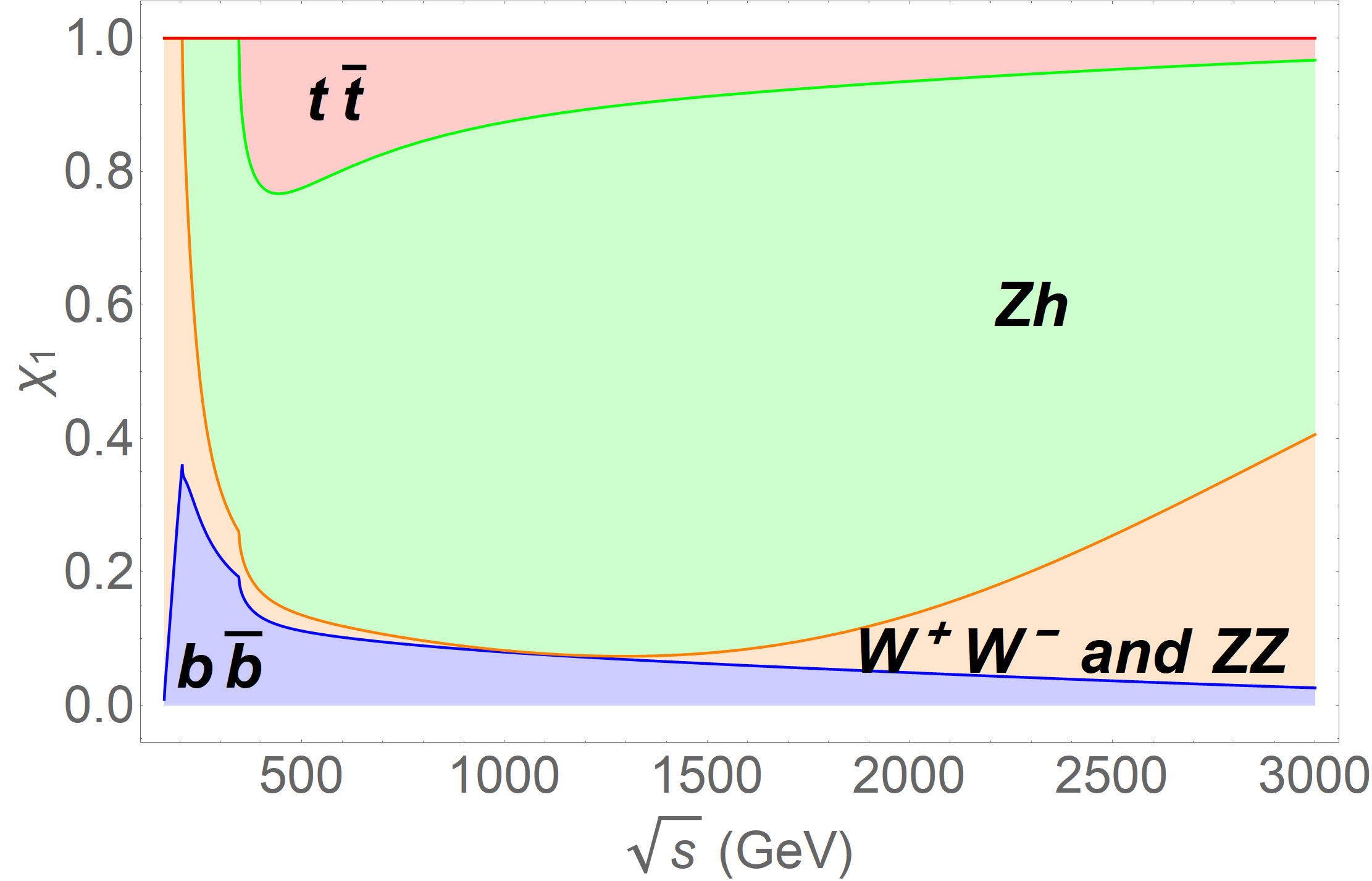}  
  \caption{ }
  \label{fig:r1_ratio_a_11_isospin}
\end{subfigure}
\hspace*{0.75cm}
\begin{subfigure}{.4 \columnwidth}
  \centering
  \includegraphics[width=1 \linewidth]{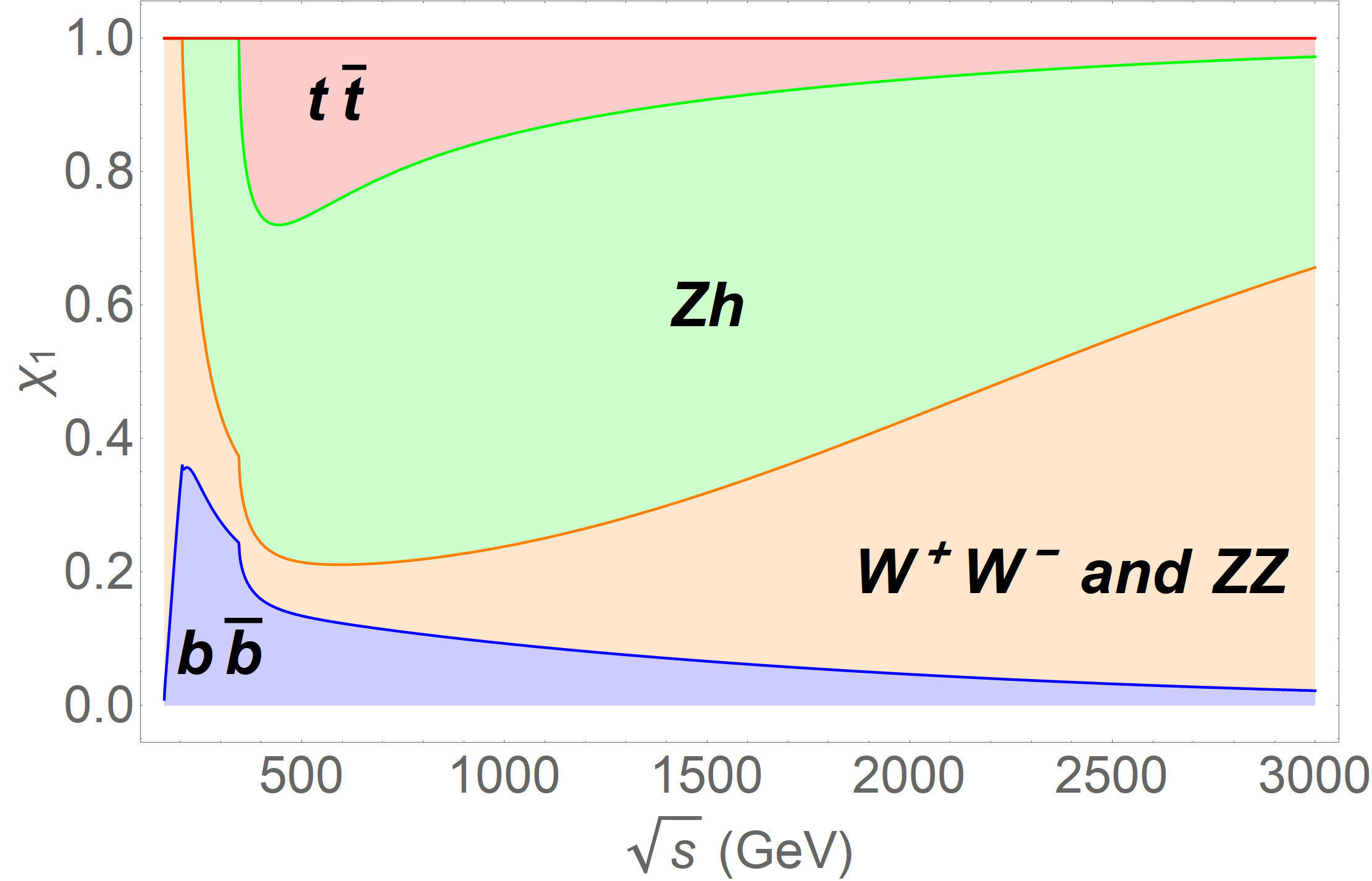}  
  \caption{}
  \label{fig:r1_ratio_a_09_isospin}
\end{subfigure}
\caption{\small Cumulative relative contribution of each channel to the $J=1$ PWA for $a=1.100$ (left) and $a=0.900$ (right).}
\label{fig:ratior1}
\end{figure}

In addition to $WW$ and $ZZ$ cuts, this $J=1$ PWA also receives contributions from the $Zh$ absorptive channel, even for $g'=0$.  
The present work completes previous preliminary studies~\cite{Dobado:2021ozt,Dobado:2020lil},   
which neglected the $Zh$ channel on the basis of nET and custodial symmetry arguments. 
This channel yields, indeed,  
a large contribution to the amplitude, as can be seen in Figs.~\ref{fig:r1ratio_isospin_SM} and~\ref{fig:ratior1}. This outcome is notable as the $Zh$ channel is usually not included when studying $WW$ scattering. Since the only available HEFT parameter is $a$, we can easily describe the dependence of $R_1$ on NP. Values of $a$ close to 1 minimize the boson contribution, thus yielding a high $R_1$. 
Since the $W^+W^-\to Zh$ tree-level scattering vanishes in the the na\"ive ET at lowest order in the chiral expansion, one needs to go beyond it to actually address this important boson-loop contribution. 

For this partial wave we can see in Figs.~\ref{fig:r1ratio_isospin_SM} and~\ref{fig:ratior1} that the $b \bar{b}$ cut provides a significant contribution to the total amplitude. Especially at large energies, the contribution from both cuts, $t \bar{t}$, and $b \bar{b}$ are similar. In order to obtain a relevant fermion-loop contribution at high energies to the $J=1$ channel from a quark doublet, at least one of the fermions needs to be heavy.

\begin{figure}[!t]    
\centering
\begin{subfigure}{.4 \columnwidth}
  \centering
  \includegraphics[width=1\linewidth]{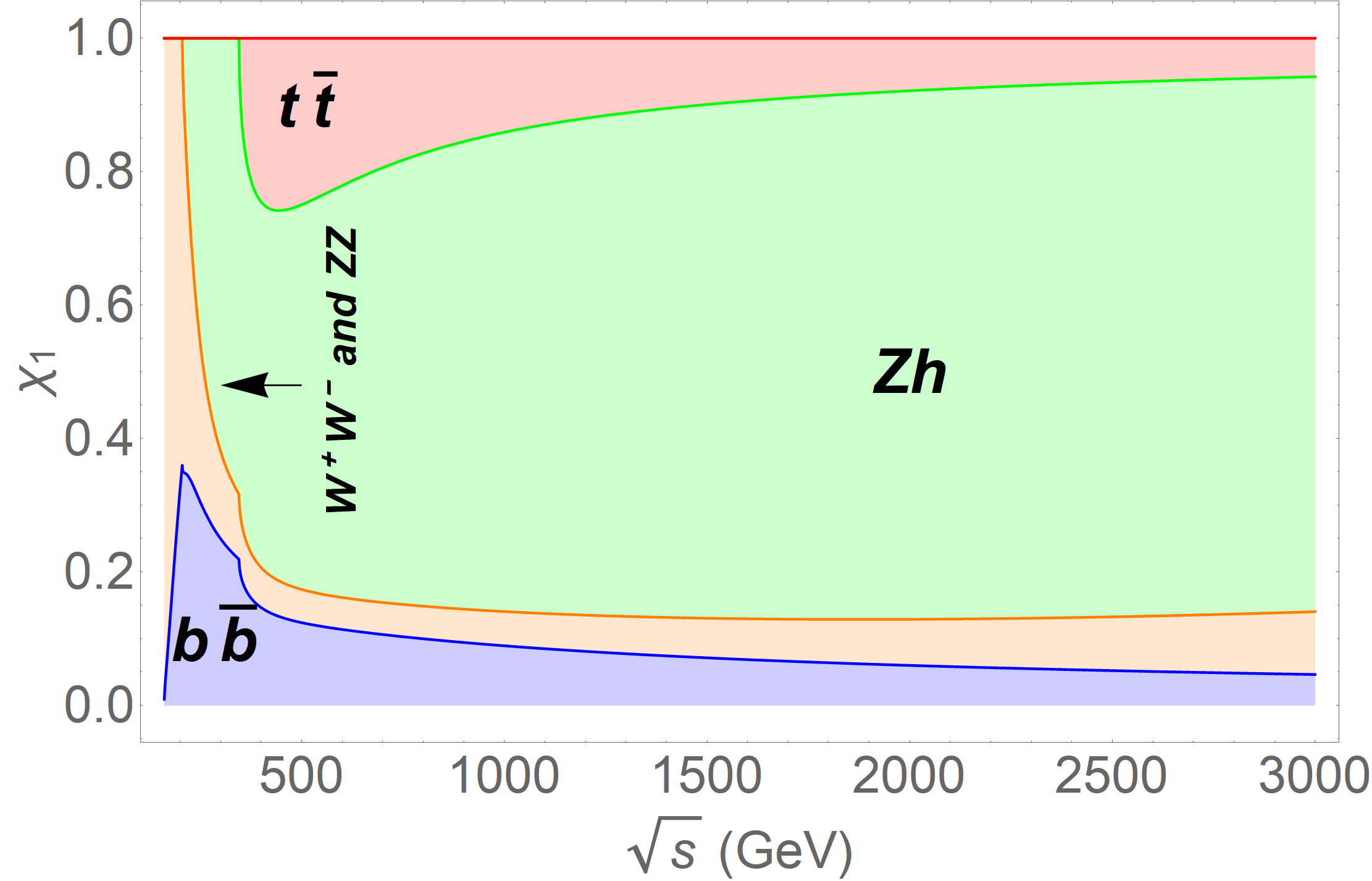}  
  \caption{$J=1$ PWA: largest fermion-loop contribution of 17\% 
  at 1.5 TeV for $a=0.991$ . }
  \label{fig:ratior1prime_optimal_1500gev}
\end{subfigure}
\hspace*{0.75cm}
\begin{subfigure}{.4 \columnwidth}
  \centering
  \includegraphics[width=1 \linewidth]{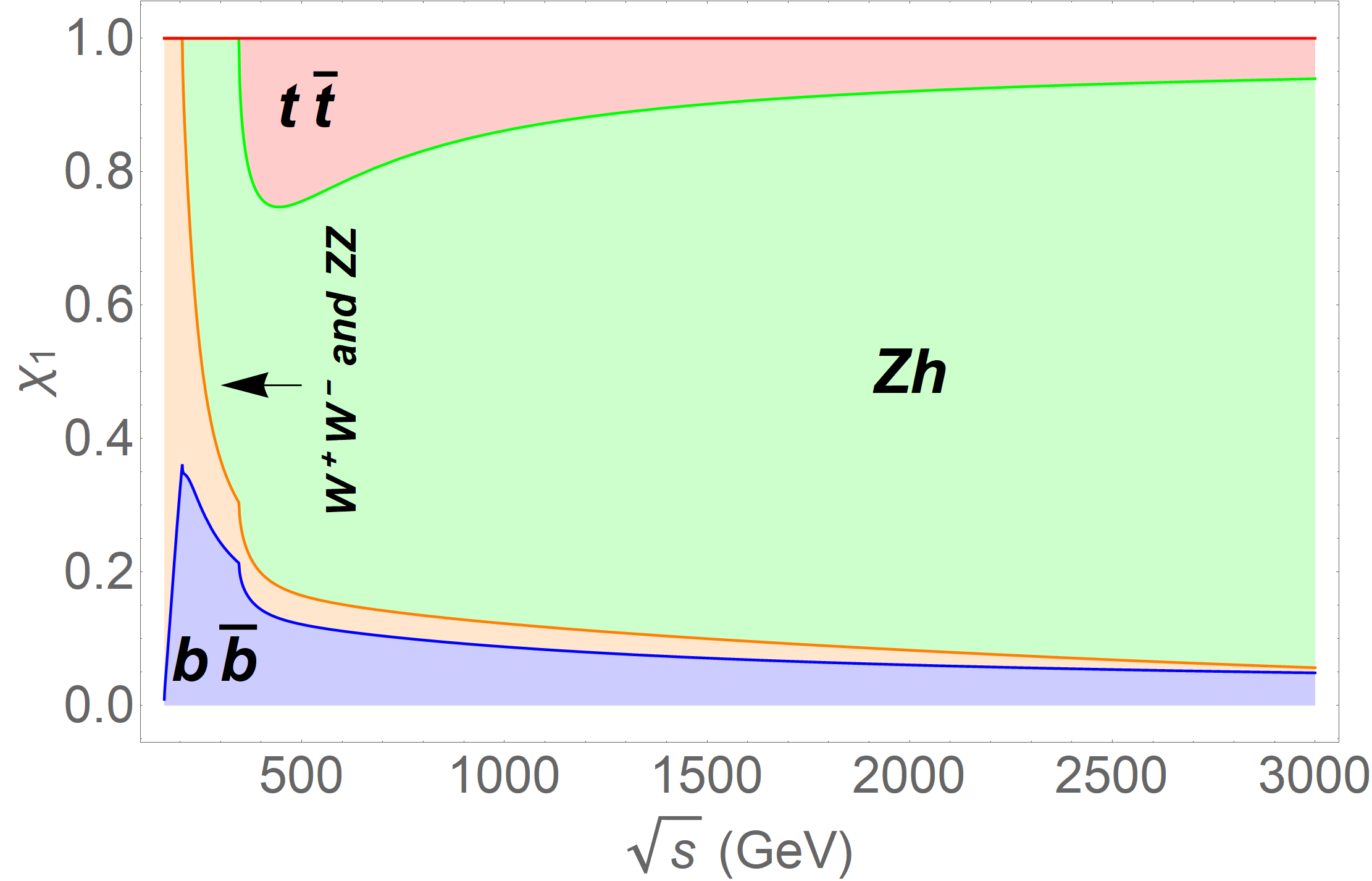}  
  \caption{$J=1$ PWA: largest fermion-loop contribution of 11\% 
  at 3 TeV for $a=1.013$ . }
  \label{fig:ratior1prime_bestfit_1500gev}
\end{subfigure}
\caption{}
\label{fig:ratior1_optimat_isospin}
\end{figure}

If we look for the optimal value of $a$ that  maximizes $R_1$ at the same benchmark energies as before, we encounter that 
$a=0.991$ yields $R_1=17 \%$  at 1.5 TeV and $a=1.013$ yields $R_1=11 \%$ at 3 TeV. In Fig \ref{fig:ratior1_optimat_isospin} we show the cumulative relative amplitudes of each cut for these benchmark energies. These values of $a$ minimize the total boson loop contribution at the mentioned CM energies ($WW$, $ZZ$ and $Zh$ cuts), giving more relevance to the fermion cuts.  

\begin{figure}[!t]    
\includegraphics[width=10cm]{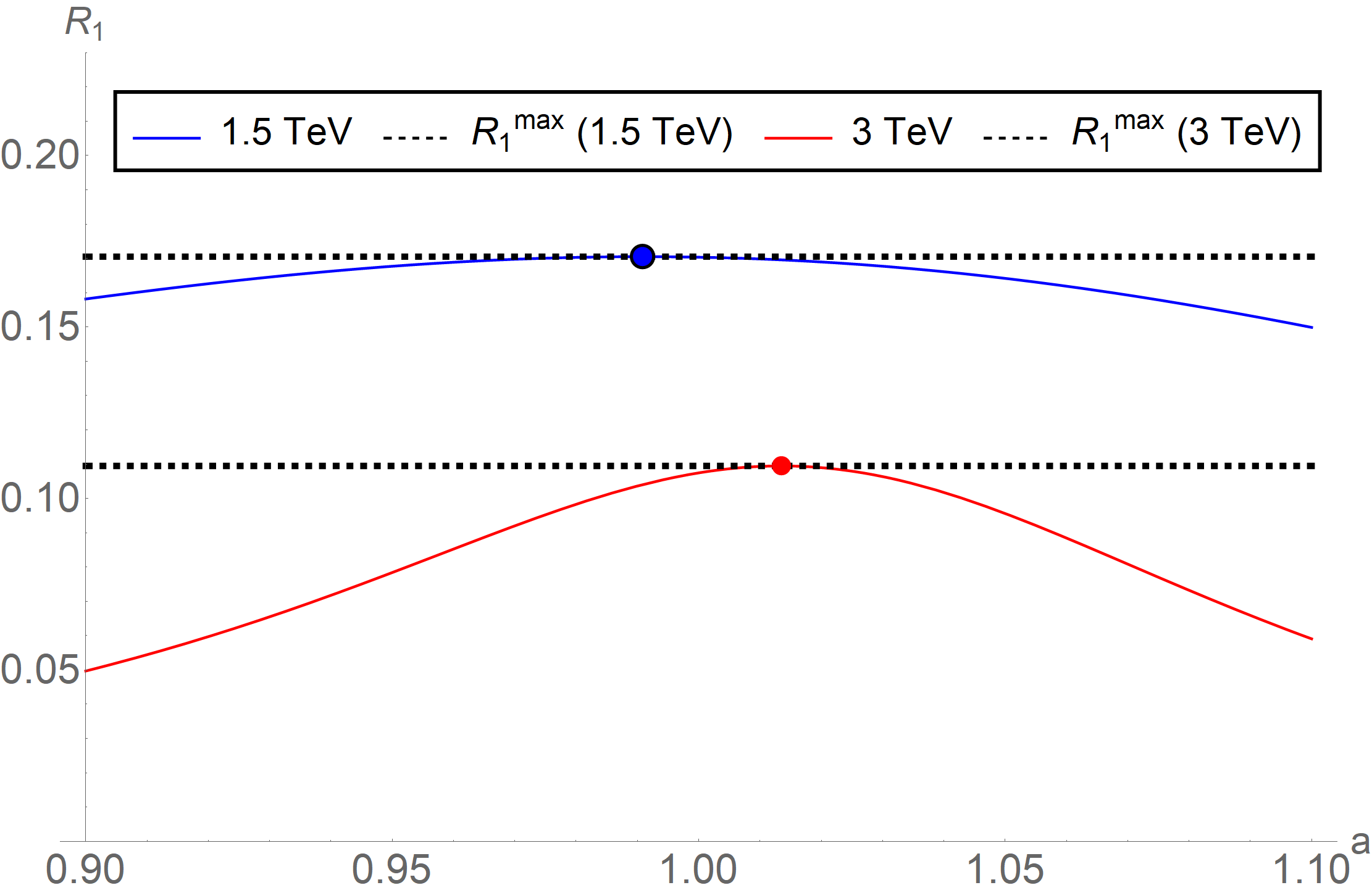}
\centering
\caption{\small Sensitivity of $R_1$ to the $a$ parameter for the highest contribution at $\sqrt{s}$ = 1.5 TeV and $\sqrt{s}$= 3 TeV.}
\label{fig:r1_sensitivity}
\end{figure}

In Fig.~\ref{fig:r1_sensitivity}  
we can see  the optimal points for both curves. The dependence on one parameter is also very revealing; even if we restrict ourselves to scenarios very close to the SM, we observe both curves do not change dramatically. This is interesting because, unlike  the $J=0$ case where we needed a fine interplay among the HEFT parameters, Fig.~\ref{fig:r1_sensitivity} shows significant fermion corrections above 5\% (15\%) for $\sqrt{s}=1.5$~TeV ($\sqrt{s}=3$~TeV) 
in the whole range of $a$ studied here.

As was the case for the previous partial wave, neglecting fermion-loop corrections is not appropriate according to our work. Even if we restrict ourselves for scenarios close to the SM one where $a\approx 1$, 
we find significant fermion corrections.

\section{Fermion-loops beyond the $g'=0$ limit: pseudo-PWA}

\subsection{$J=0$  pseudo-PWA: $R'_0$}
\label{section:r0prime}

\begin{figure}[!t]    
\begin{subfigure}{.5\textwidth}
  \centering
  \includegraphics[width=.9\linewidth]{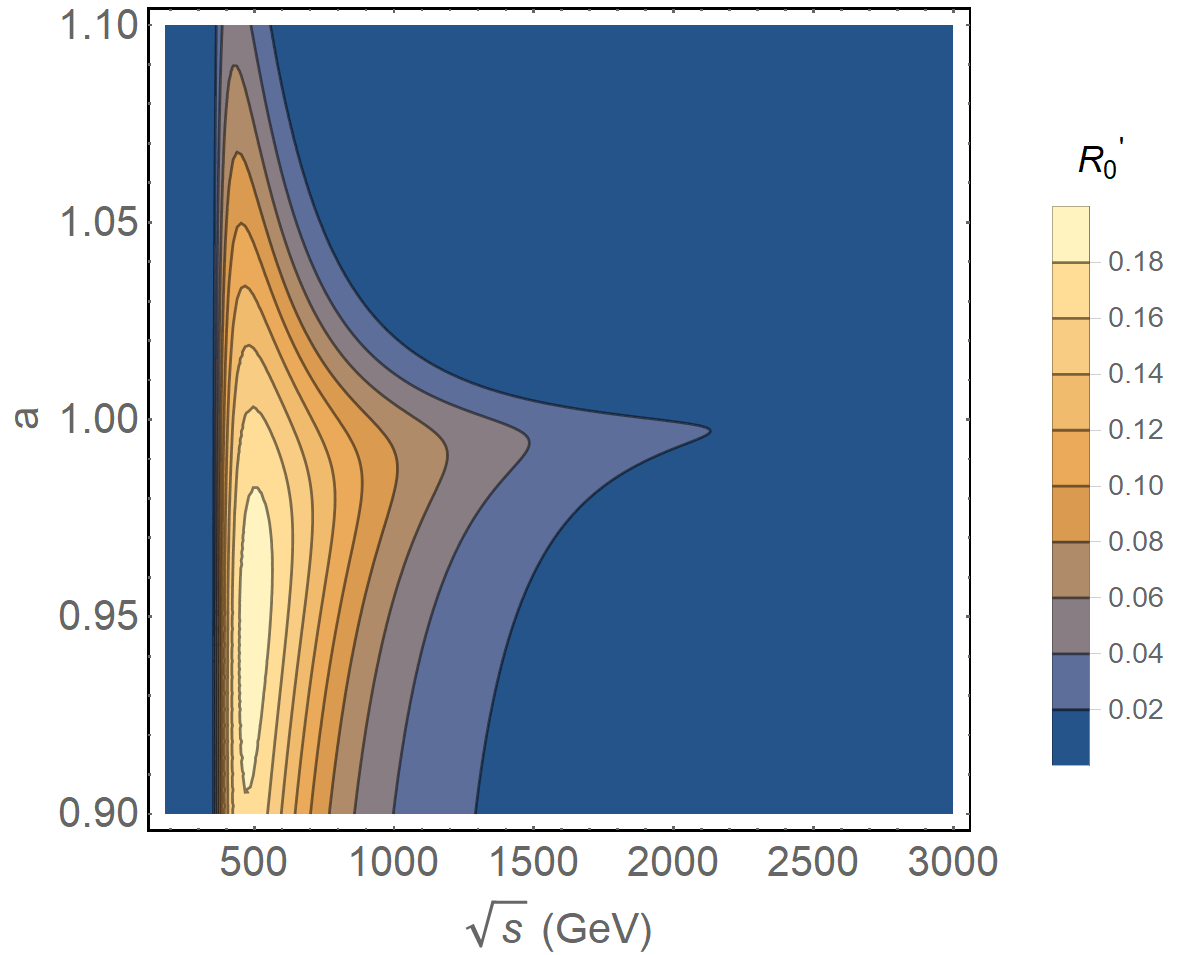}  
  \caption{$R'_0$ dependence on $a$ for $b=c_1=d_3=1$.}
  \label{fig:aj0_prime_plotpartial}
\end{subfigure}
\begin{subfigure}{.5\textwidth}
  \centering
  \includegraphics[width=.9\linewidth]{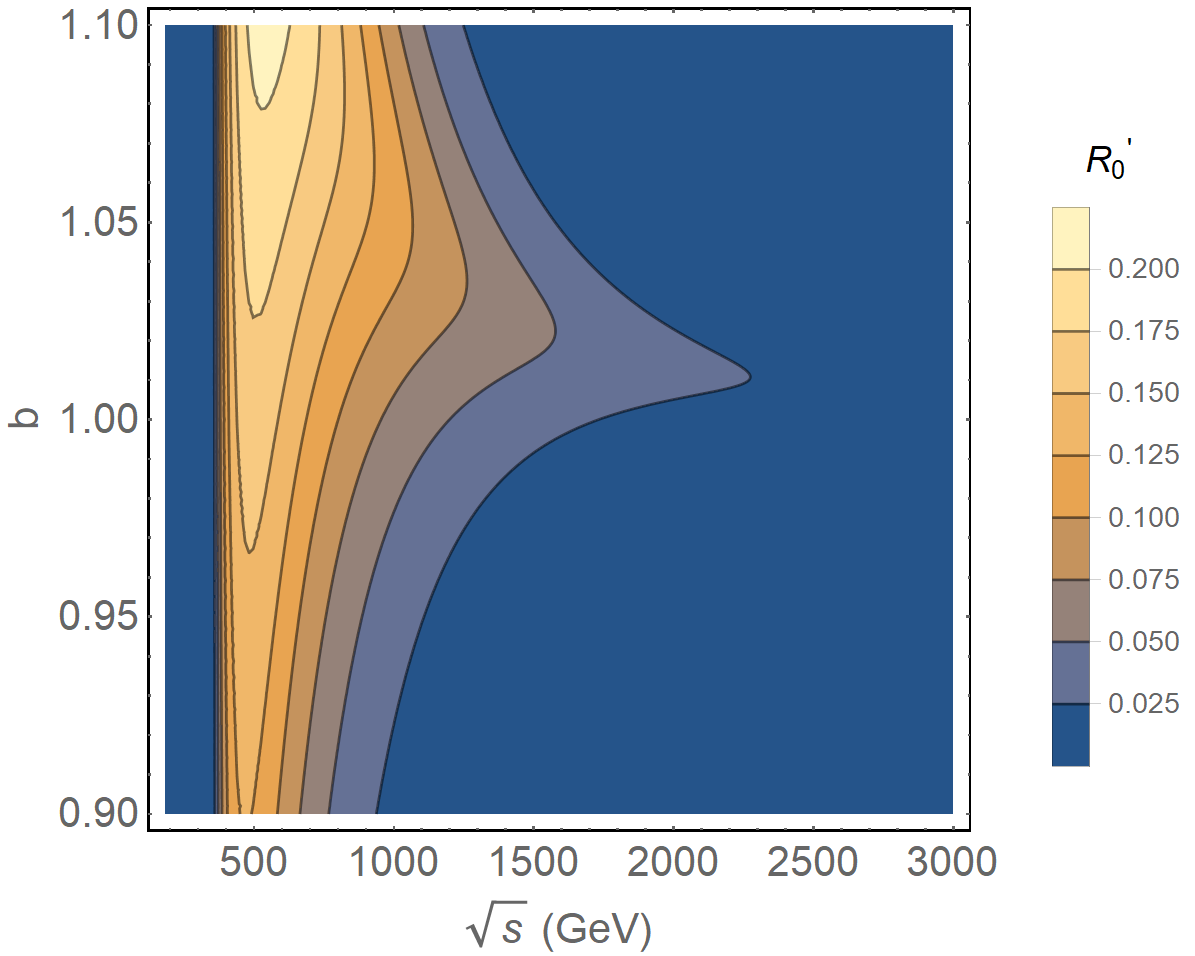}    
  \caption{$R'_0$ dependence on $b$ for $a=c_1=d_3=1$.}
  \label{fig:bj0_prime_plotpartial}
\end{subfigure}
\\[3ex]
\begin{subfigure}{.5\textwidth}
  \centering
  \includegraphics[width=.9\linewidth]{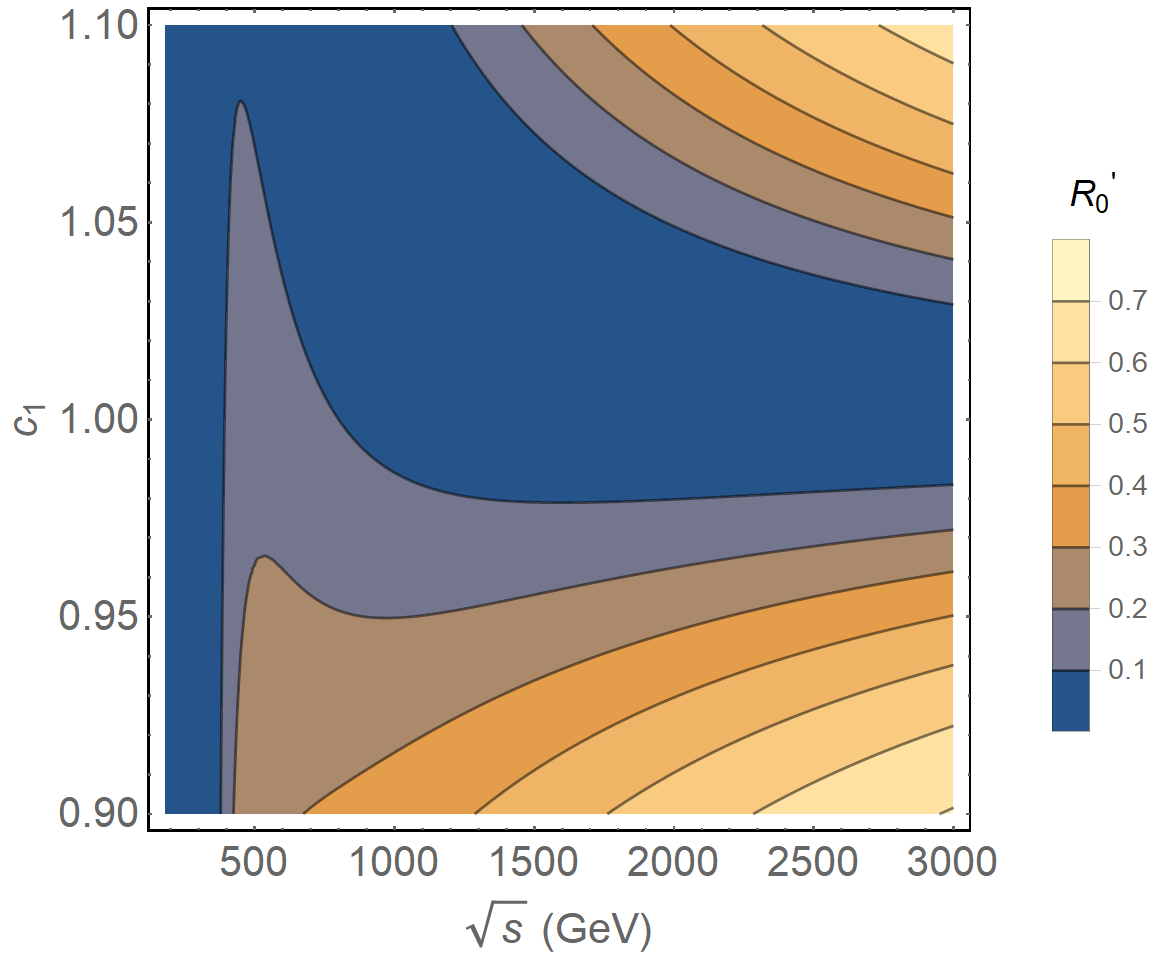}  
  \caption{$R'_0$ dependence on $c_1$ for $a=b=d_3=1$.}
  \label{fig:c1j0_prime_plotpartial}
\end{subfigure}
\begin{subfigure}{.5\textwidth}
  \centering
  \includegraphics[width=.9\linewidth]{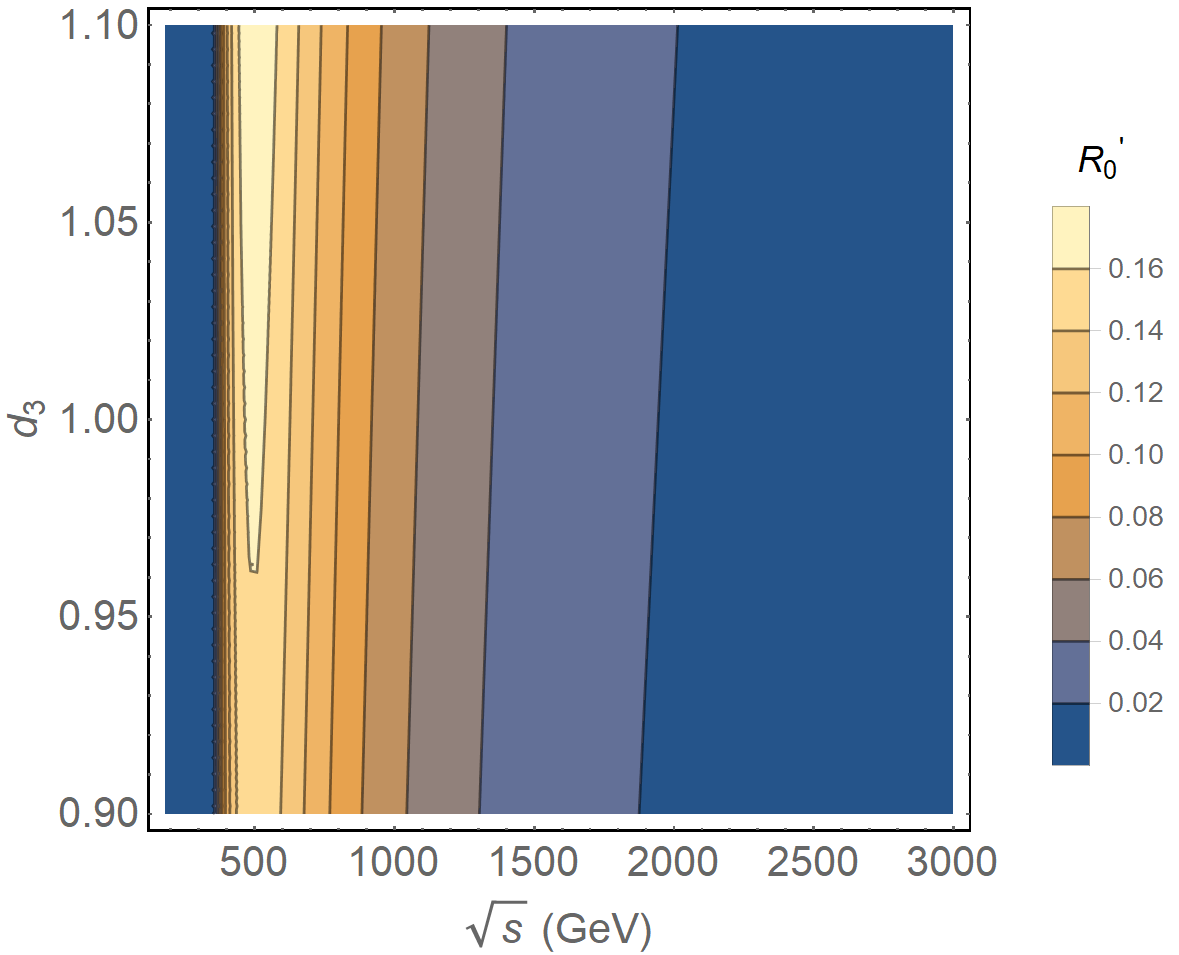}  
  \caption{$R'_0$ dependence on $d_3$ for $a=b=c_1=1$.}
  \label{fig:d3j0_prime_plotpartial}
\end{subfigure}
\caption{}
\label{fig:r0_prime_contour}
\end{figure}

Moving to the more realistic case $g' \neq0$, we have additional cuts: $\gamma \gamma$, $\gamma h$ and $\gamma Z$. As mentioned before, the integration has been performed only in the $\abs{\cos{\theta}} \leq 0.9$ region due to a divergence in the $t$-channel of the $WW$ cut.


Thus, strictly speaking these are not partial waves so we will refer to them as pseudo-Partial Wave Amplitudes (p-PWA's). Apart from this the analysis, will be analogous to the $g'=0$ case.

\begin{figure}[!t]    
\includegraphics[width=10cm]{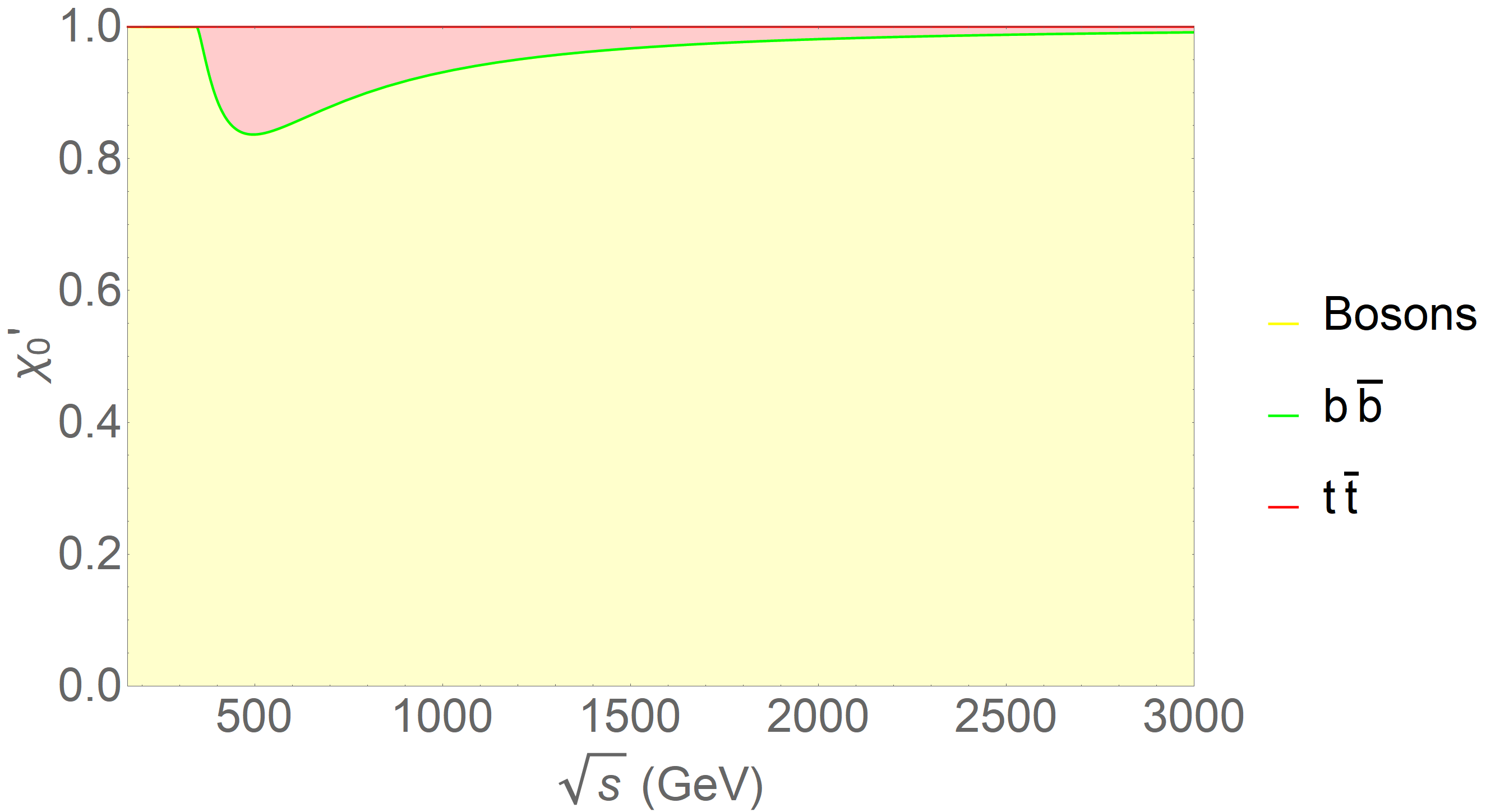}
\centering
\caption{\small Cumulative amplitude ratio for the $J=0$ p-PWA in the SM.}
\label{fig:r0ratio_prime_combined_sm}
\end{figure}

\begin{figure}[!t]    
\centering
\begin{subfigure}{0.4 \columnwidth}
  \centering
  \includegraphics[width=1\linewidth]{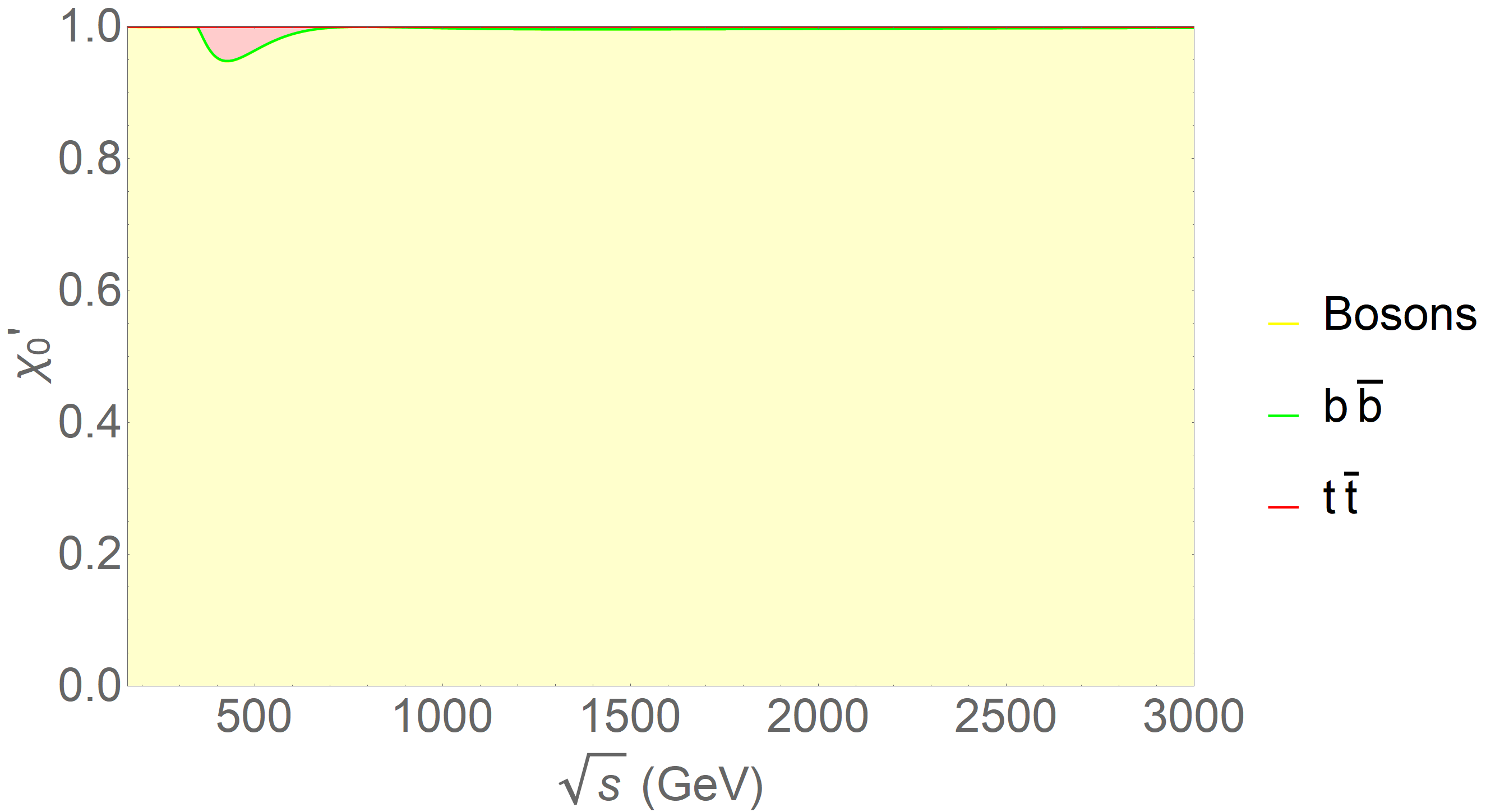}  
  \caption{$a=1.10$ and $b=c_1=d_3=1$.}
  \label{fig:r0ratio_prime_combined_a1}
\end{subfigure}
\hspace*{0.75cm}
\begin{subfigure}{0.4 \columnwidth}
  \centering
  \includegraphics[width=1 \linewidth]{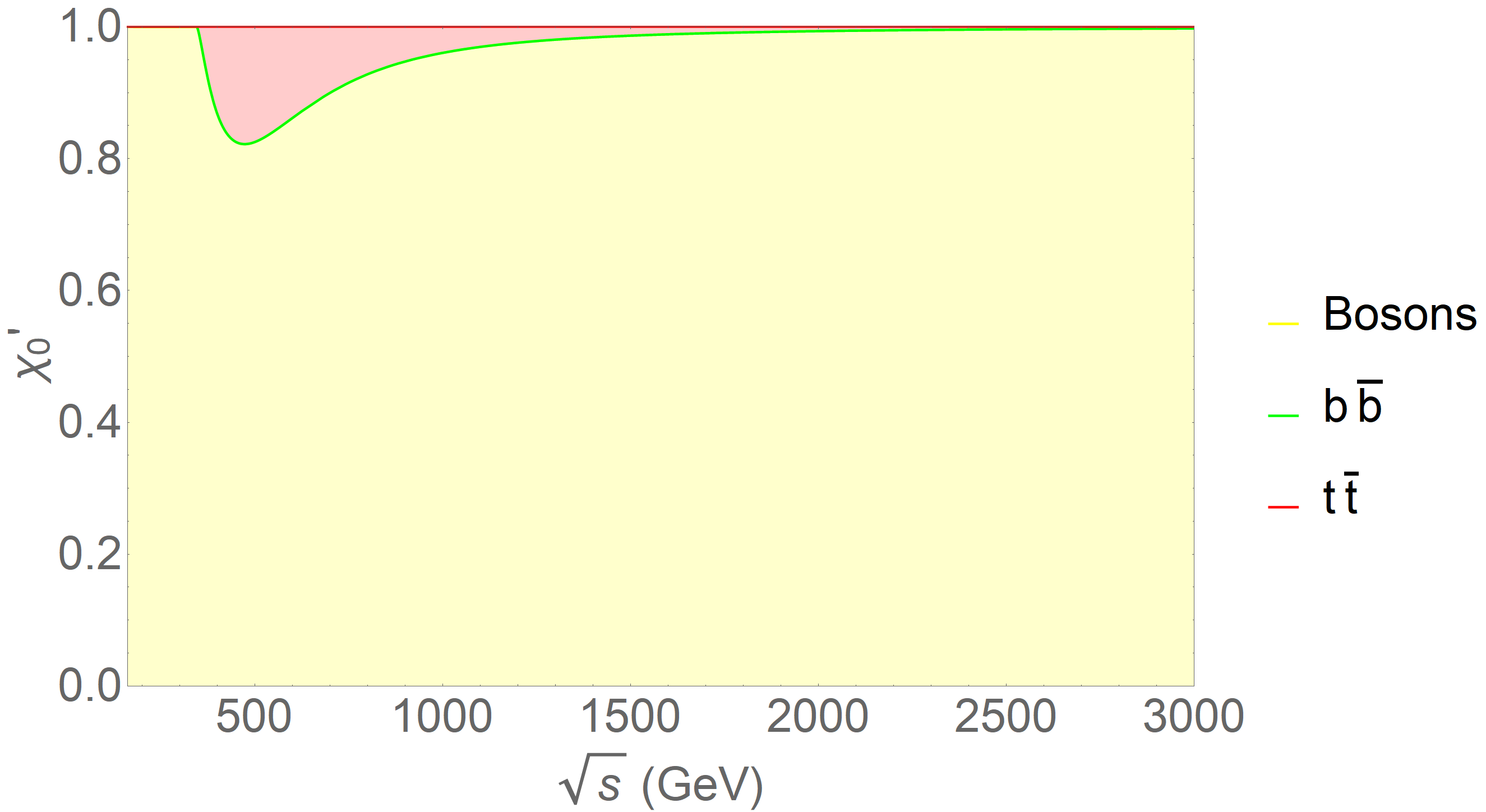}  
  \caption{$a=0.90$ and $b=c_1=d_3=1$.}
  \label{fig:r0ratio_prime_combineda09}
\end{subfigure}
%
\\[8pt]
\centering
\begin{subfigure}{0.4 \columnwidth}
  \centering
  \includegraphics[width=\linewidth]{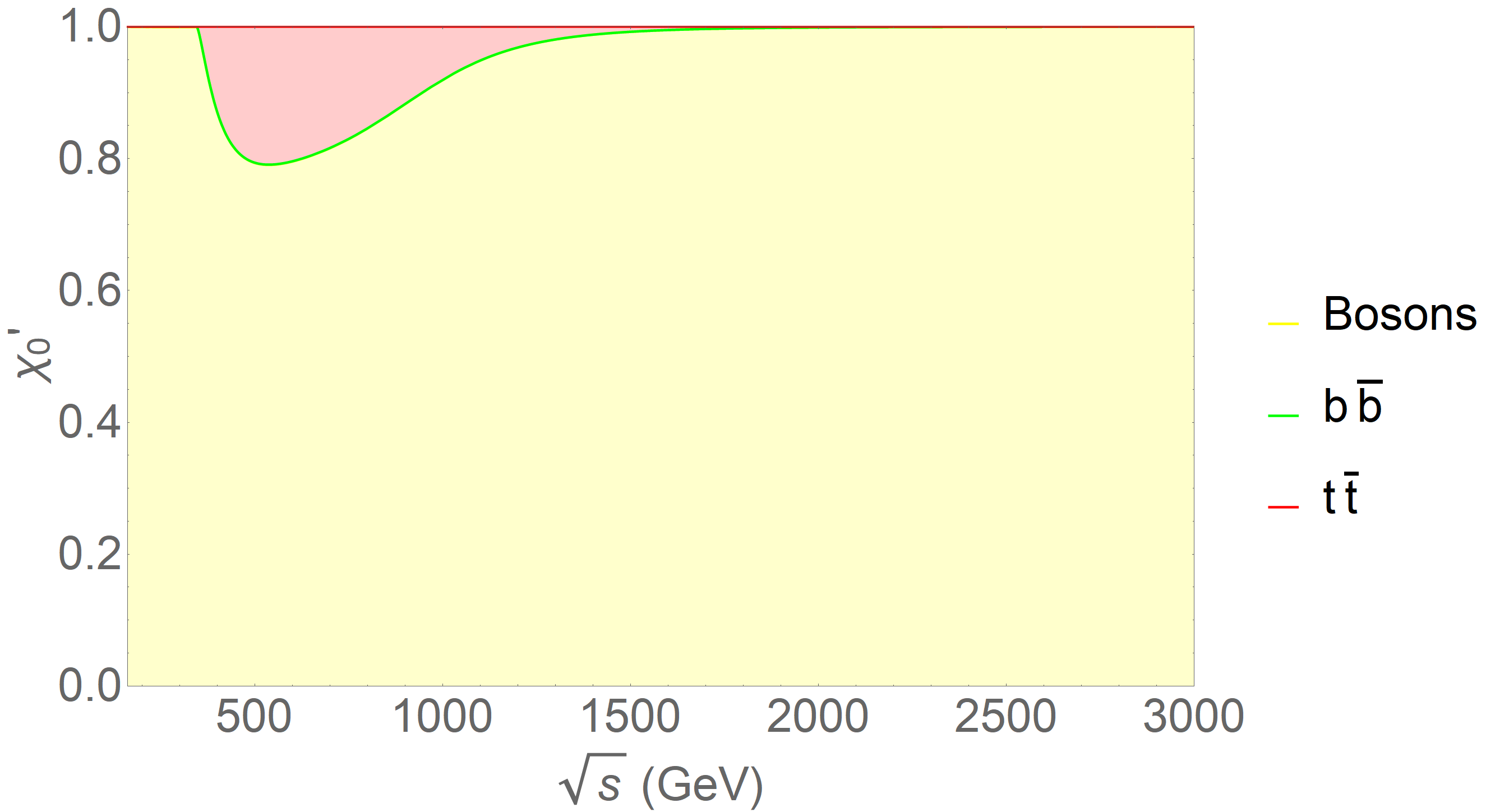}  
  \caption{ $b=1.10$ and $a=c_1=d_3=1$.}
  \label{fig:r0ratio_prime_b11_combined}
\end{subfigure}
\hspace*{0.75cm}
\begin{subfigure}{0.4 \columnwidth}
  \centering
  \includegraphics[width=\linewidth]{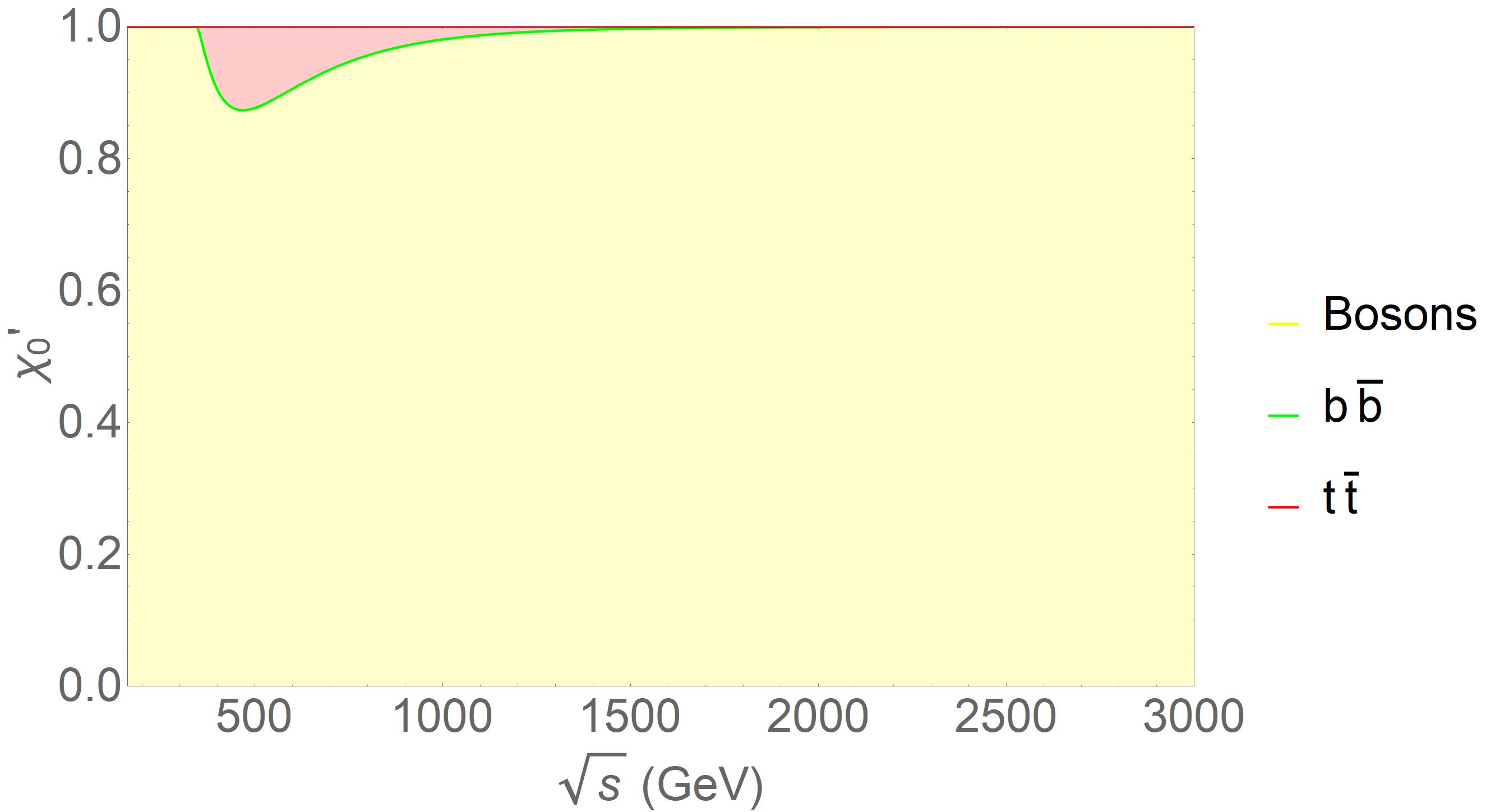}  
  \caption{ $b=0.90$ and $b=c_1=d_3=1$ .}
  \label{fig:r0ratio_prime_b09_combined}
\end{subfigure}
\\[8pt]
\centering
\begin{subfigure}{0.4 \columnwidth}
  \centering
  \includegraphics[width=\linewidth]{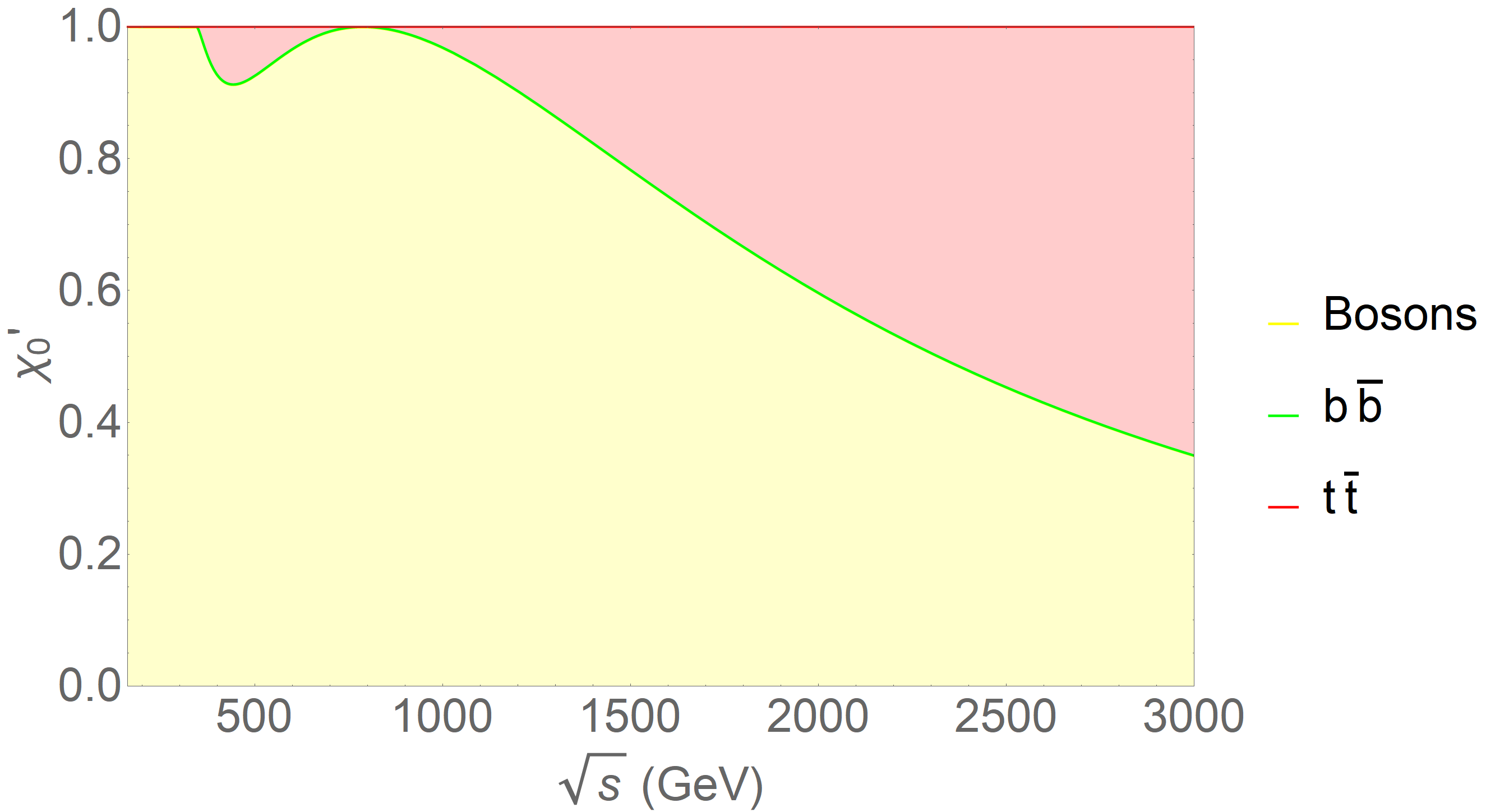}  
  \caption{$c_1=1.10$ and $a=b=d_3=1$ . }
  \label{fig:r0ratio_prime_combinedc111}
\end{subfigure}
\hspace*{0.75cm}
\begin{subfigure}{0.4 \columnwidth}
  \centering
  \includegraphics[width=\linewidth]{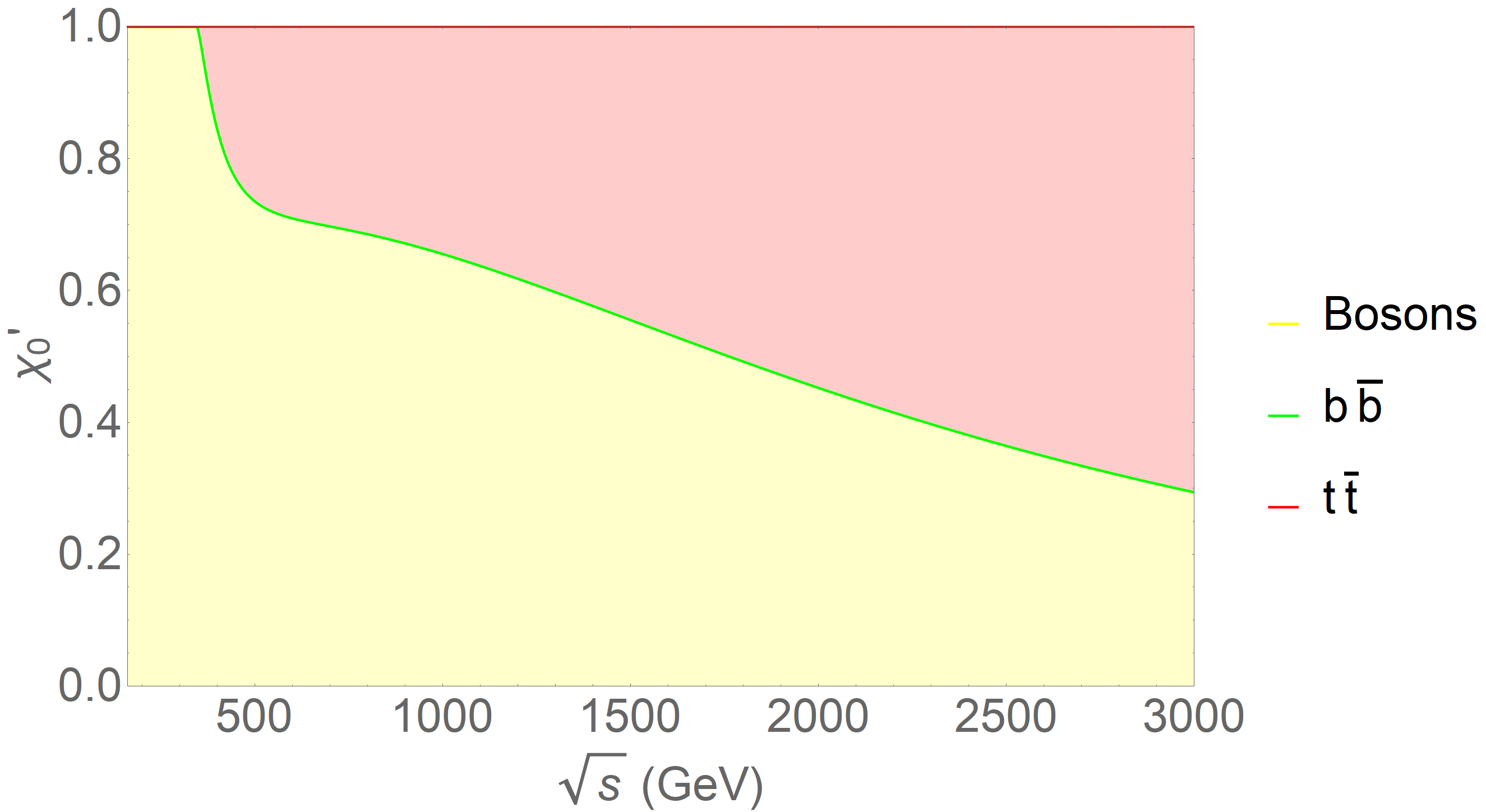}  
  \caption{$c_1=0.90$ and $a=b=d_3=1$.  }
  \label{fig:r0ratio_prime_combinedc109}
\end{subfigure}
\caption{\small Cumulative relative contributions for each absorptive cut to the $J=0$ p-PWA for $a,b$ and $c_1$ at the borders of the considered parameter space. The $b\bar{b}$ contribution is numerically  negligible for this p-PWA.}
\label{fig:ratioj0_prime}
\end{figure}

\begin{figure}[!t]    
\centering
\begin{subfigure}{.4 \columnwidth}
  \centering
  \includegraphics[width=1\linewidth]{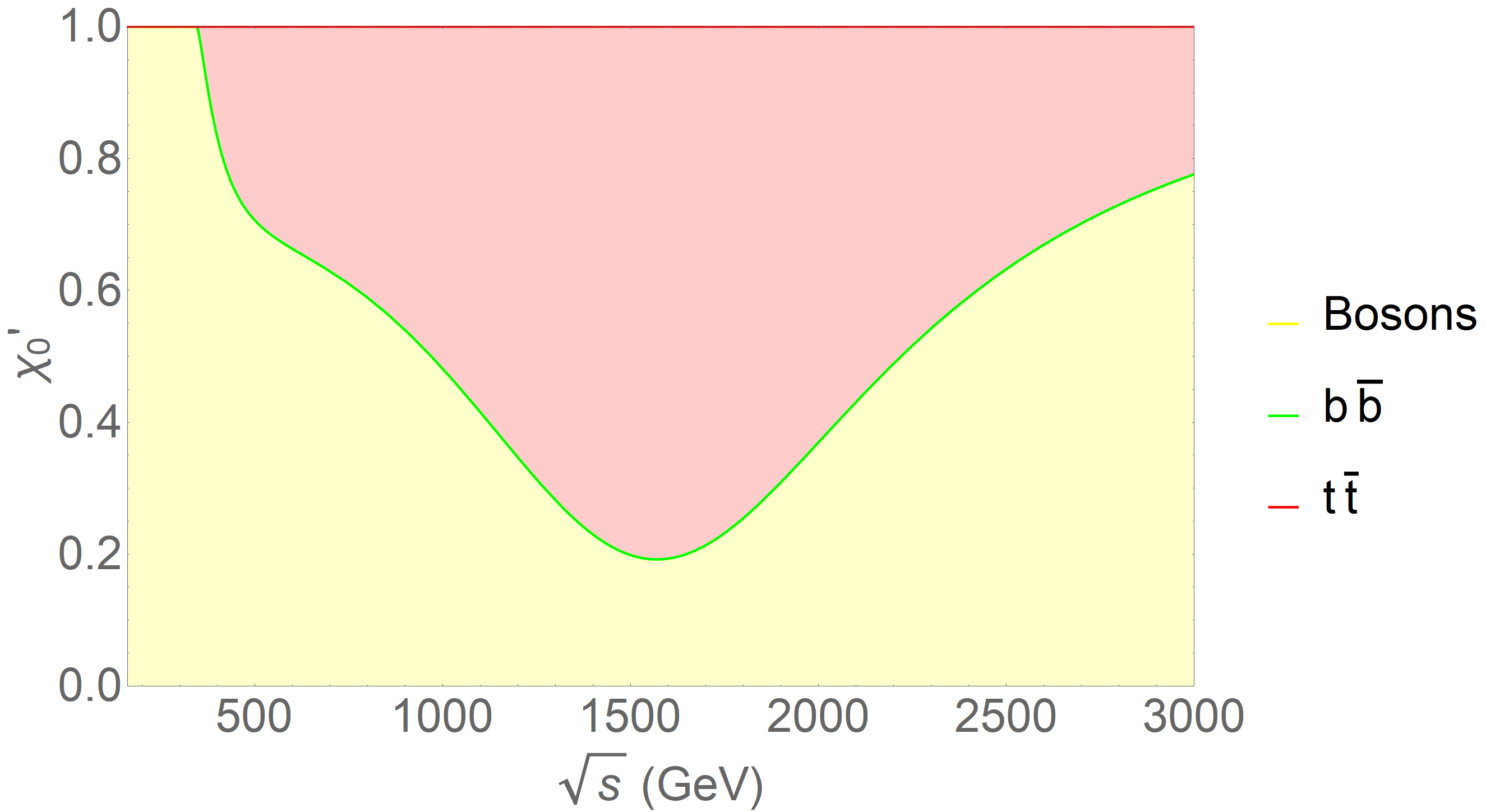}  
  \caption{$J=0$ p-PWA: largest fermion-loop contribution of 80\% for $J=0$  at 1.5 TeV  for ${a=1.011}$, $b=1.045$, $c_1=0.900$ and  ${d_3=1.094}$.  }
  \label{fig:ratior0_prime__bestfit_1500gev}
\end{subfigure}
\hspace*{0.75cm}
\begin{subfigure}{.4 \columnwidth}
  \centering
  \includegraphics[width=1 \linewidth]{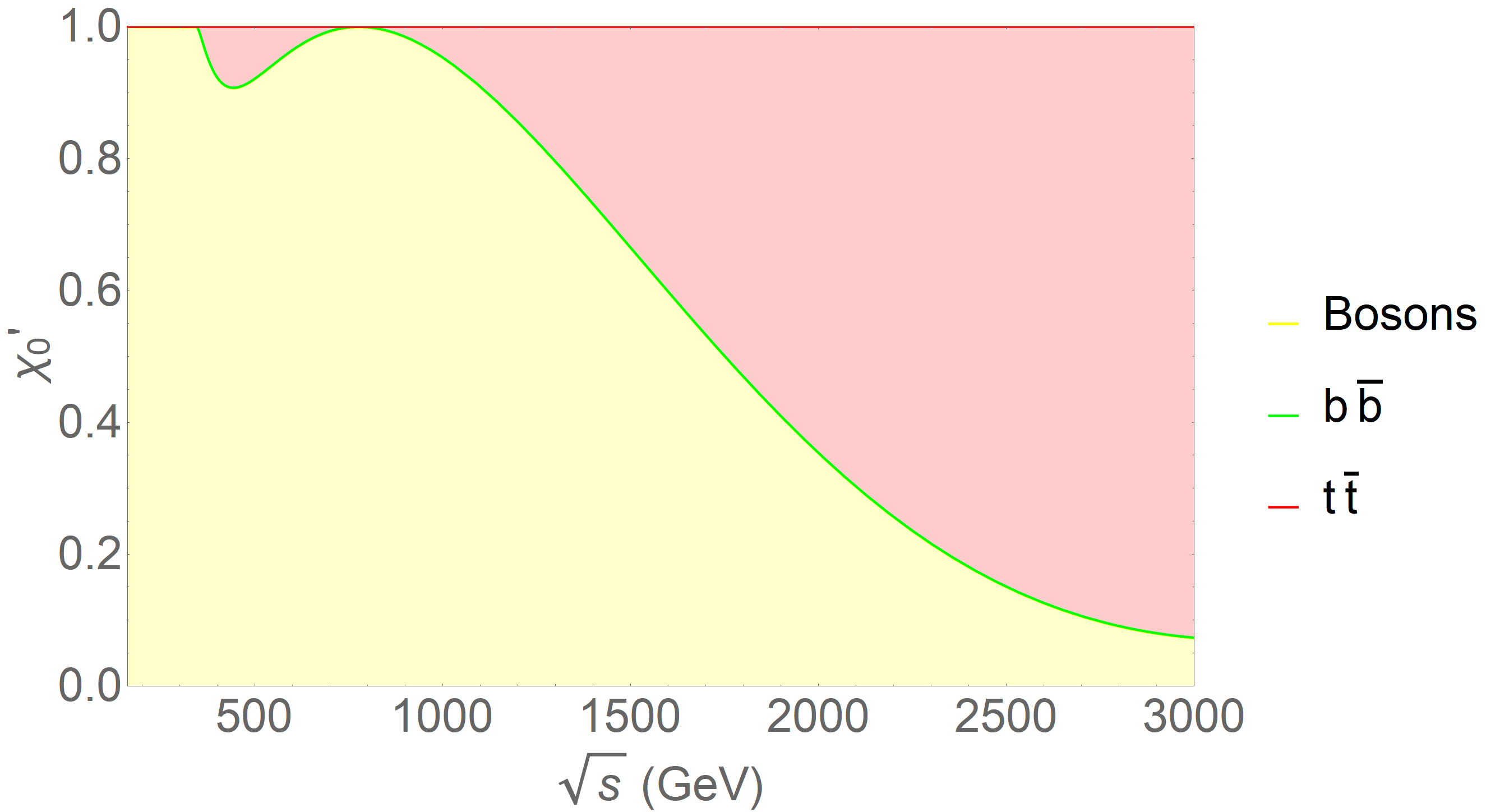}  
  \caption{$J=0$ p-PWA: largest fermion-loop contribution of 93\% for $J=0$ at 3 TeV happens  for$a=1.003$, $b=1.011$, $c_1=1.100$ and  $d_3=1.100$. }
  \label{fig:ratior0_prime__bestfit_3000gev}
\end{subfigure}
\caption{}
\label{fig:ratiobestr0_prime}
\end{figure}


As seen in Fig~\ref{fig:r0_prime_contour}, the contour plots do not dramatically change from the $g'=0$ case. 
Areas around $\sqrt{s} \sim $ 500 GeV are enhanced around  10\% for Figs.~\ref{fig:aj0_prime_plotpartial} and~\ref{fig:bj0_prime_plotpartial} and  5\% for Fig.~\ref{fig:d3j0_prime_plotpartial} (sensitivity to $a$, $b$ and $d_3$, respectively).
On the other side, when it comes to the sensitivity to $c_1$, shown in Fig.~\ref{fig:c1j0_prime_plotpartial},  we find larger contributions: from 20\% around $\sqrt{s} \sim $~500 GeV up to  70\% at 3~TeV when $c_1= 0.9$ and $c_1=1.1$. Finally, the dependence on $d_3$ is negligible just like in the $g'=0$ case, being only relevant for $\sqrt{s}\sim 500$~GeV.

Given the numerous absorptive cuts we have now (9 in total, 2 fermionic and 7 bosonic), the cumulative ratios are difficult to read from one plot, so we have subsumed all boson cuts here. In Fig.~\ref{fig:r0ratio_prime_combined_sm} we can see the corresponding cumulative relative ratios $\chi_i^{0 \, '}$ for the SM. We observe  that in the SM the fermion contributions are not relevant and can be neglected, as in the $g'=0$ case.
In Fig.\ref{fig:ratioj0_prime} we show the $\chi_i^{0\, '}$  cumulative ratios for $a$ and $c_1$ on the borders of the parameter space. Again, the most important parameter is $c_1$, giving rise to  corrections of the order of 60\% and 70\% at 3 TeV when it reaches 1.1 and 0.9, respectively.

If we find the set of parameters which maximizes the fermion corrections, we have $R_0'=80 \%$ for  $a=1.011$, $b=1.045$, $c_1=0.900$ and  $d_3=1.094$ at 1.5 TeV, and $R_0'=93 \%$ for $a=1.003$, $b=1.011$, $c_1=1.100$ and  $d_3=1.100$ at 3 TeV. The contributions for each benchmark energy are shown in Fig.~\ref{fig:ratiobestr0_prime}. 
Again, if we test the sensitivity of $R_0'$ to these optimal parameters, we find that in order to produce large fermion-loop corrections, one needs a fine interplay among the couplings. This is shown in Fig.~\ref{fig:figr0_prime_sensitivity} (in~\ref{app:sensitivity-r0pr} for the sake of clarity)  and it is essentially similar to the previous $R_0$ results in Fig.~\ref{fig:figr0sensitivity}.   

From the plot it is clear that for the $J=0$ p-PWA, fermion-loop corrections should not be neglected at high energies. They can provide a large contribution to the amplitude, even if one has several additional bosonic cuts in the $g'\neq 0$ case (e.g., $\gamma\gamma$).

\subsection{$J=1$  pseudo-PWA: $R'_1$}
\label{section:r1prime}

\begin{figure}[!t]    
\includegraphics[width=10cm]{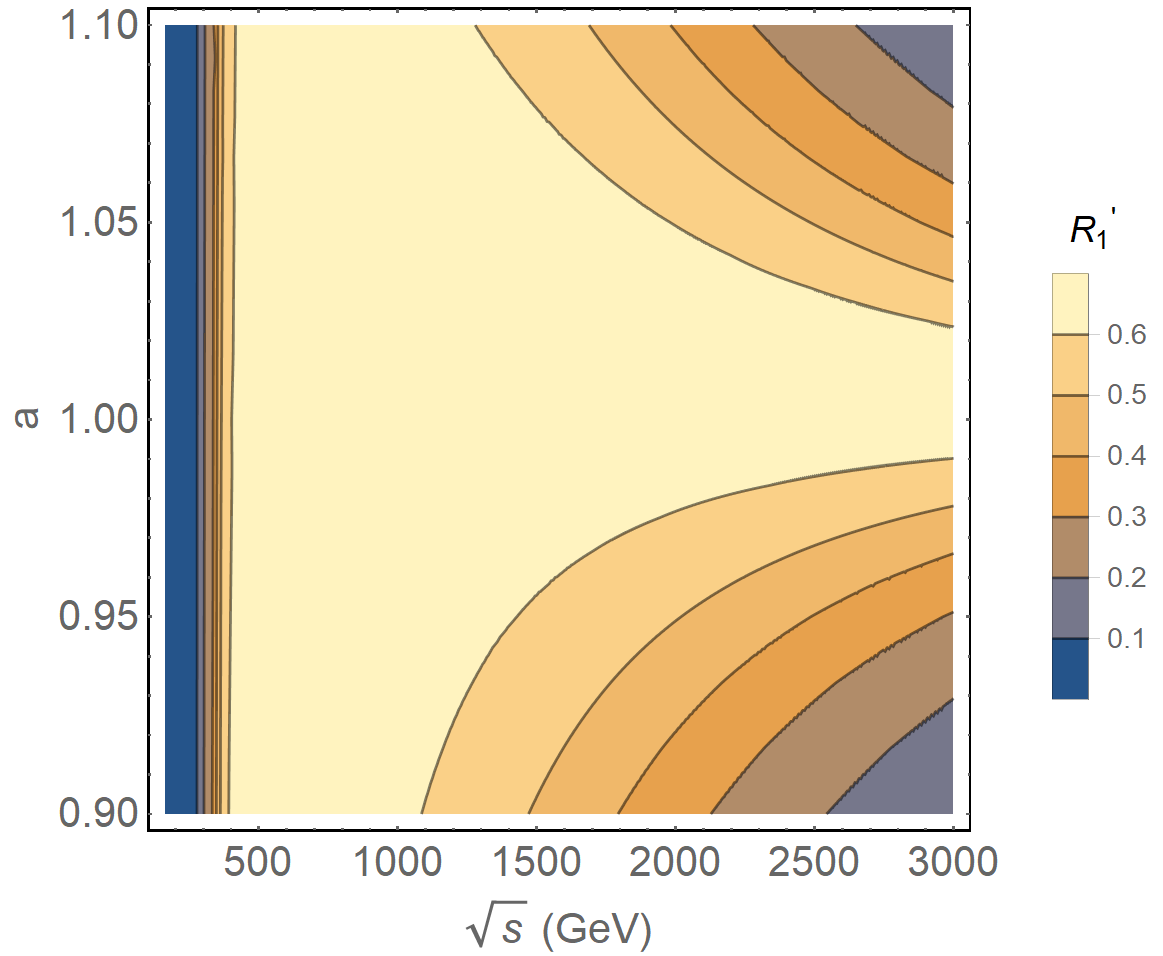}
\centering
\caption{\small $R'_1$ dependence on the $a$ parameter.}
\label{fig:r1chnga_prime}
\end{figure}

The contour plot for the next p-PWA, $J=1$, is shown in Fig \ref{fig:r1chnga_prime}. In this case, the behavior of $R'_1$ around the SM is qualitatively similar to $R_1$, but the corrections are dramatically enhanced. We find $R'_1\sim$ 60\% from 0.5 TeV to 3 TeV in the neighborhood  of the SM. 

\begin{figure}[!t]    
\includegraphics[width=10cm]{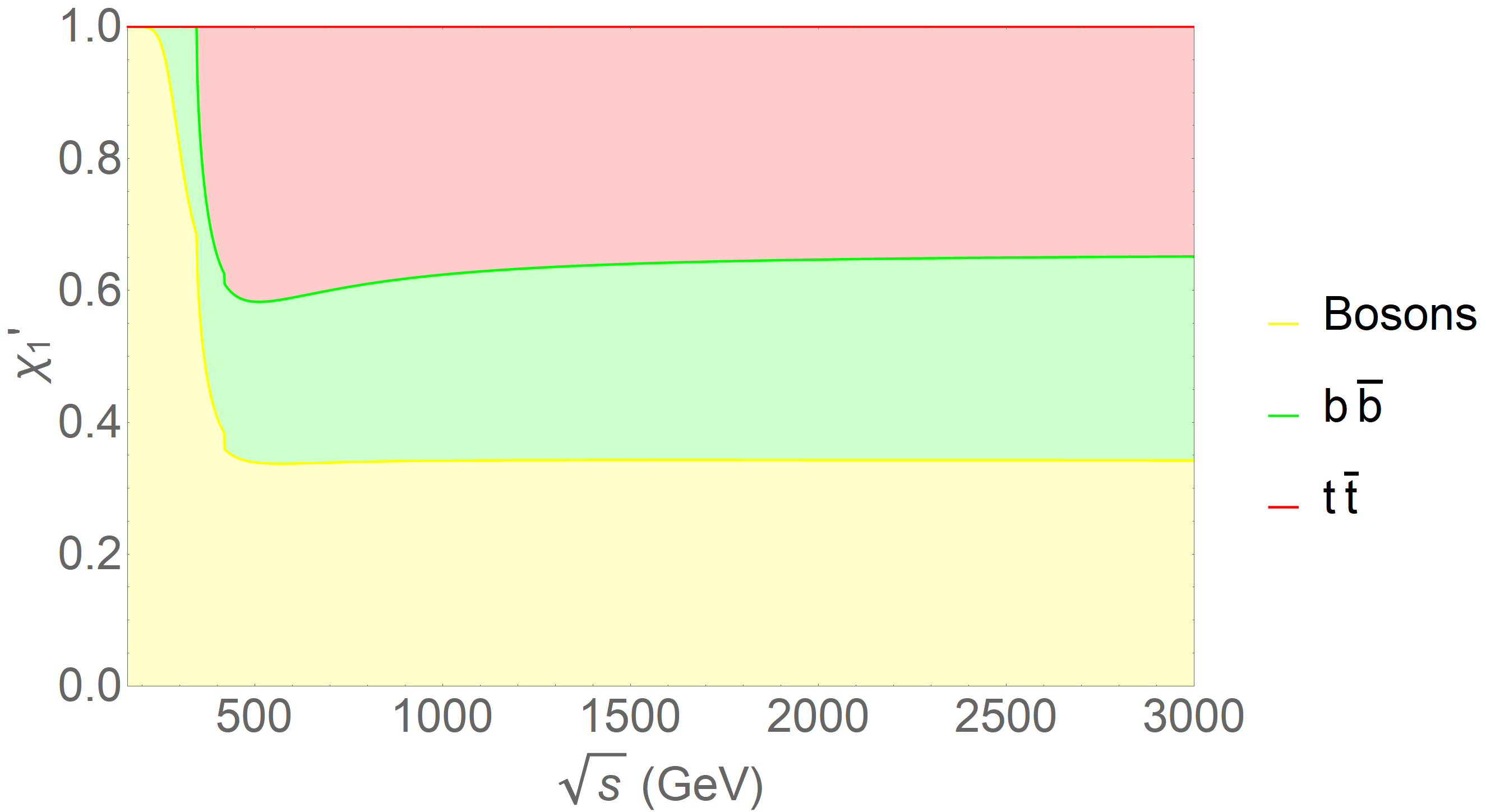}
\centering
\caption{\small Cumulative relative contribution of each channel to $J=1$ p-PWA in the SM.}
\label{fig:r1ratio_prime_sm}
\end{figure}

\begin{figure}[!t]    
\centering
\begin{subfigure}{.4 \columnwidth}
  \centering
  \includegraphics[width=1\linewidth]{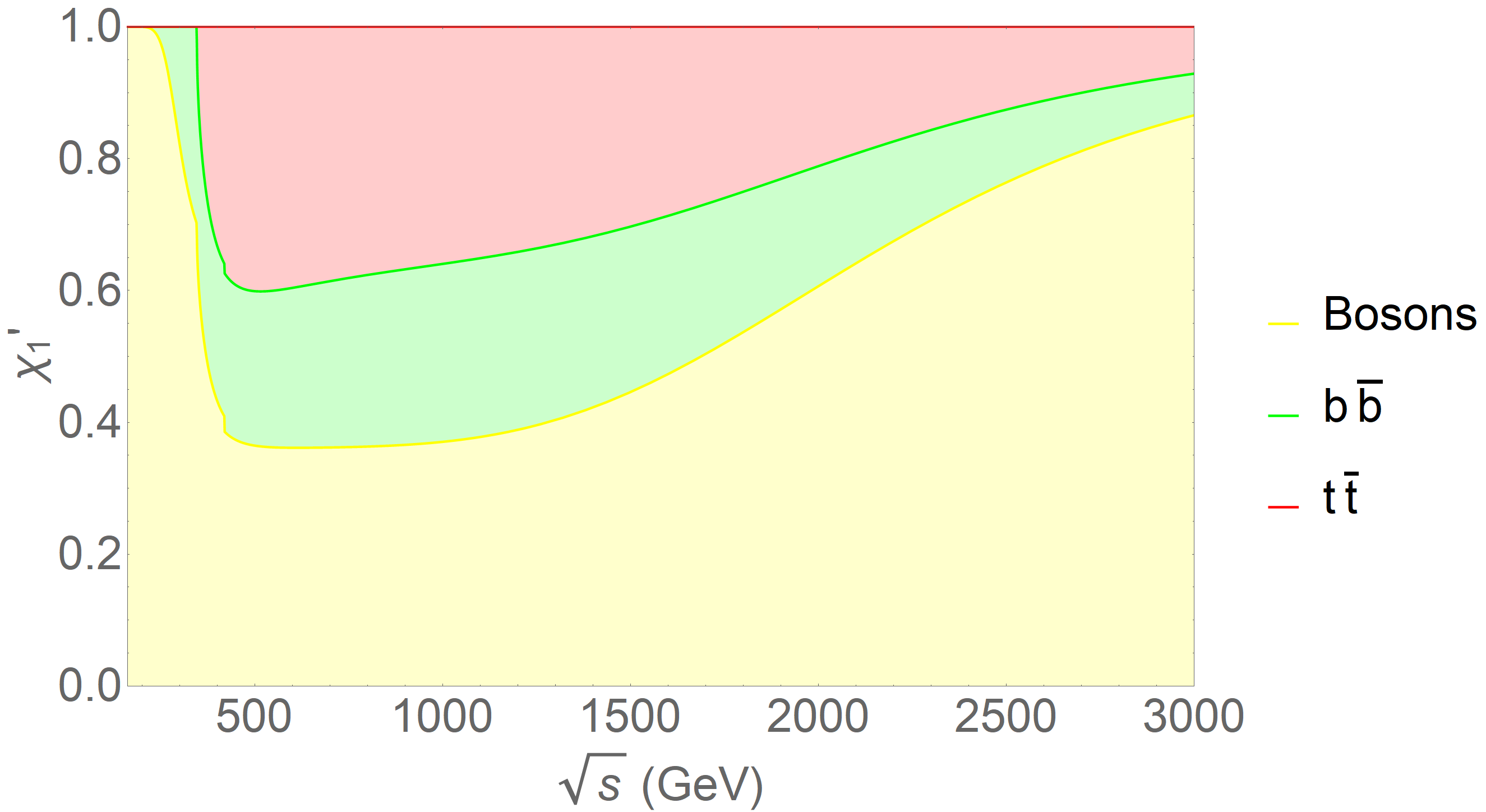}  \caption{}
  \label{fig:ratior1combined_primec}
\end{subfigure}
\hspace*{0.75cm}
\begin{subfigure}{.4 \columnwidth}
  \centering
  \includegraphics[width=1 \linewidth]{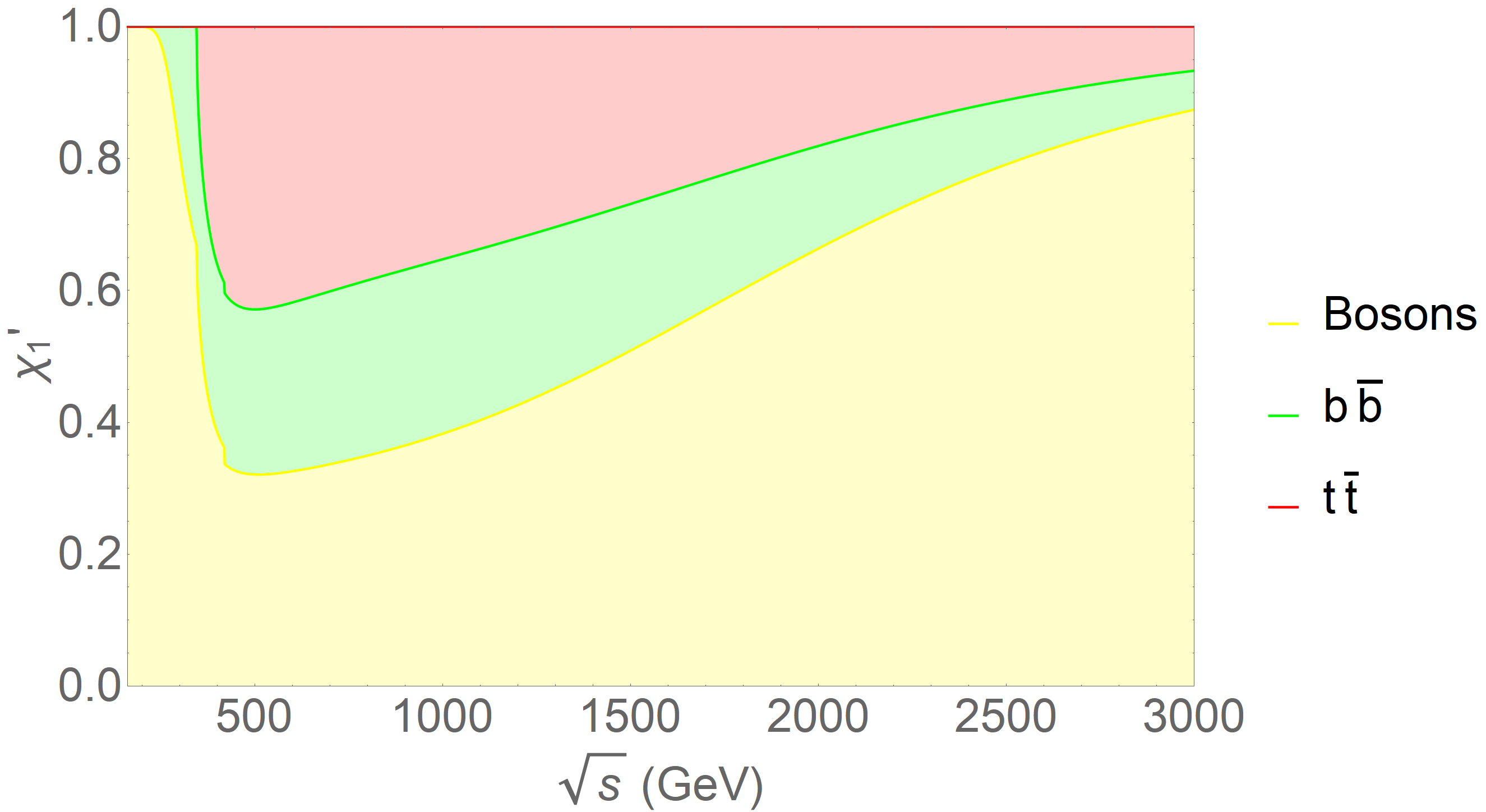}  
  \caption{}
  \label{fig:ratior1a09_prime}
\end{subfigure}
\caption{Cumulative relative contribution of each channel to the $J=1$ p-PWA for $a=1.10$ (left) and $a=0.90$ (right).}
\label{fig:ratior1partial}
\end{figure}

In Fig. \ref{fig:r1ratio_prime_sm}, we see again that in the SM both fermions provide almost 70\% of the amplitude from 500 GeV on when $a=1$. In comparison, for $a=1.1$ and $a=0.9$, they reach a maximum around 500 GeV and they rapidly decrease to around 15\% at 3 TeV, as can be seen in Fig.~\ref{fig:ratior1partial}.


\begin{figure}[!t]    
\centering
\begin{subfigure}{.4 \columnwidth}
  \centering
  \includegraphics[width=1\linewidth]{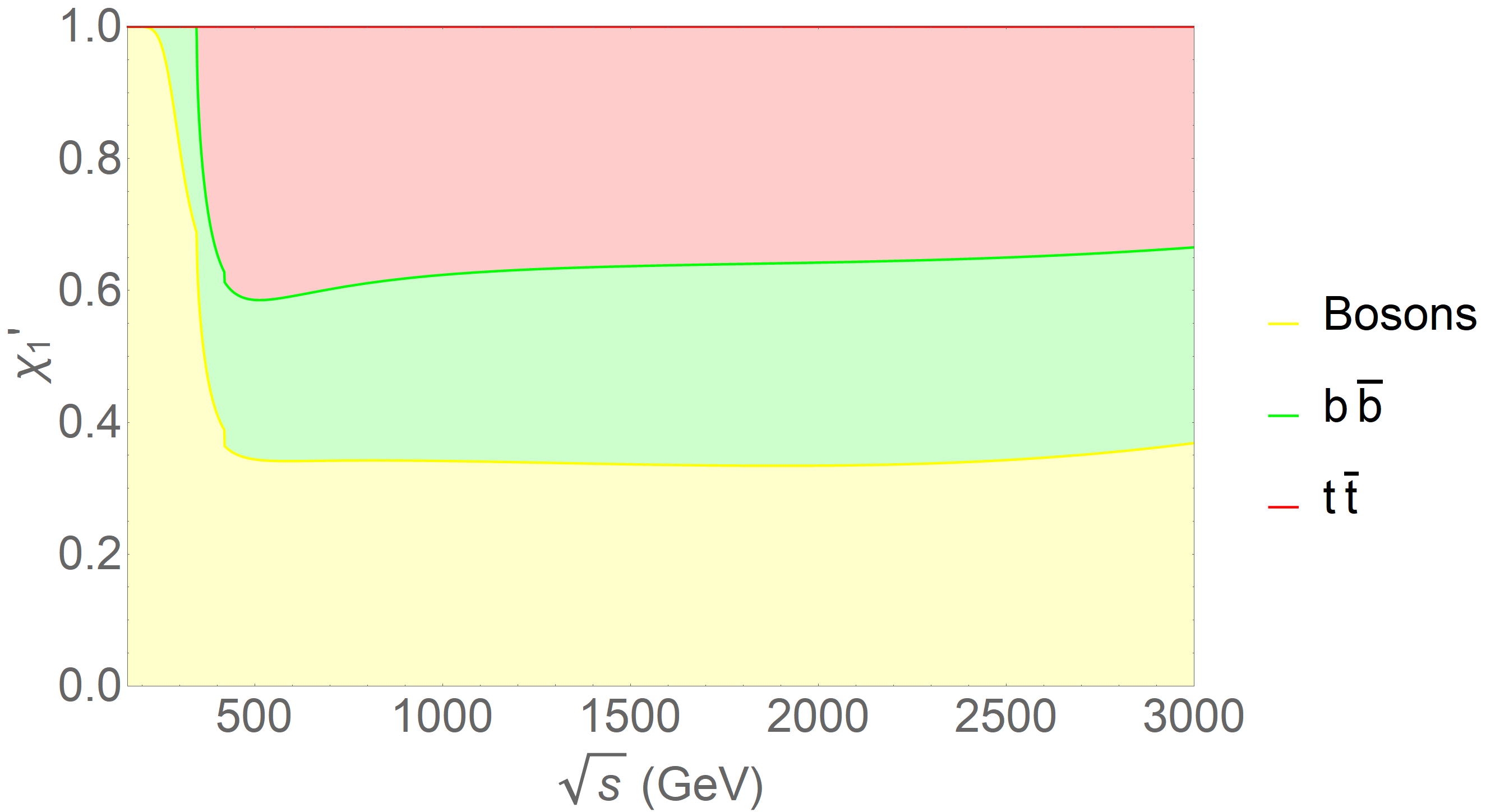}  
  \caption{$J=1$ p-PWA: largest fermion-loop contribution of 66\% at 1.5 TeV to $J=1$ for $a=1.019$. }
  \label{fig:ratior1_bestfit_1500gev}
\end{subfigure}
\hspace*{0.75cm}
\begin{subfigure}{.4 \columnwidth}
  \centering
  \includegraphics[width=1 \linewidth]{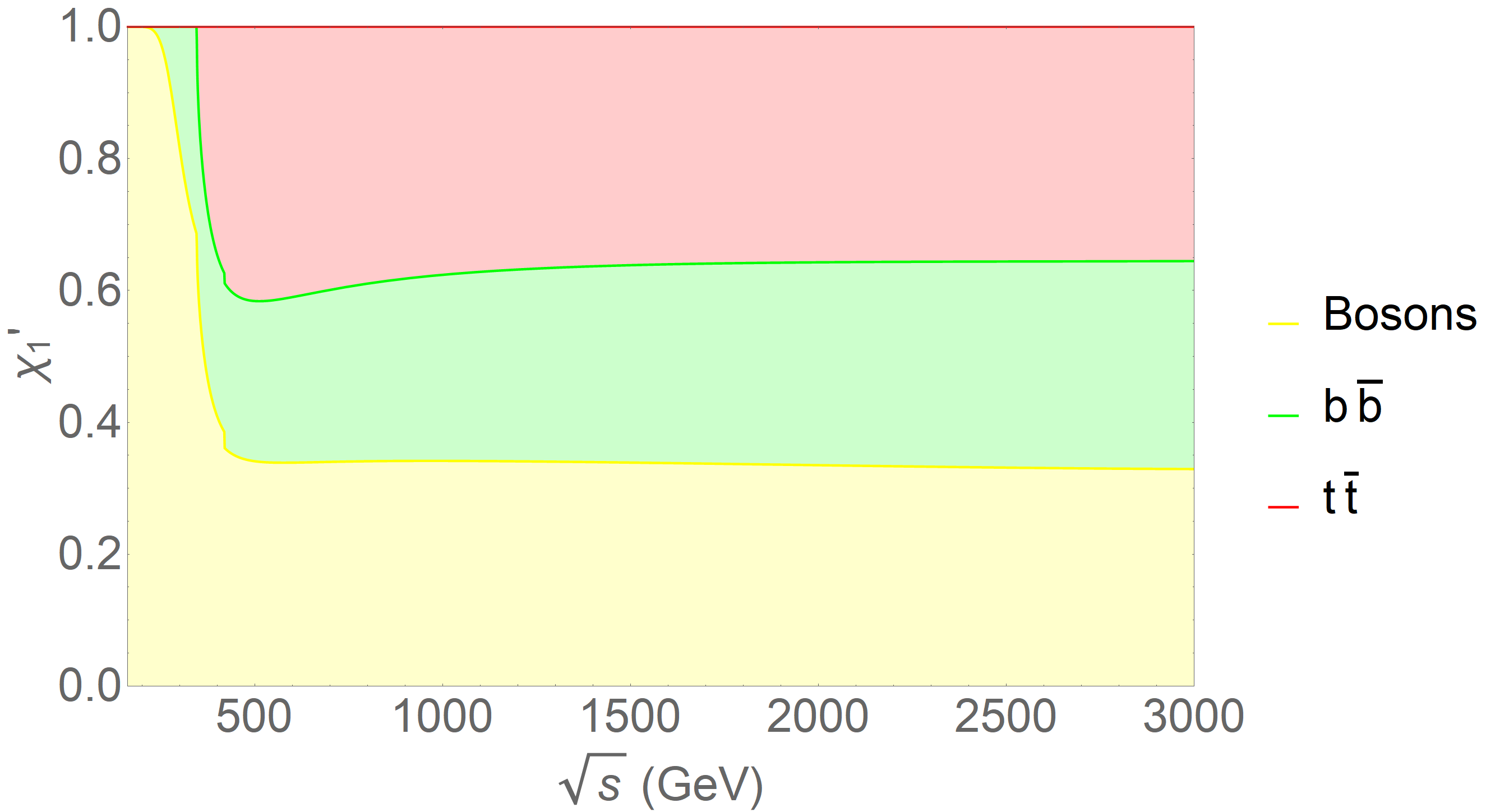}  
  \caption{$J=1$ p-PWA: largest fermion-loop contribution of 67\% at 3 TeV to $J=1$ for $a=1.007$.  }
  \label{fig:ratior1_bestfit_3000gev}
\end{subfigure}
\caption{}
\label{}
\end{figure}


As for $R_1$, the p-PWA ratio $R'_1$ only depends on $a$. We then look for the point in parameter space that maximizes $R'_1$. The optimal values of $a$ for 1.5 TeV and 3 TeV are  $a=1.019$ (with  $R'_1=66 \%$) and $a=1.007$ (with a $R'_1=67 \%$), respectively.

If we test the sensitivity of $R_1'$ to these optimal parameters, we find in Fig.~\ref{fig:r1_prime_sensitivity} (in~\ref{app:sensitivity-r1pr} for the sake of clarity) that fermion contributions remain sizable for the whole range of $a$ studied here. These are essentially the same conclusions found for $R_1$ in Fig.~\ref{fig:r1_sensitivity}.

Again, values close to $a=1$ yield significant fermion-loop corrections. These are of the order of 60\% for the optimal value of $a$, around three times larger than the optimal value for $R_1$. Hence, in the case of angular cuts (e.g., $|\cos \theta|\leq 0.9$), top and bottom intermediate channels should not be neglected.

\section{Specific Scenario: Minimal Composite Higgs Model}

When it comes to the importance of fermionic cuts, it is clear that they are relevant for some regions of the coupling space. Although these couplings could, in principle, take any value, we would like to be able to link them to specific NP scenarios where, in general, all effective parameters deviate from the SM in the particular way established by the model. 
For illustration,  
we will study here the $SO(5)/SO(4)$ Minimal Composite Higgs Model (MCHM)~\cite{MCHM}, where the Higgs is a pseudo-Goldstone boson of an underlying strongly-coupled theory. In this model, the couplings depend explicitly on the characteristic MCHM scale $f$. 

The expressions for the relevant couplings for our analysis are~\cite{MCHM,Kanemura:2014kga},
\begin{equation*}
    a^*= c_1^*=d_3^*=\sqrt{1-\xi}\, ,\quad b^*=1-2 \xi\, ,  
    \quad \text{with} \quad \xi=v^2/f^2\, .
\end{equation*}
Due to the structure of the MCHM, only values smaller than 1 are allowed for these effective  couplings.  
Note that the four HEFT couplings are determined by the NP scale $f$. We have then computed the previous PWA's and p-PWA's for various values of $f$ within this model. To ease the analysis, we will provide the corresponding value of the $hWW$ coupling $a$ together with $f$ in the labels of the different curves (Figs.~\ref{fig:r0compositeisospin}--\ref{fig:r1composite_prime}). For $0.90\leq a^*\leq 1.00$ this implies $f\geq 0.56$~TeV, with $a^*\to 1$ for $f\to\infty$.    

\subsection{Limit  $g'=0$ }

\begin{figure}[!t]    
\includegraphics[width=10cm]{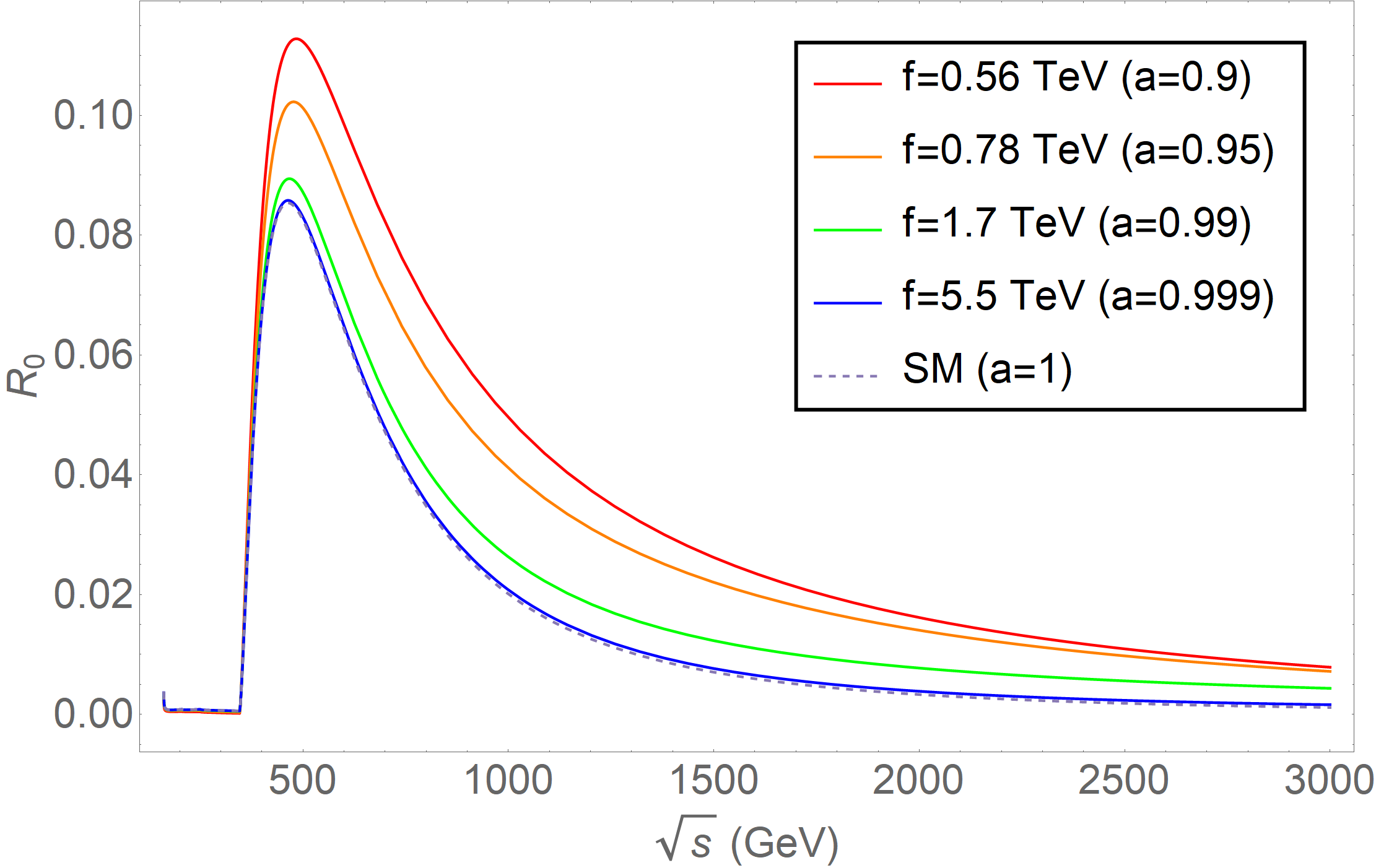}
\centering
\caption{\small Ratio for the $R_0$ PWA in the MCHM.}
\label{fig:r0compositeisospin}
\end{figure}

\begin{figure}[!t]    
\includegraphics[width=10cm]{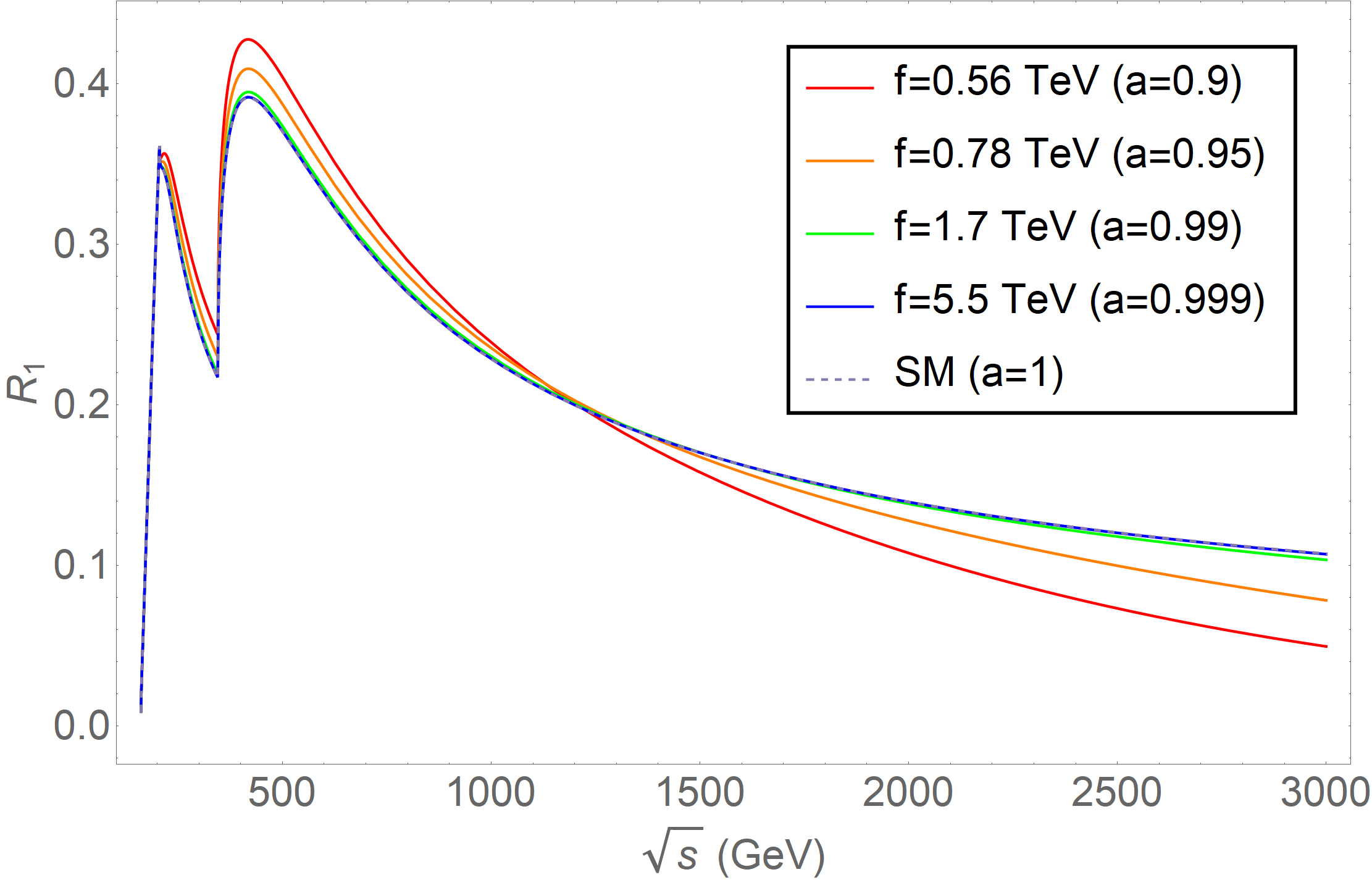}
\centering
\caption{\small Ratio for the $R_1$ PWA in the MCHM.}
\label{fig:r1compositeisospin}
\end{figure}

We have plotted the ratios $R_0$ and $R_1$ as a function of $\sqrt{s}$ for different values of $f$ 
in Figs.~\ref{fig:r0compositeisospin} and~\ref{fig:r1compositeisospin}, respectively. 

As it can be seen in Fig. \ref{fig:r0compositeisospin}, $R_0$ is drastically changed. Below the threshold of $t \bar{t}$ production only $b \bar{b}$ is present but its contribution is negligible. $R_0$ rapidly increases when top corrections enter at $\sqrt{s}\simeq $350 GeV. In the present study, we find the maximum value $R_0=11\%$ for $a^*=0.9$, at the boundary of our allowed range. 
It then quickly decreases for larger values of $a^*$ at all energies. 
The SM curve provides the lowest limit for the fermion correction, while the $a^*=0.9$ curve provides the upper bound.    

For $R_1$, (see Fig. \ref{fig:r1compositeisospin})   
we observe a similar behavior. When the $t \bar{t}$ cut appears $R_1$ reaches a maximum of 41\% around 500 GeV for an $a^*=0.9$. Again, all curves decrease rapidly but the behavior at large energies is different. In this case, the SM curve provides the largest fermion correction while the $a^*=0.9$ curve the smallest ones.

In summary, in both $R_0$ and $R_1$ cases, when we restrict ourselves to the MCHM the largest corrections appear always at $\sqrt{s}\sim 500$~GeV, with $a^*=0.9$ the value that maximizes fermion corrections at that energy point. At large energies,  $\sqrt{s} \sim$ 3 TeV,  $R_0$ becomes negligible (maximum $R_0\sim  1\%$) for $a^*=0.9$, while $R_1$ presents a significant contribution (maximum $R_1\sim  10 \%$) for $a=1$.

\subsection{Beyond the $g'=0$ limit }

\begin{figure}[!ht]    
\includegraphics[width=10cm]{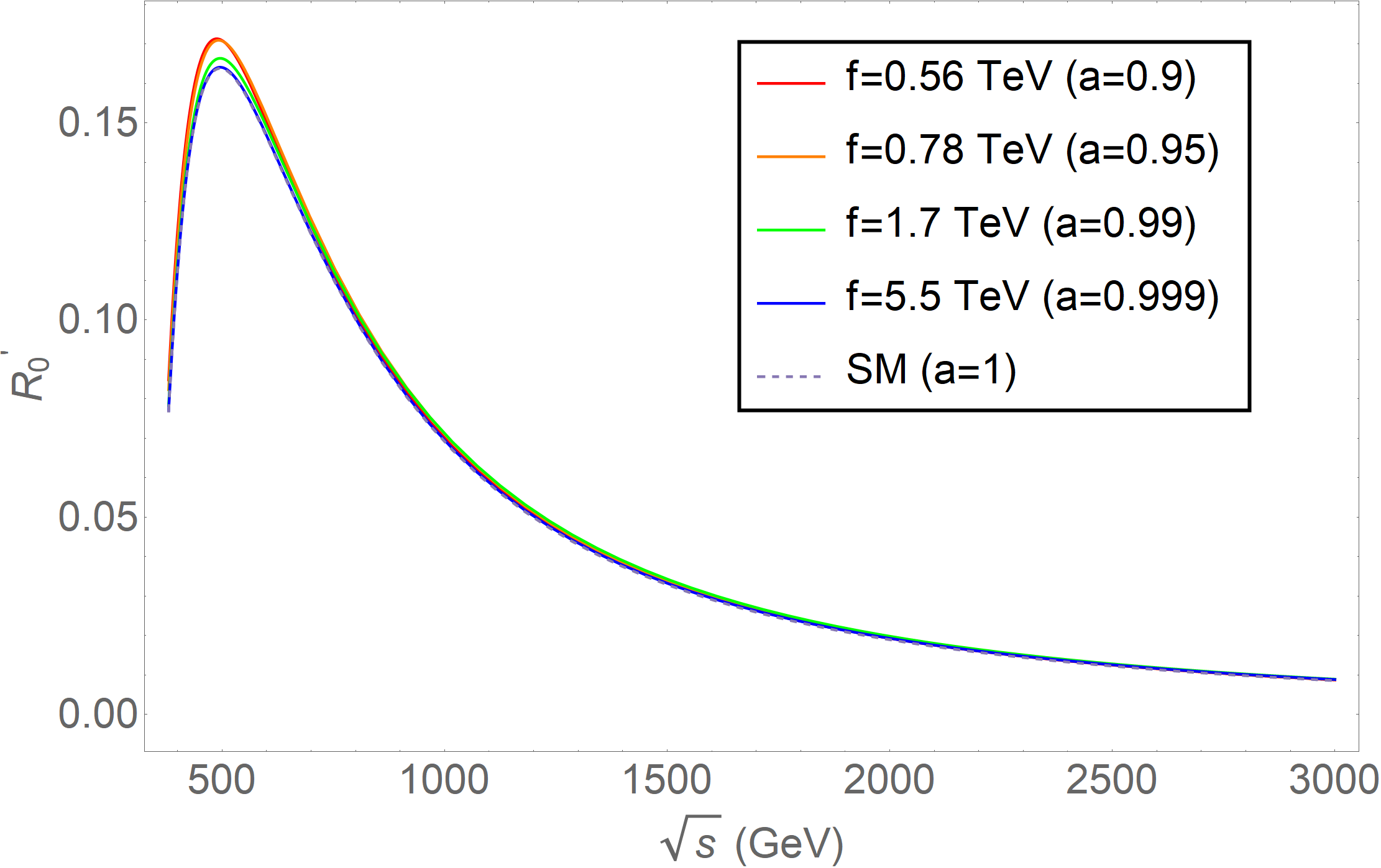}
\centering
\caption{\small Ratio for the $R'_0$ PWA in the MCHM}
\label{fig:r0compositepartial}
\end{figure}

As can be seen in Fig.~\ref{fig:r0compositepartial}, 
the result for $R_0'$ is very similar to $R_0$ in the $g'=0$ case; the largest fermion contribution is found around $\sqrt{s}\sim 500$~GeV, being around 17\%. At high energies the $R_0'$ decreases rapidly, becoming negligible. 

For $R_1'$ we note an interesting behavior. 
As in the $R_1$ analysis, we find a maximum for $R_1'$ around 500 GeV but somewhat larger ($R_1'\sim 65\%$). 
However, as we increase the energy there are curves with coupling values close to the SM which decrease very slowly with the CM energy. 
This shows the amplitudes depend highly on the angular cut, as was mentioned for the $W^+W^- \to W^+ W^-$ corrections in \cite{Denner:1997kq}. Again, the maximum contribution is found for the SM-curve at high energies and is about 65\%.

\begin{figure}[!t]    
\includegraphics[width=10cm]{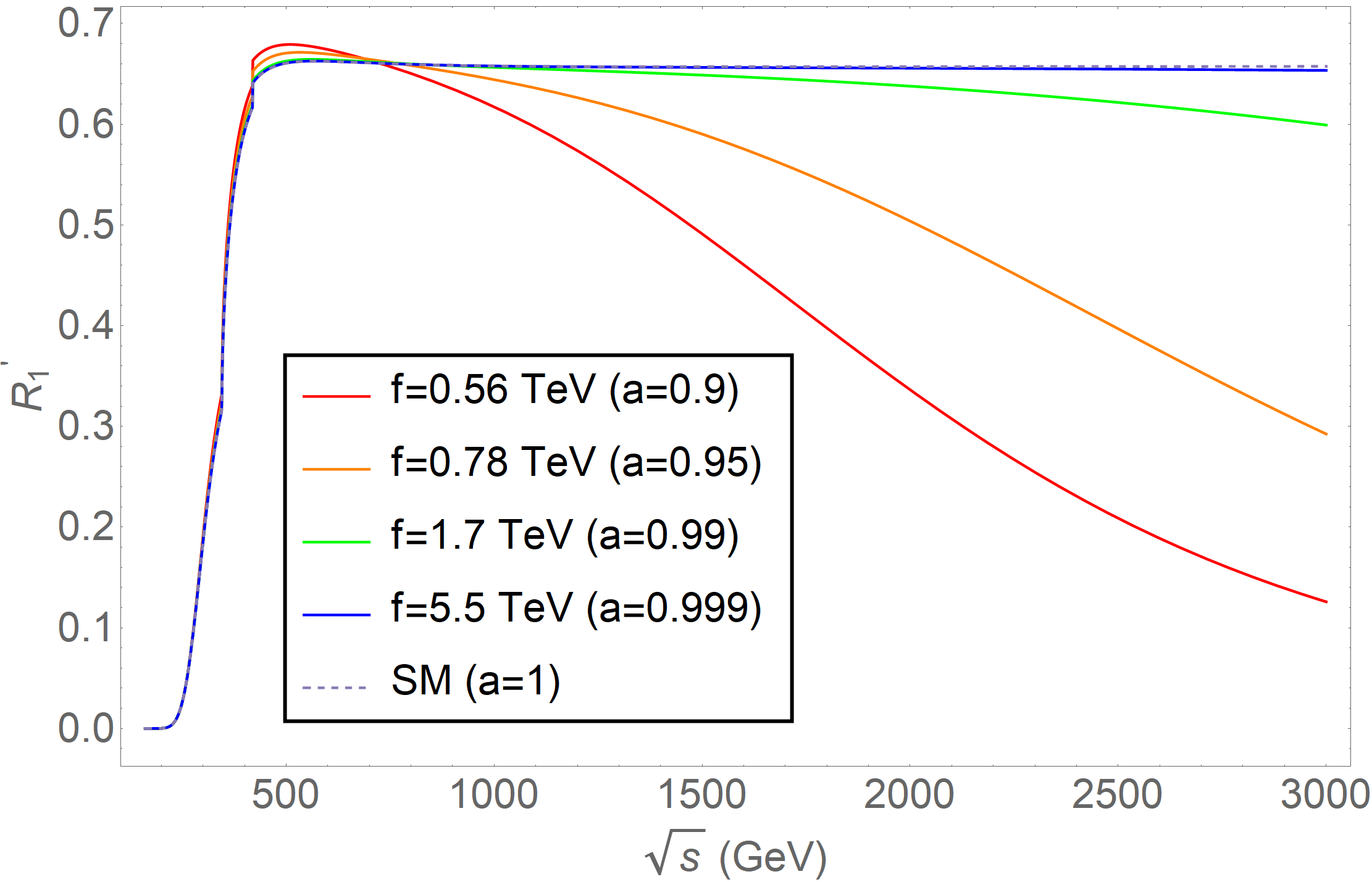}
\centering
\caption{\small Ratio for the $R'_1$ PWA in the MCHM}
\label{fig:r1composite_prime}
\end{figure}

\section{Conclusions}

In this work, we have  pondered in detail the widespread assumption that fermion-loop corrections can be neglected at high energies within the HEFT framework. For this, we have compared the imaginary part arising from top/bottom quark loops and that from boson loops in the elastic $W^+W^- \to W^+W^-$ scattering. 
We have included all intermediate channels and all possible polarization states not included  in previous preliminary works~\cite{Dobado:2021ozt,Dobado:2020lil}.

In order to analyze the importance of fermion loops, we have computed the ratios $R_0$ and $R_1$ 
for the first partial wave amplitudes, $J=0$ and $J=1$, respectively. 
$R_J$ close to 0 indicates a dominance of boson loops whereas a value close to 1 points out that fermion cuts dominate.

Due to the presence of infrared divergences in boson-loop diagrams ($WW$ cut) where the momentum of a $t$-channel photon goes to zero, a full angular projection onto partial wave amplitudes (PWA) is not possible. 
In order 
to deal with the PWA and to project onto the full angle domain, we have considered two approaches: 1) set $g'=0$, which removes photon interactions and, hence, the infrared divergent diagrams; 2) keep $g'\neq 0$ but impose an angular cut ($\abs{\cos \theta}< 0.9$), which restricts the angular integration avoiding the angular divergence.
These two approaches give rise to   
the ratios $R_J$ for the imaginary part of the PWA's with $g'=0$ (approach 1) and the $R_J'$ ratios for the imaginary part of the so-called pseudo-PWA's (approach 2). We have explored these two types of ratios, scanning the possible values of the relevant HEFT couplings. This has allowed us to assess the validity of the assumption of neglecting top/bottom quark loops. 

In the first scenario, $g'=0$, there are wide regions where the bosonic loop contributions are dominant as can be seen for the $J=0$ PWA in Fig.~\ref{fig:figr0contour}. However, this is not the case in some ranges of the parameter space; large deviations of $c_1$ ($h\bar{t}t$ coupling) from the SM yield significant top-quark contributions. The $b\bar{b}$ contributions to $J=0$ are not relevant as can be seen in Figs.~\ref{fig:r0ratio_isospin_SM} and~\ref{fig:ratiosr0}, given the fact that they are proportional to  
$M_b$. For the $J=1$ PWA, the same occurs when $a$ ($hWW$ coupling) is close to 1, as can be seen  in Fig. \ref{fig:r1_countour_a_isospin}. This minimizes the bosonic contribution and leads to a higher $R_1$. In this case, the $b \bar{b}$ cut yields a relevant contribution as it can be seen in Figs.~\ref{fig:r1ratio_prime_sm} and~\ref{fig:ratior1partial}. In particular, the amplitudes of both fermion cuts present a similar correction at high energies, showing that just one heavy quark in the EW doublet is enough 
to obtain significant corrections to the $J=1$ PWA. 

In the second scenario, $g'=0$ with angular cuts, we find that the results for the $J=0$ p-PWA $R_0'$ are similar to those for $R_0$, as one can see in Fig.~\ref{fig:r0_prime_contour}. Large deviations of $c_1$ from the SM yield important top quark contributions, again with negligible effects from the $b \bar{b}$ cut.  
For the $J=1$ p-PWA. we observe a significant raise in the ratio $R_1'$ with respect to $R_1$. This can be seen in Fig.~\ref{fig:r1chnga_prime}, indicating that the $J=1$ partial wave amplitude is highly dependent on angular cuts. 
Fig.~\ref{fig:r1ratio_prime_sm} shows a significant contribution to $R_1'$ from the $b \bar{b}$ cut, due to the same reasons discussed in the $g'=0$ case for $R_1$ in Fig.~\ref{fig:ratior1}. 

There are also configurations for the four HEFT couplings ($a$, $b$, $c_1$ and $d_3$) which make these fermion contributions even more important. We summarize the largest corrections we have found for the PWA and pseudo-PWA in Tables~\ref{tab:r0summary} and~\ref{tab:r1summary}, respectively $J=0$ and $J=1$.   We have looked for the point in parameter space that maximizes fermion contributions at two benchmark energies: $\sqrt{s}=1.5$~TeV and $\sqrt{s}=3$~TeV. 
We have computed  the sensitivity of these optimal HEFT coupling values by fixing three of them and varying one at a time. We can observe that there is a fine interplay of the couplings $a,b$ and $c_1$ which maximize $R_0$ and $R_0'$ for these benchmark energies (Figs.~\ref{fig:figr0sensitivity} and~\ref{fig:figr0_prime_sensitivity}, respectively). One can also see that the value of $d_3$ is not relevant for the analysis. For $J=1$, we find that values close to $a=1$ minimize the bosonic contribution, yielding higher $R_1$ and $R_1'$ ratios (Figs.~\ref{fig:r1_sensitivity} and~\ref{fig:r1_prime_sensitivity}, respectively).

\begin{table}[!t]
\centering
\begin{tabular}{cccccc}   
\hline
$\sqrt{s}$   (TeV) & $a-1$       & $b-1$         & $c_1-1$   & $d_3-1$     & $J=0$  \\ \hline
1.5 (PWA)       & 0.023  & 0.100       & -0.100    & 0.100      & $R_0$=76\% 
\\ \hline
3 (PWA)        & 0.008  & 0.035   & 0.100    & -0.100  & $R_0$=94\% 
\\ \hline
1.5 (p-PWA)      & 0.011  & 0.045    & -0.100 & 0.094  & $R_0'$=81\% 
\\ \hline
3 (p-PWA)        & 0.003 & 0.011 & 0.100 & 0.100      & $R_0'$=93\% 
\\ \hline
\end{tabular}
\caption{\small Corrections to $J=0$ PWA for the $g'=0$ case (first two rows) and the $J=0$ p-PWA (last two rows).  In the second, third, fourth and fifth columns, we provide, respectively, the values of $a$, $b$, $c_1$ and $d_3$ that maximize the fermion-loop contributions. 
}
\label{tab:r0summary}
\end{table}

\begin{table}[!t]
\centering
\begin{tabular}{ccc}    
\hline
$\sqrt{s}$   (TeV) & $a-1$       & $J=1$  \\ \hline
1.5 (PWA)      & -0.009 & $R_1$=18\%
\\ \hline
3 (PWA)         & 0.013 & $R_1$=12\%
\\ \hline
1.5 (p-PWA)       & 0.019   & $R_1'$= 66\% 
\\ \hline
3 (p-PWA)         & 0.007   & $R_1'$= 67\%   \\ \hline
\end{tabular}
\caption{\small Corrections to $J=1$ PWA for the $g'=0$ case (first two rows) and the $J=1$ p-PWA (last two rows). In the second column, we provide the value of $a$ that  maximizes the fermion-loop contributions.}
\label{tab:r1summary}
\end{table}

Based on what has been described above, we conclude that the assumption of neglecting the imaginary part of top-/bottom-/quark-loop contributions to $W^+ W^- \to W^+ W^-$ in favor of the imaginary part of bosonic loops does not entirely hold. For the case $J=0$, it is true there are wide ranges where fermion-loops contributions are negligible. However, this is false in some regions, where a $\pm$0.1 deviation of $c_1$ from 1 (SM) would give a 22\% and 18\% top-quark-loop contribution to $R_0$ and $R_0'$, respectivelye. Likewise, some configurations of $a,b,c_1$ and $d_3$ can make fermion loops even dominant, as shown in Table~\ref{tab:r0summary}. For $J=1$ something similar occurs since we do not need to deviate so much from $a=1$ (SM). Values of $a$ close to 1 yield significant top- and bottom-quark-loop contributions to both PWA and p-PWA for $J=1$, as shown in Table~\ref{tab:r1summary}.

For the MCHM case, 
we do not find meaningful contributions to the $J=0$ ratios, as can be seen in Figs.~\ref{fig:r0compositeisospin} and~\ref{fig:r0compositepartial}. Both plots show maximums around 500~GeV, but the ratios decay rapidly with the CM energy. For $J=1$, a value of $a=1$ ($f \rightarrow \infty$ TeV ) produces a maximum $R_1=10\%$ at 3~TeV, as can be seen in Fig.~\ref{fig:r1compositeisospin}.  When it comes to the $g'\ne 0$ case, given the strong dependence on the angular cut, $R_1'$ (Fig.~\ref{fig:r1composite_prime}) is enhanced  and 
takes an almost constant value 
$R_1'\approx 65\%$ for an $hWW$ coupling close to the SM one ($a\approx 1$). 
Therefore in the MCHM scenario, the imaginary top-/bottom-/quark- loop corrections 
would enhance the $J=1$ partial wave considerably more than the $J=0$.
Note that in the MCHM, the HEFT relevant parameters need to be smaller than 1 due to their particular dependence on the NP scale $f$.

Currently we are working on the full one-loop contribution~\cite{in-preparation}. 
We plan to complete the present computation with the real part of the one-loop amplitudes, where fermion contributions might also be important or even dominant, as we have found in some cases for the imaginary part.  
%
%
Along with this, we plan to deal with  the possibility of a strongly interacting electroweak symmetry breaking sector and the problem of unitarization of the whole amplitude for all VBS channels~\cite{in-preparation,Dobado:2019fxe}.

\section*{Acknowledgements} We would like to thank our collaborators A. Castillo, R. L. Delgado and F. Llanes-Estrada,  who participated in the earlier parts of the research presented in this article~\cite{Castillo:2016erh}. We also want to thank I.~Asi\'ain, M.J. Herrero and P.D. Ruiz-Femen\'\i a for useful discussions. 
This research is partly supported by the Ministerio de Ciencia e Inovaci\'on under research grants FPA2016-75654-C2-1-P and PID2019-108655GB-I00; by the EU STRONG-2020 project under the program H2020-INFRAIA-2018-1 [grantagreement no. 824093]; and by the STSM Grant from COST Action CA16108. C. Quezada-Calonge has been funded by the MINECO (Spain) predoctoral grant
BES-2017-082408.

\appendix

\section{Kinematics}
\label{section:kinematics}

We present the following kinematics in the center-of-mass frame used to calculate the required processes. With this and the Mandelstam variables defined as usual, one should be able to obtain the amplitudes in \ref{section:scatterin_gamplitudes}. In order to not repeat a large quantity of same polarization and momenta vectors, we will only detail the polarization and momentum of the new final states. For example, if one wants to calculate the amplitude $\mathcal{A}(W^+ (p_1,\epsilon_1^L) W^-(p_2,\epsilon_2^L) \rightarrow \gamma (p_3, \epsilon_3^a) Z(p_4,\epsilon_4^b))$ one needs to use the polarization and momentum defined in \ref{ssec:gammgammakinematics} for the photon and the  polarization and momentum defined in \ref{ssec:zzkinematics} for the $Z$ -boson. The only exception to this is the amplitude with fermions in the final state, where we use the momenta and polarizations for the W-bosons detailed in \ref{ssec:hhkinematics}.

\subsection{$W^+(p_1',\epsilon_1^{L'})W^-(p_2',\epsilon_2^{L'}) \rightarrow t(p_3,\lambda_3) \bar{t}(p_4,\lambda_4)$ }
\label{ssec:ttkinematics}

For the special case of fermions in the final state, we will set their momenta in the z-axis, facilitating the calculation of the product of spinor chains. The angular dependence hence comes from the initial states of the W-bosons.

As usual, $\theta$ is the angle between particles 1 and 3, $\phi$ is the azimuth angle and $\epsilon_1^{L'}$ and $\epsilon_2^{L'}$ refer to the longitudinal polarization of the W-bosons:
\begin{equation} 
\begin{split}
p_1'= \left( E, \abs{\vec{p}}\sin (\theta) \cos (\phi),  \abs{\vec{p}}  \sin (\theta) \sin (\phi) ,  \abs{\vec{p}}   \cos (\theta)  \right)\, , 
& \quad p_2'= \left( E , -\abs{\vec{p}} \sin (\theta) \cos (\phi),  -\abs{\vec{p}}  \sin (\theta) \sin (\phi) ,  -\abs{\vec{p}}   \cos (\theta)  \right)  \, , 
\\
p_3'= \left( E_t, 0, 0,  \abs{\vec{p_t}}    \right) \, ,  
&  \quad p_4'= \left( E_t , 0,0 ,  -\abs{\vec{p_t}}    \right) ,
\end{split}
\end{equation}
\begin{equation}
\begin{split}
\epsilon_1^{L'}=  \frac{E}{M_W}\left( \abs{\vec{p}},    E \sin (\theta) \cos (\phi), E \sin (\theta) \sin ( \phi), E \cos (\theta)  \right) \, ,  
\\
 \epsilon_2^{L'}=  \frac{E}{M_W}\left( \abs{\vec{p}},   - E \sin (\theta) \cos (\phi), -E \sin (\theta) \sin ( \phi), -E \cos (\theta)  \right),
\end{split}
\end{equation}
\begin{equation} 
\begin{split}
 u_3^+(p_3,M_t)=\left(
\begin{array}{c}
 \sqrt{\text{E}_t-\text{p}_t} \\
 0 \\
 \sqrt{\text{p}_t+\text{E}_t} \\
 0 \\
\end{array}
\right)\, , 
&  \qquad  u_3^-(p_3,M_t)=\left(
\begin{array}{c}
 0 \\
 \sqrt{\text{p}_t+\text{E}_t} \\
 0 \\
 \sqrt{\text{E}_t-\text{p}_t} \\
\end{array}
\right) \, , 
\\
  v_4^+(p_4,M_t)=\left(
\begin{array}{c}
 \sqrt{\text{p}_t+\text{E}_t} \\
 0 \\
 -\sqrt{\text{E}_t-\text{p}_t} \\
 0 \\
\end{array}
\right)\, ,  
& \qquad v_4^-(p_4,M_t)=\left(
\begin{array}{c}
 0 \\
 \sqrt{\text{E}_t-\text{p}_t} \\
 0 \\
 -\sqrt{\text{p}_t+\text{E}_t} \\
\end{array}
\right),
\end{split}
\end{equation}
where $u_3^{\lambda_3}$ and $v_4^{\lambda_4}$ are the spinors for the particle and antiparticle in the Weyl basis and $\lambda_3$ and $\lambda_4$ their polarizations, respectively.

\subsection{$W^+ (p_1,\epsilon_1^L) W^-(p_2,\epsilon_2^L) \rightarrow W^+(p_3, \epsilon_3^a) W^-(p_4,\epsilon_4^b)$}

For the full bosonic case, since there is no angular dependence on the azimuth $\phi$, we can set the momentum in the x-z plane to make the calculations easier:
\begin{equation} 
\begin{split}
p_1= (E,0,0,\abs{\vec{p}})\, ,  
&  \qquad p_2= (E,0,0,-\abs{\vec{p}}) \, , 
\\
p_3=(E,\abs{\vec{p_3}} \sin (\theta),0,\abs{\vec{p_3}} \cos (\theta) \, , 
& \qquad p_4=(E,-\abs{\vec{p_3}} \sin (\theta),0,-\abs{\vec{p_3}} \cos (\theta)) \, ,
\end{split}
\end{equation}
\begin{equation}
\begin{split}
\epsilon_1^L=\frac{1}{M_W}( \abs{\vec{p}},0,0,E )\, ,  
& \qquad \epsilon_2^L=\frac{1}{M_W}( \abs{\vec{p}},0,0,-E ) ,
\end{split}
\end{equation}
\begin{equation} 
\begin{split}
\epsilon_3^L=\frac{1}{M_W}( \abs{\vec{p}},E \sin (\theta),0,E \cos (\theta) ) \, , 
& \qquad \epsilon_4^L=\frac{1}{M_W}( \abs{\vec{p}},-E \sin(\theta),0,-E \cos (\theta) ) ,
\end{split}
\end{equation}
\begin{equation}  \begin{split} \epsilon_3^+= \frac{1}{\sqrt{2}}( 0, \cos (\theta), i , -\sin (\theta) ) & \qquad   \epsilon_3^-= {\epsilon_3^+}^{*} \, , 
\\ \epsilon_4^+= \frac{1}{\sqrt{2}}( 0, \cos (\theta), -i , -\sin (\theta) ) & \qquad   \epsilon_4^-= {\epsilon_4^+}^{*} \, ,\end{split} \end{equation}
where $\epsilon_{1,2}^L$ refer to the longitudinal polarization of the initial particles and $\epsilon_{3,4}^{+/-/L}$ refer to the polarization of the particle 3 or 4 with positive, negative or longitudinal polarization, respectively. 

\subsection{$W^+ (p_1,\epsilon_1^L) W^-(p_2,\epsilon_2^L) \rightarrow Z(p_3, \epsilon_3^a) Z(p_4,\epsilon_4^b)$}
\label{ssec:zzkinematics}

For $ZZ$ scattering, the positive and negative polarizations are given by the same vectors as for the $WW$ case, except for the longitudinal modes which depend on the mass:
\begin{equation} 
\begin{split}
\epsilon_3^L=\frac{1}{M_Z}( \abs{\vec{p_3}},E_3 \sin (\theta),0,E \cos (\theta) ) \, , 
& \qquad \epsilon_4^L=\frac{1}{M_Z}( \abs{\vec{p_3}},-E_3 \sin(\theta),0,-E \cos (\theta) ) 
\end{split}\, ,
\end{equation}
\begin{equation} 
\begin{split}
p_3=( E_3, \abs{\vec{p_3}}\sin (\theta),0,\abs{\vec{p_3}} \cos (\theta) ) \, , 
& \qquad p_4=( E_3,- \abs{\vec{p_3}}\sin (\theta),0,-\abs{\vec{p_3}} \cos (\theta) )\, ,
\end{split}
\end{equation}
where $E_3$ is the energy of the particle 3.

\subsection{$W^+ (p_1,\epsilon_1^L) W^-(p_2,\epsilon_2^L) \rightarrow \gamma (p_3,\epsilon_3^a) \gamma (p_4,\epsilon_4^b)$}
\label{ssec:gammgammakinematics}

The polarizations  $\epsilon_{3,4}^{+-}$ refer to the positive and negative polarization of the particle 3 and 4, respectively
\begin{equation}  \begin{split} \epsilon_3^+= \frac{1}{\sqrt{2}}( 0, \cos (\theta), -i , -\sin (\theta) )\, ,  & \qquad   \epsilon_3^-= {\epsilon_3^+}^{*}\, ,  
\\ \epsilon_4^+= \frac{1}{\sqrt{2}}( 0, \cos (\theta), i , -\sin (\theta) ) \, , 
& \qquad   \epsilon_4^-= {\epsilon_4^+}^{*}\, ,\end{split} \end{equation}
\begin{equation} 
\begin{split}
p_3=( E_3, \abs{\vec{p_3}}\sin (\theta),0,\abs{\vec{p_3}} \cos (\theta) )\, ,  & \qquad p_4=( E_3,- \abs{\vec{p_3}}\sin (\theta),0,-\abs{\vec{p_3}} \cos (\theta) )\, .
\end{split}
\end{equation}

\subsection{$W^+ (p_1,\epsilon_1^L) W^-(p_2,\epsilon_2^L) \rightarrow h (p_3,) h(p_4)$}
\label{ssec:hhkinematics}

\begin{equation} 
\begin{split}
p_3=( E_3, \abs{\vec{p_3}}\sin (\theta),0,\abs{\vec{p_3}} \cos (\theta) ) \, , 
& \qquad p_4=( E_3,- \abs{\vec{p_3}}\sin (\theta),0,-\abs{\vec{p_3}} \cos (\theta) )
\end{split}\, .
\end{equation}

\section{Scattering amplitudes}
\label{section:scatterin_gamplitudes}

For the calculation of the $\mathcal{O}(p^4)$ one-loop $W_LW_L$ elastic scattering beyond nET, we will need the $\mathcal{O}(p^2)$ (tree-level) $W_LW_L$ amplitudes to all possible intermediate states, which are provided below.  Since we are always dealing with longitudinal polarized electroweak gauge bosons in the initial state, we will only label the amplitudes with the polarization state of the final particles. Given the length of the analytic expression of some amplitudes, we will write the amplitude (without contracting) for each diagram in terms of the particle exchanged and the channels $s,t,$ and $u$. For example, $\mathcal{A}_{\pi, t}$ means this diagram is exchanging a Goldstone-$\pi$ in the channel $t$. Since we have performed the calculation in an arbitrary gauge, the various contributions to amplitudes contain the gauge parameters $\xi_W$, $\xi_Z$ and $\xi_A$. We have checked that the full amplitudes are gauge independent, but for the sake of achieving compact expression when it comes to the polarized amplitudes,the expressions are shown in the unitary gauge. 

All calculations have been performed by hand and checked via FeynArts~\cite{Hahn}, which generates all diagrams considered and evaluated using FeynCalc~\cite{FeynCalc}. For compactness the amplitudes are written in terms of $x=\cos \theta$ and $\beta_X=\sqrt{1-4M_X^{2}/s}$.

\subsection{$\mathcal{A}(W^+(p_1,\epsilon_1^L)W^-(p_2,\epsilon_2^L) \rightarrow t(p_3,\lambda_3) \bar{t}(p_4,\lambda_4))$ }

We will provide $t \bar{t}$ amplitude in terms of the diagrams involved, where $P_R$ and $P_L$ are the right and left chirality projectors and  and $N_C$ is the number of colors. 

\subsubsection{Amplitudes in terms of the polarization}


\begin{align}
    \begin{split}
    \mathcal{A}_{H,s} = & -a c_1\frac{\left(i \sqrt{N_c} g M_W \epsilon _1{}^{\mu } \epsilon _2{}^{\nu } \eta ^{\mu  \nu }\right)}{\left(p_3+p_4\right){}^2-M_H^2} \times \left(\bar{u_3}^{\lambda_3}(p_3,M_t)\right) .\left(-\frac{i g P_R M_t}{2 M_W}-\frac{i g P_L M_t}{2 M_W}\right).\left(v_4^{\lambda_4}
   (p_4,M_t)\right),
    \end{split}
    \\
    \begin{split}
        \mathcal{A}_{\gamma,s}= & -i \sqrt{N_C} g S_W \epsilon _1{}^{\mu } \epsilon _2{}^{\nu }\left [\left(-p_2-p_3-p_4\right){}^{\mu } \eta ^{\nu  \rho  }+\left(p_2-p_1\right){}^{\rho } \eta ^{\mu  \nu }+\left(p_1+p_3+p_4\right){}^{\nu } \eta ^{\mu  \rho }\right] \times \\ & \times  \left[\frac{\left(1-\xi
   _A\right) \left(-p_3-p_4\right){}^{\rho } \left(p_3+p_4\right){}^{\sigma }}{s^2}+\frac{\eta ^{\rho  \sigma }}{s}\right] \times \\ & \times 
  \left[ \left(\bar{u_3}^{\lambda_3}(p_3,M_t)\right).\left(-\frac{2}{3} i g S_W \bar{\gamma }^{\sigma }.P_R-\frac{2}{3} i g S_W \bar{\gamma
   }^{\sigma }.P_L\right).\left(v_4^{\lambda_4}
   (p_4,M_t)\right)\right],
    \end{split}
    \\
    \begin{split}
        \mathcal{A}_{Z,s}= & i \sqrt{N_C} g C_W \epsilon _1{}^{\mu } \epsilon _2{}^{\nu } \left[\left(-p_2-p_3-p_4\right){}^{\mu } \eta ^{\nu  \rho }+\left(p_2-p_1\right){}^{\rho } \eta ^{\mu  \nu }+\left(p_1+p_3+p_4\right){}^{\nu } \eta ^{\mu  \rho }\right] \times \\ & \times 
\left[\frac{\eta ^{\rho 
   \sigma }}{\left(p_3+p_4\right){}^2-M_Z^2}+\frac{\left(-p_3-p_4\right){}^{\rho } \left(p_3+p_4\right){}^{\sigma } \left(1-\xi _Z\right)}{\left(\left(p_3+p_4\right){}^2-M_Z^2\right) \left(\left(p_3+p_4\right){}^2-M_Z^2 \xi _Z\right)}\right] \times \\ & \times 
   \left [\left(\bar{u_3}^{\lambda_3}(p_3,M_t)\right).\left(\frac{i g \left(\frac{1}{2}-\frac{2 S_W^2}{3}\right) \bar{\gamma }^{\sigma }.P_L}{C_W}-\frac{2 i g S_W^2 \bar{\gamma }^{\sigma }.P_R}{3 C_W}\right).\left(v_4^{\lambda_4}
   (p_4,M_t)\right) \right],
    \end{split}
\end{align}

\begin{align}
    \begin{split}
        \mathcal{A}_{Z,s}= &i \sqrt{N_C} g C_W \epsilon _1{}^{\mu } \epsilon _2{}^{\nu } \left[\left(-p_2-p_3-p_4\right){}^{\mu } \eta ^{\nu  \rho }+\left(p_2-p_1\right){}^{\rho } \eta ^{\mu  \nu }+\left(p_1+p_3+p_4\right){}^{\nu } \eta ^{\mu  \rho }\right] \times \\ & \times 
\left[\frac{\eta ^{\rho 
  \sigma }}{\left(p_3+p_4\right){}^2-M_Z^2}+\frac{\left(-p_3-p_4\right){}^{\rho } \left(p_3+p_4\right){}^{\sigma } \left(1-\xi _Z\right)}{\left(\left(p_3+p_4\right){}^2-M_Z^2\right) \left(\left(p_3+p_4\right){}^2-M_Z^2 \xi _Z\right)}\right] \times \\ & \times 
  \left [\left(\bar{u_3}^{\lambda_3}(p_3,M_t)\right).\left(\frac{i g \left(\frac{1}{2}-\frac{2 S_W^2}{3}\right) \bar{\gamma }^{\sigma }.P_L}{C_W}-\frac{2 i g S_W^2 \bar{\gamma }^{\sigma }.P_R}{3 C_W}\right).\left(v_4^{\lambda_4}
  (p_4,M_t)\right) \right],
    \end{split}
    \\
    \begin{split}
        \mathcal{A}_{b,t}= & -\frac{\sqrt{N_c} \epsilon _1{}^{\mu } \epsilon _2{}^{\nu }}{\left(p_4-p_2\right){}^2-M_b^2} \left(\bar{u_3}^{\lambda_3}(p_3,M_t)\right).\frac{i g \bar{\gamma }^{\mu }.P_L}{\sqrt{2}}.\left(\bar{\gamma }\cdot \left(\overline{p}_2-\overline{p}_4\right)+M_b\right).\frac{i g \bar{\gamma }^{\nu
  }.P_L}{\sqrt{2}}.\left(v_4^{\lambda_4}
  (p_4,M_t)\right).
    \end{split}
\end{align}

\subsubsection{Polarized amplitudes}

We have checked these amplitudes via the unitarity relation finding an agreement with the imaginary part of top-quark loops $\mathcal{A}(W^+ W^- \rightarrow W^+ W^-)$ given in \cite{Dawson:1990ux} when we do not consider the exchange of $Z-/\gamma$-bosons. We provide the polarized amplitudes $\mathcal{A}(W_L^+W_L^-\rightarrow t(\lambda_3)\bar{t}(\lambda_4))=Q^{\lambda_3 \lambda_4}$, with definite helicities $\lambda_3$ and $\lambda_4$ for $N_C=3$:

\begin{align}
  \begin{split}
Q^{++} = & \frac{g^2 M_t}{16 \sqrt{3} M_W^2} \bigg [ \frac{12 a c_1 \sqrt{s} \beta _t \left(s-2 M_W^2\right)}{M_H^2-s}+\frac{3 \left(4 \sqrt{s} x M_W^2 \beta _W+\sqrt{s} \beta _t \left(s \beta _W^2-s \left(2 x^2-1\right)\right)\right)}{M_b^2-t}- \\
& \frac{32 S_W^2 x \beta _W \left(2
   M_W^2+s\right)}{\sqrt{s}}+\frac{4 \sqrt{s} \left(8 S_W^2-3\right) x \beta _W \left(2 M_W^2+s\right)}{s-M_Z^2}\bigg],
   \end{split}
   \\
   \begin{split}
        Q^{--}= & -Q^{++},
   \end{split}
   \\
   \begin{split}
        Q^{+-}= & \frac{g^2 \sqrt{1-x^2} e^{-i \phi }}{16 \sqrt{3} M_W^2} \bigg [-\frac{3 s \left(\beta _t-1\right) \left(s \beta _t \left(\beta _W+x\right)-2 M_W^2 \beta _W\right)}{t-M_b^2}+ \\ 
       & +\frac{2 s \beta _W \left(2 M_W^2+s\right) \left(8 S_W^2+3 \beta _t-3\right)}{s-M_Z^2}-16 S_W^2 \beta _W \left(2 M_W^2+s\right) \bigg ],
   \end{split}
   \\
   \begin{split}
       Q^{-+} = &\frac{g^2 \sqrt{1-x^2} e^{i \phi }}{16 \sqrt{3} M_W^2} \bigg[ \frac{3 s \left(\beta _t+1\right) \left(s \beta _t \left(x-\beta _W\right)-2 M_W^2 \beta _W\right)}{t-M_b^2}+ \\
       & +\frac{2 s \beta _W \left(2 M_W^2+s\right) \left(8 S_W^2-3 \beta _t-3\right)}{s-M_Z^2}-16 S_W^2 \beta _W \left(2 M_W^2+s\right)  \bigg].
   \end{split}
\end{align}




\subsection{$\mathcal{A}(W^+(p_1,\epsilon_1^L)W^-(p_2,\epsilon_2^L) \rightarrow b(p_3,\lambda_3) \bar{b}(p_4,\lambda_4))$ }

\subsubsection{Polarized amplitudes}


Since the top and bottom quark form a doublet with the same weak hypercharge, we can relate the amplitudes of the last subsection with the amplitudes $\mA(W_L^+W_L^-\to b(\lambda_3)\bar{b}(\lambda_4))=Q^{'\lambda_3 \lambda_4}$ for $b\bar{b}$ scattering, where $\lambda$ and $\lambda'$ are the polarization of particle 3 and 4. Then, the amplitude  $Q^{'\lambda_3 \lambda_4}$ is obtained by applying the following substitutions: $S_W \rightarrow S_W/ \sqrt{2}$, $\beta_t \leftrightarrow  \beta_b$, $M_t \leftrightarrow M_b$, $ u \leftrightarrow t$ and $\cos \theta \rightarrow - \cos \theta$ on the amplitudes $Q^{\lambda_3 \lambda_4}$.





%

\begin{align}
 Q^{'++} & =  Q^{++} \qquad (  S_W \rightarrow S_W/ \sqrt{2}, \beta_t \leftrightarrow  \beta_b, M_t \leftrightarrow M_b,  u \leftrightarrow t \  \text{and} \cos \theta \rightarrow - \cos \theta ), \\
  Q^{'+-} & =  -Q^{+-} \qquad (  S_W \rightarrow S_W/ \sqrt{2}, \beta_t \leftrightarrow  \beta_b, M_t \leftrightarrow M_b,  u \leftrightarrow t \  \text{and} \cos \theta \rightarrow - \cos \theta ), \\
   Q^{'-+} & = -Q^{-+} \qquad (  S_W \rightarrow S_W/ \sqrt{2}, \beta_t \leftrightarrow  \beta_b, M_t \leftrightarrow M_b,  u \leftrightarrow t \  \text{and} \cos \theta \rightarrow - \cos \theta ), \\
    Q^{'--} & = Q^{--} \qquad (  S_W \rightarrow S_W/ \sqrt{2}, \beta_t \leftrightarrow  \beta_b, M_t \leftrightarrow M_b,  u \leftrightarrow t \  \text{and} \cos \theta \rightarrow - \cos \theta ) .
\end{align}

\subsection{$\mathcal{A}((W^+ (p_1,\epsilon_1^L) W^-(p_2,\epsilon_2^L) \rightarrow W^+(p_3, \epsilon_3^a) W^-(p_4,\epsilon_4^b))$}

\subsubsection{Amplitudes in terms of the polarization}

\begin{align}
    \begin{split}
    \mathcal{A}_{contact} = &-i \epsilon _1{}^{\mu } \epsilon _2{}^{\nu } \epsilon _3{}^*{}^{\rho } \epsilon _4{}^*{}^{\sigma } \left(2 i g^2 \eta ^{\mu  \sigma } \eta ^{\nu  \rho }-i g^2 \eta ^{\mu  \rho } \eta ^{\nu  \sigma }-i g^2 \eta ^{\mu  \nu } \eta ^{\rho  \sigma }\right),
    \end{split}
    \\
    \begin{split}
          \mathcal{A}_{H,s} = &-a^2\frac{g^2 M_W^2 \epsilon _1{}^{\mu } \epsilon _2{}^{\nu } \epsilon _3{}^*{}^{\rho } \epsilon _4{}^*{}^{\sigma } \eta ^{\mu  \nu } \eta ^{\rho  \sigma }}{\left(p_3+p_4\right){}^2-M_H^2},
    \end{split}
\end{align}

\begin{align}
    \begin{split}
            A_{\gamma,s}= & -g^2 S_W^2 \epsilon _1{}^{\mu } \epsilon _2{}^{\nu } \epsilon _3{}^*{}^{\rho } \epsilon _4{}^*{}^{\sigma } \left[\left(p_3-p_4\right){}^{\gamma } \eta ^{\rho  \sigma }+\left(-2 p_3-p_4\right){}^{\sigma } \eta ^{\gamma  \rho }+\left(p_3+2
  p_4\right){}^{\rho } \eta ^{\gamma  \sigma }\right] \\ 
  & \left[\left(p_2-p_1\right){}^{\delta } \eta ^{\mu  \nu }+\left(p_1+p_3+p_4\right){}^{\nu } \eta ^{\delta  \mu }+\left(-p_2-p_3-p_4\right){}^{\mu } \eta ^{\delta  \nu }\right] \\
  & \left[\frac{\left(1-\xi
  _A\right) \left(p_3+p_4\right){}^{\gamma } \left(-p_3-p_4\right){}^{\delta }}{s^2}+\frac{1}{\left(p_3+p_4\right){}^2} \eta ^{\gamma  \delta }\right],
    \end{split}
    \\
    \begin{split}
        \mathcal{A}_{Z,s} = & -C_W^2 g^2 \epsilon _1{}^{\mu } \epsilon _2{}^{\nu } \epsilon _3{}^*{}^{\rho } \epsilon _4{}^*{}^{\sigma } \left[\left(p_3-p_4\right){}^{\gamma } \eta ^{\rho  \sigma }+\left(-2 p_3-p_4\right){}^{\sigma } \eta ^{\gamma  \rho }+\left(p_3+2
  p_4\right){}^{\rho } \eta ^{\gamma  \sigma }\right] \big[\left(p_2-p_1\right){}^{\delta } \eta ^{\mu  \nu }+ \\
  & \left(p_1+p_3+p_4\right){}^{\nu } \eta ^{\delta  \mu }+\left(-p_2-p_3-p_4\right){}^{\mu } \eta ^{\delta  \nu } \big]  \\
  & { \left[\frac{\eta ^{\gamma 
  \delta }}{\left(p_3+p_4\right){}^2-M_Z^2}+\frac{\left(p_3+p_4\right){}^{\gamma } \left(-p_3-p_4\right){}^{\delta } \left(1-\xi _Z\right)}{\left(\left(p_3+p_4\right){}^2-M_Z^2\right) \left(\left(p_3+p_4\right){}^2-M_Z^2 \xi _Z\right)}\right]},
    \end{split}
    \\
    \begin{split}
        \mathcal{A}_{H,t} = &-a^2 \frac{g^2 M_W^2 \epsilon _1{}^{\mu } \epsilon _2{}^{\nu } \epsilon _3{}^*{}^{\rho } \epsilon _4{}^*{}^{\sigma } \eta ^{\mu  \rho } \eta ^{\nu  \sigma }}{\left(p_4-p_2\right){}^2-M_H^2},
    \end{split}
    \\
    \begin{split}
        \mathcal{A}_{\gamma , t}  = & -g^2 S_W^2 \epsilon _1{}^{\mu } \epsilon _2{}^{\nu } \epsilon _3{}^*{}^{\rho } \epsilon _4{}^*{}^{\sigma } \left[\left(-p_2-p_4\right){}^{\gamma } \eta ^{\nu  \sigma }+\left(2 p_2-p_4\right){}^{\sigma } \eta ^{\gamma  \nu }+\left(2
  p_4-p_2\right){}^{\nu } \eta ^{\gamma  \sigma }\right] \\
  & {\left[\left(-p_1-p_3\right){}^{\delta } \eta ^{\mu  \rho }+\left(p_1-p_2+p_4\right){}^{\rho } \eta ^{\delta  \mu }+\left(p_2+p_3-p_4\right){}^{\mu } \eta ^{\delta  \rho }\right]} \\ & 
  \left[\bigg(\frac{1}{\left(p_4-p_2\right){}^2} \bigg)^2 \left(1-\xi _A\right) \left(p_4-p_2\right){}^{\gamma } \left(p_2-p_4\right){}^{\delta }+\frac{1}{\left(p_4-p_2\right){}^2} \eta ^{\gamma  \delta }\right],
    \end{split}
    \\
    \begin{split}
        \mathcal{A}_{Z, t}  = & -C_W^2 g^2 \epsilon _1{}^{\mu } \epsilon _2{}^{\nu } \epsilon _3{}^*{}^{\rho } \epsilon _4{}^*{}^{\sigma } \left[\left(-p_2-p_4\right){}^{\gamma } \eta ^{\nu  \sigma }+\left(2 p_2-p_4\right){}^{\sigma } \eta ^{\gamma  \nu }+\left(2
  p_4-p_2\right){}^{\nu } \eta ^{\gamma  \sigma }\right] \\
  & \left[\left(-p_1-p_3\right){}^{\delta } \eta ^{\mu  \rho }+\left(p_1-p_2+p_4\right){}^{\rho } \eta ^{\delta  \mu }+\left(p_2+p_3-p_4\right){}^{\mu } \eta ^{\delta  \rho }\right] \\
  & \left[\frac{\eta
  ^{\gamma  \delta }}{\left(p_4-p_2\right){}^2-M_Z^2}+\frac{\left(p_4-p_2\right){}^{\gamma } \left(p_2-p_4\right){}^{\delta } \left(1-\xi _Z\right)}{\left(\left(p_4-p_2\right){}^2-M_Z^2\right) \left(\left(p_4-p_2\right){}^2-M_Z^2 \xi _Z\right)}\right].
    \end{split}
\end{align}

\subsubsection{Polarized amplitudes}

We will label the 9 polarized amplitudes with $\mathcal{A}_{\epsilon_3 \epsilon_4}$, where $\epsilon_3$ and $\epsilon_4$ refer to the polarization of particle 3 and 4, respectively. We have checked the $\mathcal{A}_{LL}\rightarrow \mathcal{A}_{LL}$ amplitude with\cite{espriu}. We only have 4 independent amplitudes; the other 4 can be found through the relations $\mathcal{A}_{NN}=\mathcal{A}_{PP}, \mathcal{A}_{NP}=\mathcal{A}_{PN}, \mathcal{A}_{PL}=\mathcal{A}_{NL}=-\mathcal{A}_{LP}=-\mathcal{A}_{NP}$.

\begin{align}
    \begin{split}
         \mathcal{A}_{LL}= & \frac{a^2 g^2 \left(4 M_W^2+s (x-1)\right){}^2}{8 M_W^2 \left(2 M_H^2-s (x-1) \beta _W^2\right)}+\frac{a^2 g^2 \left(s-2 M_W^2\right){}^2}{4 M_W^2 \left(M_H^2-s\right)}+ \\
         & + \frac{g^2 C_W^2 \left(-4 s^2 (x-1) (x (x+10)-3) M_W^2+16 s (x (10 x-7)+1) M_W^4-64
   (x+1) M_W^6+s^3 (x-1)^2 (x+3)\right)}{16 M_W^4 \left(4 (x-1) M_W^2+2 M_Z^2-s x+s\right)}- \\
   & - \frac{g^2 s x C_W^2 \beta _W^2 \left(2 M_W^2+s\right){}^2}{4 M_W^4 \left(s-M_Z^2\right)}- \\
   & - \frac{g^2 S_W^2 \left(-4 s^2 (x-1) (x (x+10)-3) M_W^2+16 s (x (10 x-7)+1)
   M_W^4-64 (x+1) M_W^6+s^3 (x-1)^2 (x+3)\right)}{16 s (x-1) M_W^4 \beta _W^2}- \\ & - \frac{g^2 x S_W^2 \beta _W^2 \left(2 M_W^2+s\right){}^2}{4 M_W^4}+\frac{g^2 s \left((8-24 x) M_W^2+s (x (x+6)-3)\right)}{16 M_W^4}
    \end{split}
\end{align}

\begin{align}
    \begin{split}
         \mathcal{A}_{PP}= & \frac{a^2 g^2 s \left(x^2-1\right)}{4 \left(2 M_H^2-s (x-1) \beta _W^2\right)}-\frac{a^2 g^2 \left(s-2 M_W^2\right)}{2 \left(M_H^2-s\right)}+ \\
         & +\frac{g^2 (x-1) C_W^2 \left(-4 s (x-3) (x-1) M_W^2+32 (x+1) M_W^4+s^2 (x-3) (x-1)\right)}{8 M_W^2 \left(4 (x-1)
  M_W^2+2 M_Z^2-s x+s\right)}-\\
  & -\frac{g^2 s x C_W^2 \beta _W^2 \left(2 M_W^2+s\right)}{2 M_W^2 \left(s-M_Z^2\right)}+ \\
  & + \frac{g^2 S_W^2 \left(-4 s (x-3) (x-1) M_W^2+32 (x+1) M_W^4+s^2 (x-3) (x-1)\right)}{32 M_W^4-8 s M_W^2}- \\
  & - \frac{g^2 x S_W^2 \left(s-4
  M_W^2\right) \left(2 M_W^2+s\right)}{2 s M_W^2}+\frac{g^2 \left(s \left(x^2+3\right)-8 M_W^2\right)}{8 M_W^2},
    \end{split}
    \\
        \begin{split}
         \mathcal{A}_{PN}= & \frac{a^2 g^2 s \left(x^2-1\right)}{4 \left(2 M_H^2-s (x-1) \beta _W^2\right)}+\frac{g^2 \left(x^2-1\right) C_W^2 \left(-4 s (x-1) M_W^2+32 M_W^4+s^2 (x-1)\right)}{8 M_W^2 \left(4 (x-1) M_W^2+2 M_Z^2-s x+s\right)}+ \\
         & + \frac{g^2 (x+1) S_W^2 \left(-4 s (x-1)
  M_W^2+32 M_W^4+s^2 (x-1)\right)}{32 M_W^4-8 s M_W^2}+\frac{g^2 s \left(x^2-1\right)}{8 M_W^2},
    \end{split}
    \\
    \begin{split}
         \mathcal{A}_{LP}= & \frac{a^2 g^2 \sqrt{s-s x^2} \left(4 M_W^2+s (x-1)\right)}{4 \sqrt{2} M_W \left(2 M_H^2-s (x-1) \beta _W^2\right)}+ \\
         & + \frac{g^2 C_W^2 \sqrt{s-s x^2} \left(-4 s \left(x^2+x-2\right) M_W^2+16 (5 x-3) M_W^4+s^2 (x-1)^2\right)}{8 \sqrt{2} M_W^3 \left(4 (x-1)
  M_W^2+2 M_Z^2-s x+s\right)} - \\
  & -\frac{g^2 C_W^2 \sqrt{s-s x^2} \left(s-4 M_W^2\right) \left(2 M_W^2+s\right)}{2 \sqrt{2} M_W^3 \left(s-M_Z^2\right)}- \\
  & - \frac{g^2 \sqrt{s-s x^2} S_W^2 \left(-4 s \left(x^2+x-2\right) M_W^2+16 (5 x-3) M_W^4+s^2 (x-1)^2\right)}{8
  \sqrt{2} s (x-1) M_W^3 \beta _W^2}- \\
  & -\frac{g^2 s \sqrt{\frac{2-2 x^2}{s}} S_W^2 \beta _W^2 \left(2 M_W^2+s\right)}{4 M_W^3}+\frac{g^2 \sqrt{s-s x^2} \left(s (x+3)-12 M_W^2\right)}{8 \sqrt{2} M_W^3}.
    \end{split}
\end{align}

\subsection{$\mathcal{A}(W^+ (p_1,\epsilon_1^L) W^-(p_2,\epsilon_2^L) \rightarrow Z(p_3, \epsilon_3^a) Z(p_4,\epsilon_4^b))$}

\subsubsection{Amplitudes in terms of the polarization}

\begin{align}
    \begin{split}
        \mathcal{A}_{contact}= & -i \epsilon _1{}^{\mu } \epsilon _2{}^{\nu } \epsilon _3{}^*{}^{\rho } \epsilon _4{}^*{}^{\sigma } \left(i C_W^2 g^2 \eta ^{\mu  \sigma } \eta ^{\nu  \rho }+i C_W^2 g^2 \eta ^{\mu  \rho } \eta ^{\nu  \sigma }-2 i C_W^2 g^2 \eta ^{\mu  \nu
   } \eta ^{\rho  \sigma }\right),
    \end{split}
    \\
    \begin{split}
        \mathcal{A}_{H,s}= & -a^2\frac{g^2 M_W^2 \epsilon _1{}^{\mu } \epsilon _2{}^{\nu } \epsilon _3{}^*{}^{\rho } \epsilon _4{}^*{}^{\sigma } \eta ^{\mu  \nu } \eta ^{\rho  \sigma }}{C_W^2 \left(\left(p_3+p_4\right){}^2-M_H^2\right)},
    \end{split}
    \\
    \begin{split}
        \mathcal{A}_{\pi,t}= & -\frac{g^2 S_W^4 M_W^2 \epsilon _1{}^{\mu } \epsilon _2{}^{\nu } \epsilon _3{}^*{}^{\rho } \epsilon _4{}^*{}^{\sigma } \eta ^{\mu  \rho } \eta ^{\nu  \sigma }}{C_W^2 \left(\left(p_4-p_2\right){}^2-M_W^2 \xi _W\right)},
    \end{split}
    \\
    \begin{split}
        \mathcal{A}_{W ,t} = & -C_W^2 g^2 \epsilon _1{}^{\mu } \epsilon _2{}^{\nu } \epsilon _3{}^*{}^{\rho } \epsilon _4{}^*{}^{\sigma } \left[\left(-p_2-p_4\right){}^{\gamma } \eta ^{\nu  \sigma }+\left(2 p_2-p_4\right){}^{\sigma } \eta ^{\gamma  \nu }+\left(2
  p_4-p_2\right){}^{\nu } \eta ^{\gamma  \sigma }\right] \\
  & \left[\left(-p_1-p_3\right){}^{\delta } \eta ^{\mu  \rho }+\left(p_1-p_2+p_4\right){}^{\rho } \eta ^{\delta  \mu }+\left(p_2+p_3-p_4\right){}^{\mu } \eta ^{\delta  \rho }\right] \\
  & \left[\frac{\eta
  ^{\gamma  \delta }}{\left(p_4-p_2\right){}^2-M_W^2}+\frac{\left(p_4-p_2\right){}^{\gamma } \left(p_2-p_4\right){}^{\delta } \left(1-\xi _W\right)}{\left(\left(p_4-p_2\right){}^2-M_W^2\right) \left(\left(p_4-p_2\right){}^2-M_W^2 \xi _W\right)}\right],
    \end{split}
    \\
    \begin{split}
        \mathcal{A}_{\pi, u}= & -\frac{g^2 S_W^4 M_W^2 \epsilon _1{}^{\mu } \epsilon _2{}^{\nu } \epsilon _3{}^*{}^{\rho } \epsilon _4{}^*{}^{\sigma } \eta ^{\mu  \sigma } \eta ^{\nu  \rho }}{C_W^2 \left(\left(p_3-p_2\right){}^2-M_W^2 \xi _W\right)},
    \end{split}
\end{align}

\begin{align}
    \begin{split}
        \mathcal{A}_{W ,u}= & -C_W^2 g^2 \epsilon _1{}^{\mu } \epsilon _2{}^{\nu } \epsilon _3{}^*{}^{\rho } \epsilon _4{}^*{}^{\sigma } \left[\left(-p_2-p_3\right){}^{\gamma } \eta ^{\nu  \rho }+\left(2 p_2-p_3\right){}^{\rho } \eta ^{\gamma  \nu }+\left(2 p_3-p_2\right){}^{\nu }
  \eta ^{\gamma  \rho }\right] \\
  & \left[\left(-p_1-p_4\right){}^{\delta } \eta ^{\mu  \sigma }+\left(p_1-p_2+p_3\right){}^{\sigma } \eta ^{\delta  \mu }+\left(p_2-p_3+p_4\right){}^{\mu } \eta ^{\delta  \sigma }\right] \\
  & \left[\frac{\eta ^{\gamma  \delta
  }}{\left(p_3-p_2\right){}^2-M_W^2}+\frac{\left(p_3-p_2\right){}^{\gamma } \left(p_2-p_3\right){}^{\delta } \left(1-\xi _W\right)}{\left(\left(p_3-p_2\right){}^2-M_W^2\right) \left(\left(p_3-p_2\right){}^2-M_W^2 \xi _W\right)}\right].
    \end{split}
\end{align}

\subsubsection{Polarized Amplitudes}

We will label the 9 polarized amplitudes with $\mathcal{A}_{\epsilon_3 \epsilon_4}$, where $\epsilon_3$ and $\epsilon_4$ refer to the polarization of particle 3 and 4, respectively. We have checked $\mathcal{A}_{LL} \rightarrow \mathcal{A}_{LL}$ with reference \cite{Denner:1996ug}. Just like in the previous case, we only have 4 independent amplitudes: $\mathcal{A}_{NN}=\mathcal{A}_{PP}$, $\mathcal{A}_{NP}=\mathcal{A}_{PN},\mathcal{A}_{LP}=\mathcal{A}_{NL}=-\mathcal{A}_{PL}=-\mathcal{A}_{NL}$.

\begin{align}
    \begin{split}
    \mathcal{A}_{LL}= & \frac{a^2 g^2 \left(2 M_W^2-s\right) \left(s-2 M_Z^2\right)}{4 C_W^2 M_Z^2 \left(s-M_H^2\right)}+\frac{g^2 s C_W^2 \left(4 M_W^2+4 M_Z^2+s \left(x^2-3\right)\right)}{8 M_W^2 M_Z^2}+ \\
    & + \frac{1}{16 M_W^4 M_Z^2 \left(2 M_Z^2+s x \beta _W \beta _Z-s\right)} \bigg[ -g^2 C_W^2 \bigg(-4 M_W^4 \bigg(4 s x^2 \left(s-6 M_Z^2\right) - 2 s M_Z^2+24 M_Z^4+s^2\bigg)+ \\
    & + 2 s^2 x^2 M_Z^4+M_W^2 \bigg(-2 s^2 \left(11 x^2+5\right) M_Z^2+16 M_Z^4 \left(2 s x^2+s\right)+32 M_Z^6+ s^3 \left(x^2+3\right)\bigg) + \\
    & + s \beta _Z \left(x \beta _W
   \left(s M_W^2 \left(24 M_Z^2+s \left(x^2-5\right)\right)+4 M_W^4 \left(4 M_Z^2+3 s\right)-4 s M_Z^4\right)+2 s M_Z^4 \beta _Z\right) + 32 s x^2 M_W^6\bigg) \bigg] - \\
   & -\frac{1}{16 M_W^4 M_Z^2 \left(2 M_Z^2-s x \beta _W \beta _Z-s\right)} \bigg[ -g^2 C_W^2 \bigg(-4 M_W^4 \bigg(4 s x^2 \left(s-6 M_Z^2\right) - 2 s M_Z^2+24 M_Z^4+s^2\bigg)+ \\
    & + 2 s^2 x^2 M_Z^4+M_W^2 \bigg(-2 s^2 \left(11 x^2+5\right) M_Z^2+16 M_Z^4 \left(2 s x^2+s\right)+32 M_Z^6+ s^3 \left(x^2+3\right)\bigg) + \\
    & + s \beta _Z \left(-x \beta _W
   \left(s M_W^2 \left(24 M_Z^2+s \left(x^2-5\right)\right)+4 M_W^4 \left(4 M_Z^2+3 s\right)-4 s M_Z^4\right)+2 s M_Z^4 \beta _Z\right) + 32 s x^2 M_W^6\bigg) \bigg],
    \end{split}
    \\
    \begin{split}
      \mathcal{A}_{PP}= & \frac{g^2 C_W^2 \left(s \left(x^2+3\right)-8 M_W^2\right)}{4 M_W^2}-\frac{a^2 g^2 \left(s-2 M_W^2\right)}{2 C_W^2 \left(M_H^2-s\right)}+ \\
      & +\frac{1}{8 M_W^4 \left(2 M_Z^2-s+s x \beta _W \beta _Z\right)} \bigg[-g^2 C_W^2 \bigg(32 \left(x^2-1\right) M_W^6+8 s \left(x^2+1\right) M_W^4+s \bigg(2 \left(5 x^2+3\right) M_Z^2- \\
      & - s \left(5 x^2+3\right)+s x \left(x^2+7\right) \beta _W \beta _Z\bigg) M_W^2+2 s \left(x^2-1\right) M_Z^4\bigg) \bigg],
  \end{split}
  \\
  \begin{split}
        \mathcal{A}_{PN}= & \frac{ g^2 s \left(x^2-1\right) C_W^2}{4 M_W^2}- \\
        & -\frac{ g^2 \left(x^2-1\right) C_W^2} {8 M_W^4 \left(2
  M_Z^2+s x \beta _W \beta _Z-s\right)} \left(-s M_W^2 \left(6 M_Z^2-s x \beta _W \beta _Z+s\right)+8 s M_W^4+2 s M_Z^4+32 M_W^6\right)+ \\ 
  & + \frac{ g^2 \left(x^2-1\right) C_W^2}{8 M_W^4 \left(-2 M_Z^2+s x \beta _W
  \beta _Z+s\right)} \left(-s M_W^2 \left(6 M_Z^2+s x \beta _W \beta _Z+s\right)+8 s M_W^4+2 s M_Z^4+32 M_W^6\right),
    \end{split}
\end{align}

\begin{align}
    \begin{split}
        \mathcal{A}_{LP}  = & \frac{g^2 s^{3/2} x \sqrt{1-x^2} C_W^2}{4 \sqrt{2} M_W^2 M_Z}+ 
  \\
  & +\frac{1}{8 \sqrt{2} M_W^4 M_Z \left(2 M_Z^2+s \left(x \beta _W \beta _Z-1\right)\right)} \left[ -g^2 C_W^2 \sqrt{s-s x^2} \bigg(s M_W^2 \bigg(\beta _W \beta _Z \left(6 M_Z^2+s x^2+s\right)-  \right. \\
  & -2 x \left(M_Z^2+s\right)\bigg) +M_W^4 \left(48 x M_Z^2+8 s \beta _W \beta _Z-4 s x\right)+2 s M_Z^4 \left(x-\beta _W \beta _Z\right)+32 x M_W^6\bigg)\bigg] +
  \\
  &+ \frac{1}{8 \sqrt{2} M_W^4 M_Z \left(-2 M_Z^2+s x \beta _W \beta _Z+s\right)} \bigg[-g^2 C_W^2 \sqrt{s-s x^2} \bigg(s M_W^2 \bigg(\beta _W \beta _Z \left(6 M_Z^2+s x^2+s\right)+ \\
  & + 2 x \left(M_Z^2+s\right)\bigg) +4 M_W^4 \left(s \left(2 \beta _W \beta _Z+x\right)-12 x M_Z^2\right)-2 s M_Z^4 \left(\beta _W \beta _Z+x\right)-32 x M_W^6\bigg) \bigg].
    \end{split}
\end{align}

\subsection{$\mathcal{A}(W^+ (p_1,\epsilon_1^L) W^-(p_2,\epsilon_2^L) \rightarrow \gamma (p_3,\epsilon_3^a) \gamma (p_4,\epsilon_4^b))$}

\subsubsection{Amplitudes in terms of the polarization}

\begin{align}
    \begin{split}
        \mathcal{A}_{contact}= & -i \epsilon _1{}^{\mu } \epsilon _2{}^{\nu } \epsilon _3{}^*{}^{\rho } \epsilon _4{}^*{}^{\sigma } S_W^2 \left(i g^2   \eta ^{\mu  \sigma } \eta ^{\nu  \rho }+i g^2   \eta ^{\mu  \rho } \eta ^{\nu  \sigma }-2 i g^2  \eta ^{\mu  \nu
   } \eta ^{\rho  \sigma }\right),
    \end{split}
    \\
    \begin{split}
        \mathcal{A}_{\pi, t}= & -\frac{g^2 S_W^2 M_W^2 \epsilon _1{}^{\mu } \epsilon _2{}^{\nu } \epsilon _3{}^*{}^{\rho } \epsilon _4{}^*{}^{\sigma } \eta ^{\mu  \rho } \eta ^{\nu  \sigma }}{\left(p_4-p_2\right){}^2-M_W^2 \xi _W},
    \end{split}
    \\
    \begin{split}
        \mathcal{A}_{W ,t}= & -g^2 S_W^2 \epsilon _1{}^{\mu } \epsilon _2{}^{\nu } \epsilon _3{}^*{}^{\rho } \epsilon _4{}^*{}^{\sigma } \left[\left(-p_2-p_4\right){}^{\gamma } \eta ^{\nu  \sigma }+\left(2 p_2-p_4\right){}^{\sigma } \eta ^{\gamma  \nu }+\left(2
   p_4-p_2\right){}^{\nu } \eta ^{\gamma  \sigma }\right] \\
   & \left[\left(-p_1-p_3\right){}^{\delta } \eta ^{\mu  \rho }+\left(p_1-p_2+p_4\right){}^{\rho } \eta ^{\delta  \mu }+\left(p_2+p_3-p_4\right){}^{\mu } \eta ^{\delta  \rho }\right] \\
   & \left[\frac{\eta
   ^{\gamma  \delta }}{\left(p_4-p_2\right){}^2-M_W^2}+\frac{\left(p_4-p_2\right){}^{\gamma } \left(p_2-p_4\right){}^{\delta } \left(1-\xi _W\right)}{\left(\left(p_4-p_2\right){}^2-M_W^2\right) \left(\left(p_4-p_2\right){}^2-M_W^2 \xi _W\right)}\right],
    \end{split}
    \\
    \begin{split}
        \mathcal{A}_{\pi , u}= & -\frac{g^2 S_W^2 M_W^2 \epsilon _1{}^{\mu } \epsilon _2{}^{\nu } \epsilon _3{}^*{}^{\rho } \epsilon _4{}^*{}^{\sigma } \eta ^{\mu  \sigma } \eta ^{\nu  \rho }}{\left(p_3-p_2\right){}^2-M_W^2 \xi _W},
    \end{split}
    \\
    \begin{split}
        \mathcal{A}_{W ,u}=& -g^2 S_W^2 \epsilon _1{}^{\mu } \epsilon _2{}^{\nu } \epsilon _3{}^*{}^{\rho } \epsilon _4{}^*{}^{\sigma } \left[\left(-p_2-p_3\right){}^{\gamma } \eta ^{\nu  \rho }+\left(2 p_2-p_3\right){}^{\rho } \eta ^{\gamma  \nu }+\left(2 p_3-p_2\right){}^{\nu
   } \eta ^{\gamma  \rho }\right] \\
   & \left[\left(-p_1-p_4\right){}^{\delta } \eta ^{\mu  \sigma }+\left(p_1-p_2+p_3\right){}^{\sigma } \eta ^{\delta  \mu }+\left(p_2-p_3+p_4\right){}^{\mu } \eta ^{\delta  \sigma }\right] \\
  & \left[\frac{\eta ^{\gamma  \delta
   }}{\left(p_3-p_2\right){}^2-M_W^2}+\frac{\left(p_3-p_2\right){}^{\gamma } \left(p_2-p_3\right){}^{\delta } \left(1-\xi _W\right)}{\left(\left(p_3-p_2\right){}^2-M_W^2\right) \left(\left(p_3-p_2\right){}^2-M_W^2 \xi _W\right)}\right].
    \end{split}
\end{align}

\subsubsection{Polarized amplitudes}

We have checked the polarized amplitudes with \cite{Denner:1995jv}, 
finding an agreement.
We only have two independent amplitudes since $\mathcal{A}_{++}=\mathcal{A}_{--},\mathcal{A}_{+-}=\mathcal{A}_{-+}$, where $+$ and $-$ refer to the positive and negative polarization of the photons:
%
%
%
\begin{align}
 \mathcal{A}_{++} & =-\frac{8 g^2 M_W^2 S_W^2}{x^2 \left(4 M_W^2-s\right)+s}\, , \\
  \mathcal{A}_{+-} & =\frac{2 g^2 \left(x^2-1\right) S_W^2 \left(4 M_W^2+s\right)}{x^2 \left(4 M_W^2-s\right)+s}\, .
\end{align}

\subsection{$\mathcal{A}(W^+ (p_1,\epsilon_1^L) W^-(p_2,\epsilon_2^L) \rightarrow h (p_3) h(p_4))$}

\subsubsection{Amplitudes}

\begin{align}
    \begin{split}
        \mathcal{A}_{contact}= & \frac{1}{2}b g^2 \epsilon _1{}^{\mu } \epsilon _2{}^{\nu } \eta ^{\mu  \nu },
    \end{split}
    \\
    \begin{split}
        \mathcal{A}_{H, s}= & \frac{3 g^2 M_H^2 \epsilon _1{}^{\mu } \epsilon _2{}^{\nu } \eta ^{\mu  \nu }}{2 \left(\left(p_3+p_4\right){}^2-M_H^2\right)}a d_3,
    \end{split}
    \\
    \begin{split}
        \mathcal{A}_{\pi, t}= & a^2\frac{g^2 \left(-p_2-p_3+p_4\right){}^{\mu } \left(p_2-2 p_4\right){}^{\nu } \epsilon _1{}^{\mu } \epsilon _2{}^{\nu }}{4 \left(\left(p_4-p_2\right){}^2-M_W^2 \xi _W\right)},
    \end{split}
    \\
    \begin{split}
        \mathcal{A}_{W, t}= & a^2 g^2 M_W^2 \epsilon _1{}^{\mu } \epsilon _2{}^{\nu } \eta ^{\mu  \rho } \eta ^{\nu  \sigma } \left[\frac{\eta ^{\rho  \sigma }}{\left(p_4-p_2\right){}^2-M_W^2}+\frac{\left(p_2-p_4\right){}^{\rho } \left(p_4-p_2\right){}^{\sigma } \left(1-\xi
  _W\right)}{\left(\left(p_4-p_2\right){}^2-M_W^2\right) \left(\left(p_4-p_2\right){}^2-M_W^2 \xi _W\right)}\right],
    \end{split}
    \\
    \begin{split}
        \mathcal{A}_{\pi , u}= & a^2\frac{g^2 \left(-p_2+p_3-p_4\right){}^{\mu } \left(p_2-2 p_3\right){}^{\nu } \epsilon _1{}^{\mu } \epsilon _2{}^{\nu }}{4 \left(\left(p_3-p_2\right){}^2-M_W^2 \xi _W\right)},
    \end{split}
    \\
    \begin{split}
        \mathcal{A}_{W, u}= & a^2 g^2 M_W^2 \epsilon _1{}^{\mu } \epsilon _2{}^{\nu } \eta ^{\mu  \rho } \eta ^{\nu  \sigma } \left[\frac{\eta ^{\rho  \sigma }}{\left(p_3-p_2\right){}^2-M_W^2}+\frac{\left(p_2-p_3\right){}^{\rho } \left(p_3-p_2\right){}^{\sigma } \left(1-\xi
  _W\right)}{\left(\left(p_3-p_2\right){}^2-M_W^2\right) \left(\left(p_3-p_2\right){}^2-M_W^2 \xi _W\right)}\right].
    \end{split}
\end{align}

We have checked the amplitude with references \cite{wwhh,Kallianpur:1988cs}. Since the Higgs boson $h$ is a scalar particle, the only amplitude is:

\[
    \mathcal{A}_{W^+W^- \rightarrow hh}=\frac{a^2 \left(s^2 \left(\beta _W-x \beta _H\right){}^2+8 M_W^2 \left(s-2 M_W^2\right)\right)}{2 v^2 \left(2 M_H^2+s x \beta _H \beta _W-s\right)}-\frac{a^2 \left(s^2 \left(x \beta _H+\beta _W\right){}^2+8 M_W^2 \left(s-2 M_W^2\right)\right)}{2 v^2
   \left(-2 M_H^2+s x \beta _H \beta _W+s\right)}-\frac{3 a d_3 M_H^2 \left(s-2 M_W^2\right)}{v^2 \left(M_H^2-s\right)}+\frac{b \left(s-2 M_W^2\right)}{v^2}.
\]

\subsection{$\mathcal{A}(W^+ (p_1,\epsilon_1^L) W^-(p_2,\epsilon_2^L) \rightarrow Z (p_3,\epsilon_3^a) h(p_4))$}

\subsubsection{Amplitudes in terms of the polarization}

\begin{align}
    \begin{split}
       \mathcal{A}_{Z, s} = & a \frac{g^2 M_W}{C_W} \epsilon _1{}^{\mu } \epsilon _2{}^{\nu } \epsilon _3{}^*{}^{\rho } \eta ^{\delta  \rho } \left[\left(-p_2-p_3-p_4\right){}^{\mu } \eta ^{\nu  \sigma }+\left(p_2-p_1\right){}^{\sigma } \eta ^{\mu  \nu }+\left(p_1+p_3+p_4\right){}^{\nu } \eta
   ^{\mu  \sigma }\right] \\
   & \left[\frac{\eta ^{\delta  \sigma }}{\left(p_3+p_4\right){}^2-M_Z^2}+\frac{\left(p_3+p_4\right){}^{\delta } \left(-p_3-p_4\right){}^{\sigma } \left(1-\xi _Z\right)}{\left(\left(p_3+p_4\right){}^2-M_Z^2\right)
   \left(\left(p_3+p_4\right){}^2-M_Z^2 \xi _Z\right)}\right], 
    \end{split}
    \\
    \begin{split}
        \mathcal{A}_{\pi, t}= & -a \frac{g^2 S_W^2 M_W \left(p_2-2 p_4\right){}^{\nu } \epsilon _1{}^{\mu } \epsilon _2{}^{\nu } \epsilon _3{}^*{}^{\rho } \eta ^{\mu  \rho }}{2 C_W \left(\left(p_4-p_2\right){}^2-M_W^2 \xi _W\right)},
    \end{split}
    \\
    \begin{split}
        \mathcal{A}_{W, t} = & -a C_W g^2 M_W \epsilon _1{}^{\mu } \epsilon _2{}^{\nu } \epsilon _3{}^*{}^{\rho } \eta ^{\delta  \nu } \left[\left(p_2+p_3-p_4\right){}^{\mu } \eta ^{\rho  \sigma }+\left(-p_1-p_3\right){}^{\sigma } \eta ^{\mu  \rho }+\left(p_1-p_2+p_4\right){}^{\rho }
   \eta ^{\mu  \sigma }\right] \\
   & \left[\frac{\eta ^{\delta  \sigma }}{\left(p_4-p_2\right){}^2-M_W^2}+\frac{\left(p_4-p_2\right){}^{\delta } \left(p_2-p_4\right){}^{\sigma } \left(1-\xi _W\right)}{\left(\left(p_4-p_2\right){}^2-M_W^2\right)
   \left(\left(p_4-p_2\right){}^2-M_W^2 \xi _W\right)}\right],
    \end{split}
    \\
    \begin{split}
        \mathcal{A}_{\pi, u}= &a \frac{g^2 S_W^2 M_W \left(-p_2+p_3-p_4\right){}^{\mu } \epsilon _1{}^{\mu } \epsilon _2{}^{\nu } \epsilon _3{}^*{}^{\rho } \eta ^{\nu  \rho }}{2 C_W \left(\left(p_3-p_2\right){}^2-M_W^2 \xi _W\right)},
    \end{split}
    \\
    \begin{split}
        \mathcal{A}_{W , u}= &a C_W g^2 M_W \epsilon _1{}^{\mu } \epsilon _2{}^{\nu } \epsilon _3{}^*{}^{\rho } \eta ^{\mu  \sigma } \left[\left(-p_2-p_3\right){}^{\delta } \eta ^{\nu  \rho }+\left(2 p_2-p_3\right){}^{\rho } \eta ^{\delta  \nu }+\left(2 p_3-p_2\right){}^{\nu } \eta
   ^{\delta  \rho }\right] \\
   & \left[\frac{\eta ^{\delta  \sigma }}{\left(p_3-p_2\right){}^2-M_W^2}+\frac{\left(p_3-p_2\right){}^{\delta } \left(p_2-p_3\right){}^{\sigma } \left(1-\xi _W\right)}{\left(\left(p_3-p_2\right){}^2-M_W^2\right)
   \left(\left(p_3-p_2\right){}^2-M_W^2 \xi _W\right)}\right].
    \end{split}
\end{align}

\subsubsection{Polarized Amplitudes}

We will label the polarized amplitude with $\mathcal{A}_{\epsilon_3}$, where $\epsilon_3$ refers to the polarization of the Z-boson. We only have two independent amplitudes since $\mathcal{A}_{P}=\mathcal{A}_{N}$.

\begin{align}
    \begin{split}
       \mathcal{A}_{L} =&\frac{a g^2 x \beta _W \left(2 M_W^2+s\right) \left(-M_H^2+M_Z^2+s\right)}{4 C_W M_W M_Z \left(M_Z^2-s\right)}   
\\  
& + \frac{- a g^2 C_W}{8 s M_W^3 M_Z \left(M_H^2+M_Z^2-s\right) \left(x \beta _W-1\right)} 
\times \bigg[  \left(M_H^2+M_Z^2-s\right) \bigg(-8 s M_W^4+4 \left(-M_Z^4+s M_Z^2+s^2\right) M_W^2 
\\ 
&
\left.\left.
+M_H^2 \left(4 M_W^2+s \left(x^2-1\right)\right) M_Z^2-s M_Z^2 \left(\left(M_Z^2+s\right) x^2-M_Z^2+s\right)\bigg)  
\right.\right. 
+2 s x \bigg(-4 \left(M_Z^2+s\right) M_W^4+
\\ & + 2 s \left(s-3 M_Z^2\right) M_W^2+s M_Z^2 \left(M_Z^2-s\right)
\left.+M_H^2 \left(4 M_W^4-2 s M_W^2+s M_Z^2\right)\bigg) \beta _W\right) \bigg]
\\ 
& -       \frac{- a g^2 C_W}{8 s M_W^3 M_Z \left(M_H^2+M_Z^2-s\right) \left(-x \beta _W-1\right)} 
\times \bigg[  \left(M_H^2+M_Z^2-s\right) \bigg(-8 s M_W^4+4 \left(-M_Z^4+s M_Z^2+s^2\right) M_W^2 
\\ 
&
\left.\left.
+M_H^2 \left(4 M_W^2+s \left(x^2-1\right)\right) M_Z^2-s M_Z^2 \left(\left(M_Z^2+s\right) x^2-M_Z^2+s\right)\bigg)  
\right.\right. 
-2 s x \bigg(-4 \left(M_Z^2+s\right) M_W^4+
\\ & + 2 s \left(s-3 M_Z^2\right) M_W^2+s M_Z^2 \left(M_Z^2-s\right)
\left.\left.+M_H^2 \left(4 M_W^4-2 s M_W^2+s M_Z^2\right)\bigg) \beta _W\right) \right] ,
    \end{split}
    \\
        \begin{split}
        \mathcal{A}_{P}= & \frac{a g^2 \left(2 M_W^2+s\right) \sqrt{-s \left(x^2-1\right) \beta _W^2}}{2 \sqrt{2} C_W M_W \left(s-M_Z^2\right)} + \\
        & +  \frac{a}{4 \sqrt{2} M_W^3 \left(-M_H^2-M_Z^2+s\right) \left(x \beta _W-1\right)}  \bigg[  g^2 \sqrt{s \left(1-x^2\right)} C_W \bigg(x \left(M_H^2+M_Z^2-s\right) M_Z^2+ \\
        & + \left(8 M_W^4-4 M_H^2 M_W^2-M_Z^4+\left(M_H^2+4 M_W^2+s\right) M_Z^2\right) \beta _W\bigg)\bigg] \\
        &  +    \frac{a}{4 \sqrt{2} M_W^3 \left(-M_H^2-M_Z^2+s\right) \left(x \beta _W+1\right)}   \bigg[   g^2 \sqrt{s \left(1-x^2\right)} C_W \bigg(\left(x+\beta _W\right) M_Z^4-\bigg(\left(\beta _W-x\right) M_H^2+ \\
        & + s x+\left(4 M_W^2+s\right) \beta _W\bigg) M_Z^2+4 M_W^2 \left(M_H^2-2 M_W^2\right) \beta _W\bigg) \bigg].
    \end{split}
\end{align}

\subsection{$\mathcal{A}(W^+ (p_1,\epsilon_1^L) W^-(p_2,\epsilon_2^L) \rightarrow \gamma (p_3,\epsilon_3^a) h(p_4))$}

\subsubsection{Amplitudes in terms of the polarization}

\begin{align}
    \begin{split}
        A_{\pi, t}= & -a\frac{g^2 S_W M_W \left(p_2-2 p_4\right){}^{\nu } \epsilon _1{}^{\mu } \epsilon _2{}^{\nu } \epsilon _3{}^*{}^{\rho } \eta ^{\mu  \rho }}{2 \left(\left(p_4-p_2\right){}^2-M_W^2 \xi _W\right)},
    \end{split}
    \\
    \begin{split}
        \mathcal{A}_{W, t}= & ag^2 S_W M_W \epsilon _1{}^{\mu } \epsilon _2{}^{\nu } \epsilon _3{}^*{}^{\rho } \eta ^{\delta  \nu } \left[\left(p_2+p_3-p_4\right){}^{\mu } \eta ^{\rho  \sigma }+\left(-p_1-p_3\right){}^{\sigma } \eta ^{\mu  \rho }+\left(p_1-p_2+p_4\right){}^{\rho }
   \eta ^{\mu  \sigma }\right] \\
   & \left[\frac{\eta ^{\delta  \sigma }}{\left(p_4-p_2\right){}^2-M_W^2}+\frac{\left(p_4-p_2\right){}^{\delta } \left(p_2-p_4\right){}^{\sigma } \left(1-\xi _W\right)}{\left(\left(p_4-p_2\right){}^2-M_W^2\right)
   \left(\left(p_4-p_2\right){}^2-M_W^2 \xi _W\right)}\right],
    \end{split}
    \\
    \begin{split}
        \mathcal{A}_{\pi, u}= & a \frac{g^2 S_W M_W \left(-p_2+p_3-p_4\right){}^{\mu } \epsilon _1{}^{\mu } \epsilon _2{}^{\nu } \epsilon _3{}^*{}^{\rho } \eta ^{\nu  \rho }}{2 \left(\left(p_3-p_2\right){}^2-M_W^2 \xi _W\right)},
    \end{split}
    \\
    \begin{split}
        \mathcal{A}_{W , u}= & -a g^2 S_W M_W \epsilon _1{}^{\mu } \epsilon _2{}^{\nu } \epsilon _3{}^*{}^{\rho } \eta ^{\mu  \sigma } \left[\left(-p_2-p_3\right){}^{\delta } \eta ^{\nu  \rho }+\left(2 p_2-p_3\right){}^{\rho } \eta ^{\delta  \nu }+\left(2 p_3-p_2\right){}^{\nu } \eta
   ^{\delta  \rho }\right] \\
   & \left[\frac{\eta ^{\delta  \sigma }}{\left(p_3-p_2\right){}^2-M_W^2}+\frac{\left(p_3-p_2\right){}^{\delta } \left(p_2-p_3\right){}^{\sigma } \left(1-\xi _W\right)}{\left(\left(p_3-p_2\right){}^2-M_W^2\right)
   \left(\left(p_3-p_2\right){}^2-M_W^2 \xi _W\right)}\right].
    \end{split}
\end{align}

\subsubsection{Polarized amplitudes}

We will label the polarized amplitudes with $\mathcal{A}_{\epsilon_3}$, where $\epsilon_3$ refers to the polarization of the photon. The only independent amplitude is ($\mathcal{A}_{+}=\mathcal{A}_{-}$), where $+$ or $-$ refers to the polarization of the photon:
\begin{equation*}
  \mathcal{A}_{+} =  \frac{a g^2 \sqrt{s} \sqrt{2-2 x^2} S_W \beta _W \left(M_H^2-2 M_W^2\right)}{M_W \left(s-M_H^2\right) \left(x^2 \beta _W^2-1\right)}.
\end{equation*}

\subsection{$\mathcal{A}(W^+ (p_1,\epsilon_1^L) W^-(p_2,\epsilon_2^L) \rightarrow \gamma (p_3,\epsilon_3^a) Z(p_4,\epsilon_4))$}

\subsubsection{Amplitude}

\begin{align}
    \begin{split}
        \mathcal{A}_{contact}= &-i \epsilon _1{}^{\mu } \epsilon _2{}^{\nu } \epsilon _3{}^*{}^{\rho } \epsilon _4{}^*{}^{\sigma } C_W g^2 S_W  \left(-i \eta ^{\mu  \sigma } \eta ^{\nu  \rho }-i    \eta ^{\mu  \rho } \eta ^{\nu  \sigma }+2 i     \eta ^{\mu  \nu } \eta ^{\rho  \sigma }\right),
    \end{split}
    \\
    \begin{split}
        \mathcal{A}_{\pi ,t}= & -\frac{g^2 S_W^3 M_W^2 \epsilon _1{}^{\mu } \epsilon _2{}^{\nu } \epsilon _3{}^*{}^{\rho } \epsilon _4{}^*{}^{\sigma } \eta ^{\mu  \rho } \eta ^{\nu  \sigma }}{C_W \left(\left(p_4-p_2\right){}^2-M_W^2 \xi _W\right)},
    \end{split}
    \\
    \begin{split}
        \mathcal{A}_{W , t}= & C_W g^2 S_W \epsilon _1{}^{\mu } \epsilon _2{}^{\nu } \epsilon _3{}^*{}^{\rho } \epsilon _4{}^*{}^{\sigma } \left[\left(-p_2-p_4\right){}^{\gamma } \eta ^{\nu  \sigma }+\left(2 p_2-p_4\right){}^{\sigma } \eta ^{\gamma  \nu }+\left(2
   p_4-p_2\right){}^{\nu } \eta ^{\gamma  \sigma }\right] \\
   & \left[\left(-p_1-p_3\right){}^{\delta } \eta ^{\mu  \rho }+\left(p_1-p_2+p_4\right){}^{\rho } \eta ^{\delta  \mu }+\left(p_2+p_3-p_4\right){}^{\mu } \eta ^{\delta  \rho }\right] \\
   & \left[\frac{\eta
   ^{\gamma  \delta }}{\left(p_4-p_2\right){}^2-M_W^2}+\frac{\left(p_4-p_2\right){}^{\gamma } \left(p_2-p_4\right){}^{\delta } \left(1-\xi _W\right)}{\left(\left(p_4-p_2\right){}^2-M_W^2\right) \left(\left(p_4-p_2\right){}^2-M_W^2 \xi _W\right)}\right],
    \end{split}
    \\
    \begin{split}
        \mathcal{A}_{\pi , u}= & -\frac{g^2 S_W^3 M_W^2 \epsilon _1{}^{\mu } \epsilon _2{}^{\nu } \epsilon _3{}^*{}^{\rho } \epsilon _4{}^*{}^{\sigma } \eta ^{\mu  \sigma } \eta ^{\nu  \rho }}{C_W \left(\left(p_3-p_2\right){}^2-M_W^2 \xi _W\right)},
    \end{split}
    \\
    \begin{split}
        \mathcal{A}_{W , u}= & C_W g^2 S_W \epsilon _1{}^{\mu } \epsilon _2{}^{\nu } \epsilon _3{}^*{}^{\rho } \epsilon _4{}^*{}^{\sigma } \left[\left(-p_2-p_3\right){}^{\gamma } \eta ^{\nu  \rho }+\left(2 p_2-p_3\right){}^{\rho } \eta ^{\gamma  \nu }+\left(2
   p_3-p_2\right){}^{\nu } \eta ^{\gamma  \rho }\right]\\ & \left[\left(-p_1-p_4\right){}^{\delta } \eta ^{\mu  \sigma }+\left(p_1-p_2+p_3\right){}^{\sigma } \eta ^{\delta  \mu }+\left(p_2-p_3+p_4\right){}^{\mu } \eta ^{\delta  \sigma }\right] \\
   & \left[\frac{\eta
   ^{\gamma  \delta }}{\left(p_3-p_2\right){}^2-M_W^2}+\frac{\left(p_3-p_2\right){}^{\gamma } \left(p_2-p_3\right){}^{\delta } \left(1-\xi _W\right)}{\left(\left(p_3-p_2\right){}^2-M_W^2\right) \left(\left(p_3-p_2\right){}^2-M_W^2 \xi _W\right)}\right].
    \end{split}
\end{align}

\subsubsection{Polarized Amplitudes}

We will label the polarized amplitudes with $\mathcal{A}_{\epsilon_3 \epsilon_4}$, where $\epsilon_3$ and $\epsilon_4$ refer to the polarizations of the photon and the Z-boson. We only have 3 independent amplitudes since $\mathcal{A}_{+L}=\mathcal{A}_{-L}, \mathcal{A}_{+P}=\mathcal{A}_{-N},\mathcal{A}_{+N}=\mathcal{A}_{-P}$.




\begin{align}
\mathcal{A}_{+L} & = \frac{2 g^2 x \sqrt{1-x^2} C_W M_Z S_W \left(-2 s M_W^2+s M_Z^2-8 M_W^4\right)}{\sqrt{s} M_W^2 \left(s-M_Z^2\right) \left(x^2 \beta _W^2-1\right)}, \\
\mathcal{A}_{+P} & =  \frac{g^2 C_W S_W \left(-8 M_W^4 \left(s-x^2 M_Z^2\right)+2 M_W^2 M_Z^2 \left(-2 M_Z^2+s x^2+s\right)-s \left(x^2-1\right) M_Z^4\right)}{M_W^2 \left(M_Z^2-s\right) \left(x^2 \left(4 M_W^2-s\right)+s\right)}, \\
\mathcal{A}_{+N} & = -\frac{g^2 \left(x^2-1\right) C_W S_W \left(2 s M_W^2-s M_Z^2+8 M_W^4\right)}{M_W^2 \left(M_Z^2-s\right) \left(x^2 \beta _W^2-1\right)}.
\end{align}

\section{Complete contribution of each channel to the p-PWA's.}

In this Appendix we will provide the plots for the cumulative ratios $\chi_i^{J\, '}$ for the p-PWA's for $g'\neq 0$. 
For the sake of clarity, in the main text we separated the plots in bosonic cuts, $b\bar{b}$ cuts and $t\bar{t}$ cuts. Therefore, we did not specify the contributions from each individual channel, as they were very numerous for $g'\neq 0$. The plots must be read in the same way as the PWA cumulative ratios $\chi_i^J$ in Sec.~\ref{sec:gpr=0} ($g'=0$ case): 
each line contains the relative cumulative contribution of the past cuts. Hence, the shaded area accounts for the contribution of the same-color curve directly above it.

Some contributions are difficult to see because the curves are close to each other and some of them directly overlap. For the $J=0$ p-PWA, we observe this for the $b\bar{b}$ cut and $\gamma \gamma$ in Figs.~\ref{fig:r0ratio_prime_sm}, \ref{fig:ratioj0partial} and~\ref{fig:ratio_r0_prime_highest}. While the $\gamma \gamma$ contribution (bottom in pale blue) can still be appreciated (barely) between the x-axis and the orange curve, the $b \bar{b}$ channel is very suppressed and sits on top of the $\gamma \gamma$, impossible to see because of its negligible contribution. The same occurs for the $\gamma h$ and $\gamma Z$ curves in the mentioned plots, where the $\gamma h$ channel is essentially negligible and its curve sits on top of the $\gamma Z$ one. For the $J=1$ p-PWA in Figs.~\ref{fig:r1ratio_prime_allcuts_sm}, \ref{fig:ratior1_all_partial} and~\ref{fig:ratior1partialbest}, this time it is the $\gamma Z$ curve which sits on top of the $b \bar{b}$ and cannot be seen as is shown in the Figures. 
The same occurs for the $ZZ$ cut which sits on top of the $WW$ curve in the same plots mentioned.

In general, if the shaded area is not of the same color of the curve immediately above it, a second curve with a negligible contribution sits on top of the first one.

\subsection{$J=0$  pseudo-PWA: $\chi'_0$}

\begin{figure}[!t]    
\includegraphics[width=10cm]{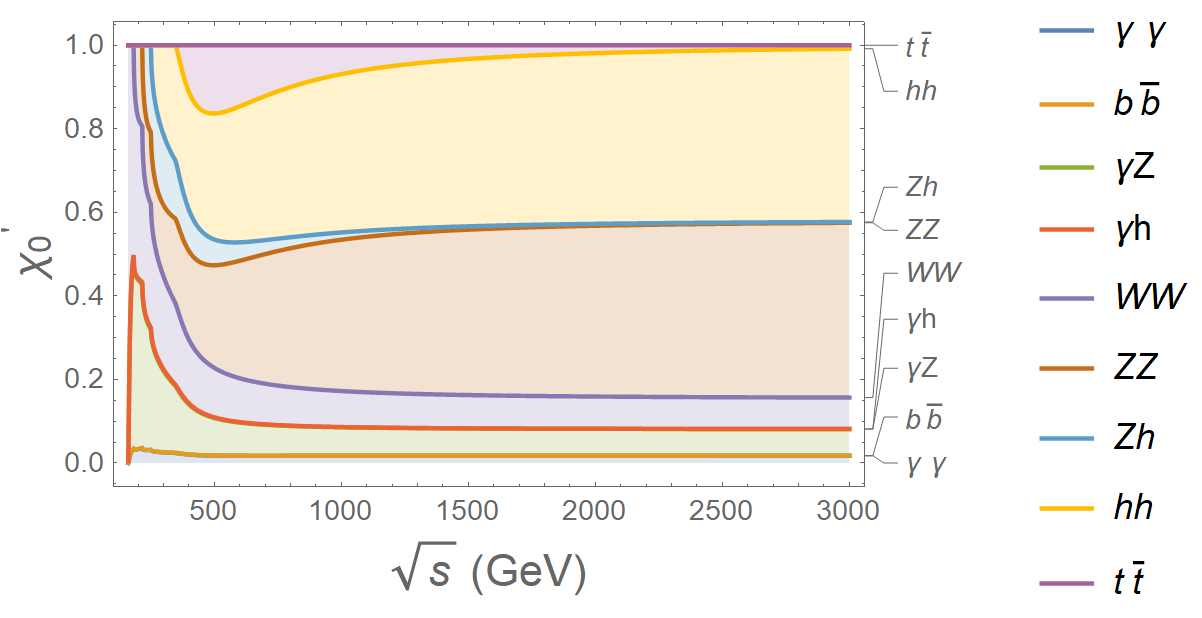}
\centering
\caption{\small Ratio for the $R'_0$ p-PWA at the SM}
\label{fig:r0ratio_prime_sm}
\end{figure}

\begin{figure}[!t]    
\centering
\begin{subfigure}{0.4 \columnwidth}
  \centering
  \includegraphics[width=1\linewidth]{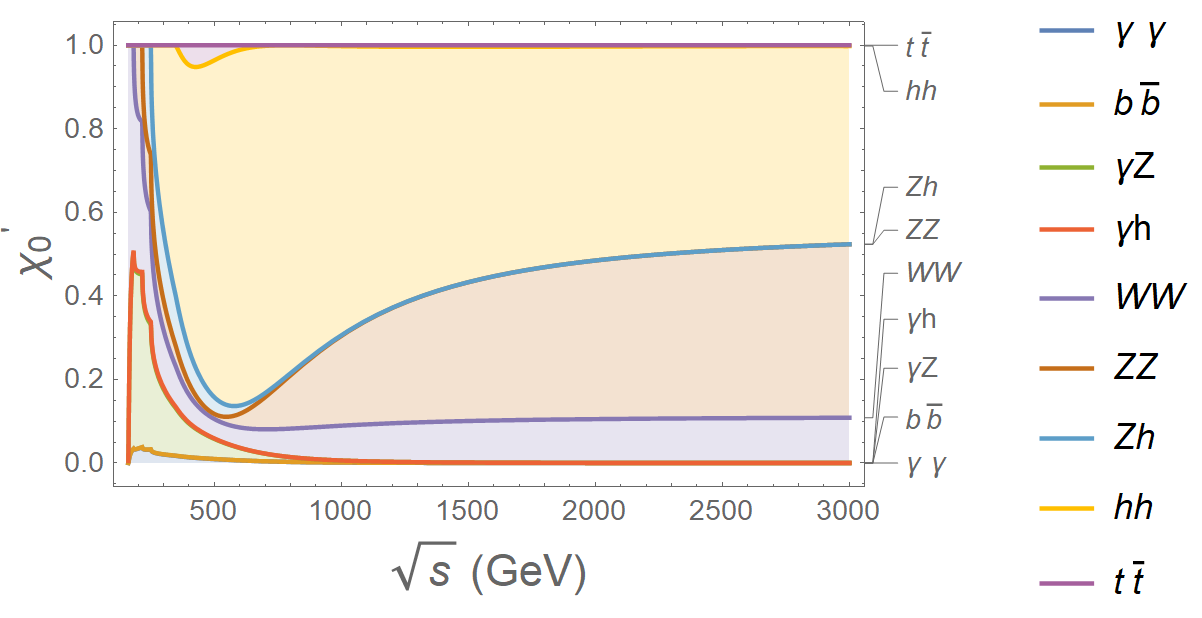}  
  \caption{$J=0$ p-PWA: contribution of each channel for $a=1.10$ and $b=c_1=d_3=1$}
  \label{fig:r0ratio_prime_a1}
\end{subfigure}
\hspace*{0.75cm}
\begin{subfigure}{0.4 \columnwidth}
  \centering
  \includegraphics[width=1 \linewidth]{Eta_09/ratioj0a11_etacut0.9.png}  
  \caption{$J=0$ p-PWA: contribution of each channel for $a=0.90$ and $b=c_1=d_3=1$}
  \label{fig:r0ratio_prime_a09}
\end{subfigure}
%
\\[8pt]
\centering
\begin{subfigure}{0.4 \columnwidth}
  \centering
  \includegraphics[width=\linewidth]{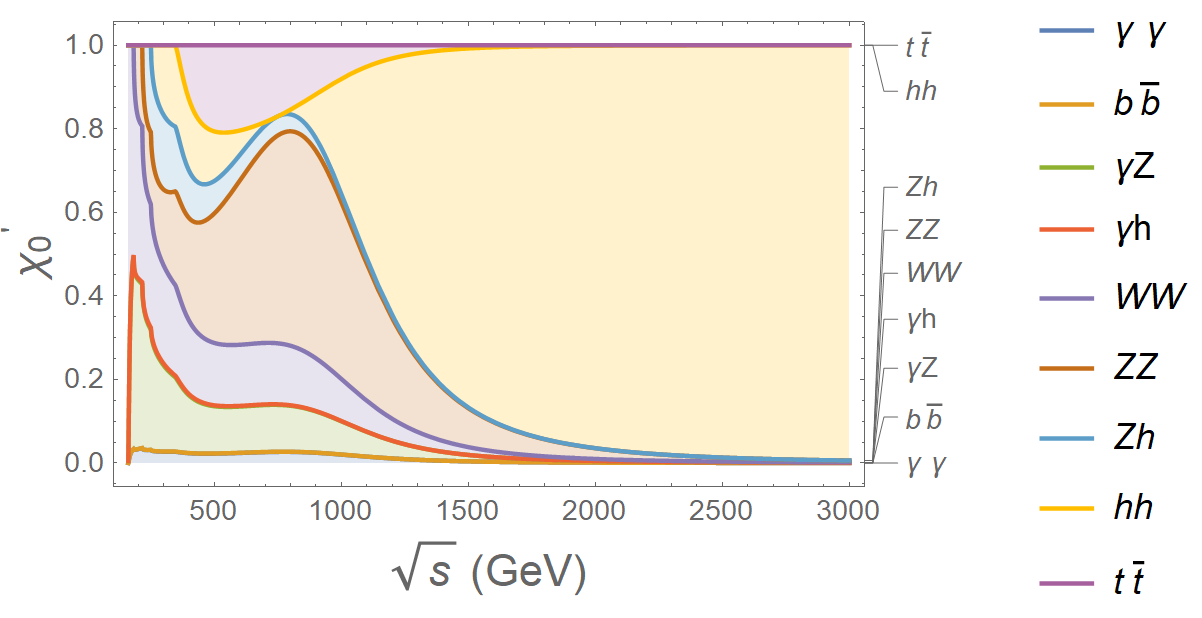}  
  \caption{$J=0$ p-PWA: contribution of each channel for $b=1.1$ and $a=d_3=1$}
  \label{fig:r0ratio_prime_b11}
\end{subfigure}
\hspace*{0.75cm}
\begin{subfigure}{0.4 \columnwidth}
  \centering
  \includegraphics[width=\linewidth]{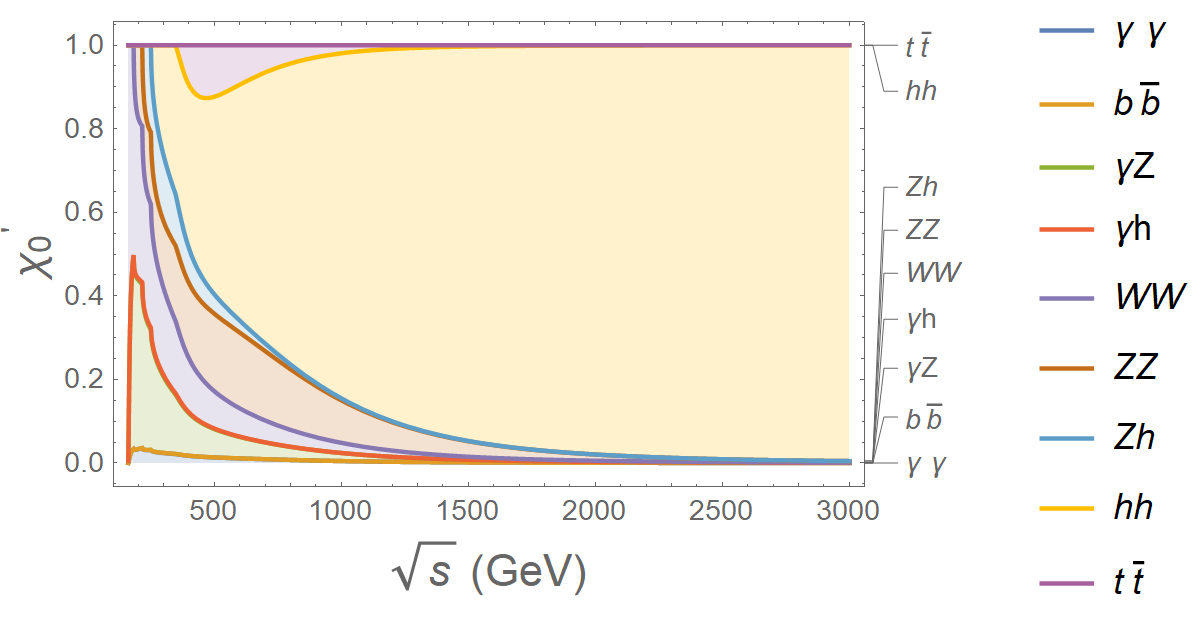}  
  \caption{$J=0$ p-PWA: contribution of each channel for $b=0.90$ and $b=d_3=1$ }
  \label{fig:r0ratio_prime_b09}
\end{subfigure}
\\[8pt]
\centering
\begin{subfigure}{0.4 \columnwidth}
  \centering
  \includegraphics[width=\linewidth]{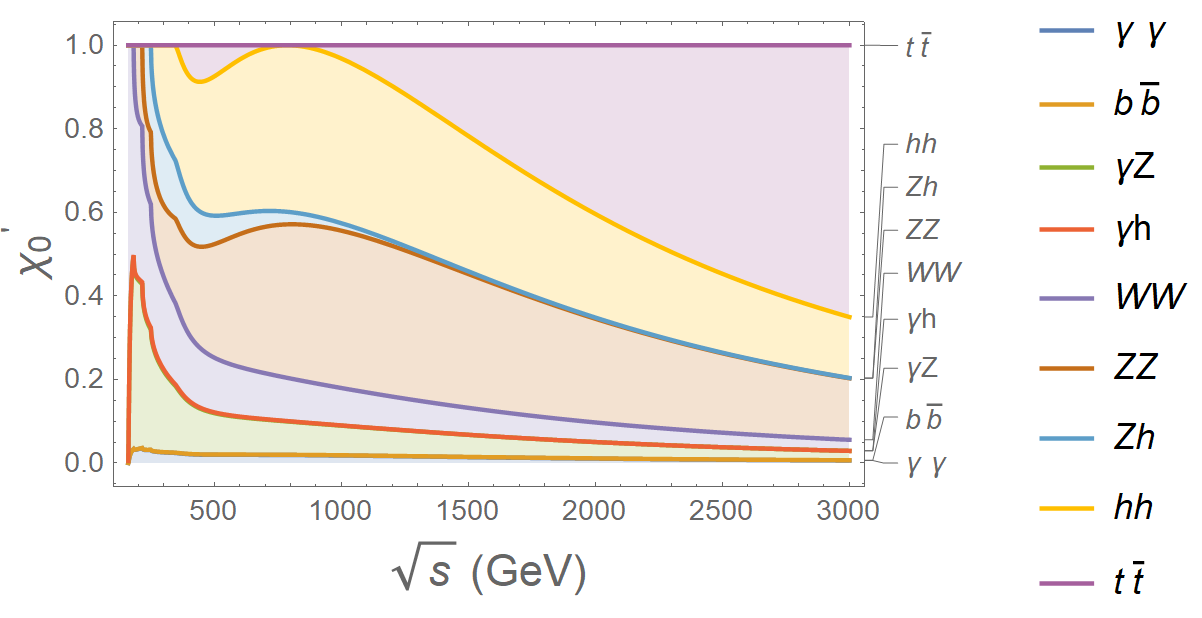}  
  \caption{$J=0$ p-PWA: contribution of each channel for $c_1=1.10$ and $a=b=d_3=1$}
  \label{fig:r0ratio_prime_c111}
\end{subfigure}
\hspace*{0.75cm}
\begin{subfigure}{0.4 \columnwidth}
  \centering
  \includegraphics[width=\linewidth]{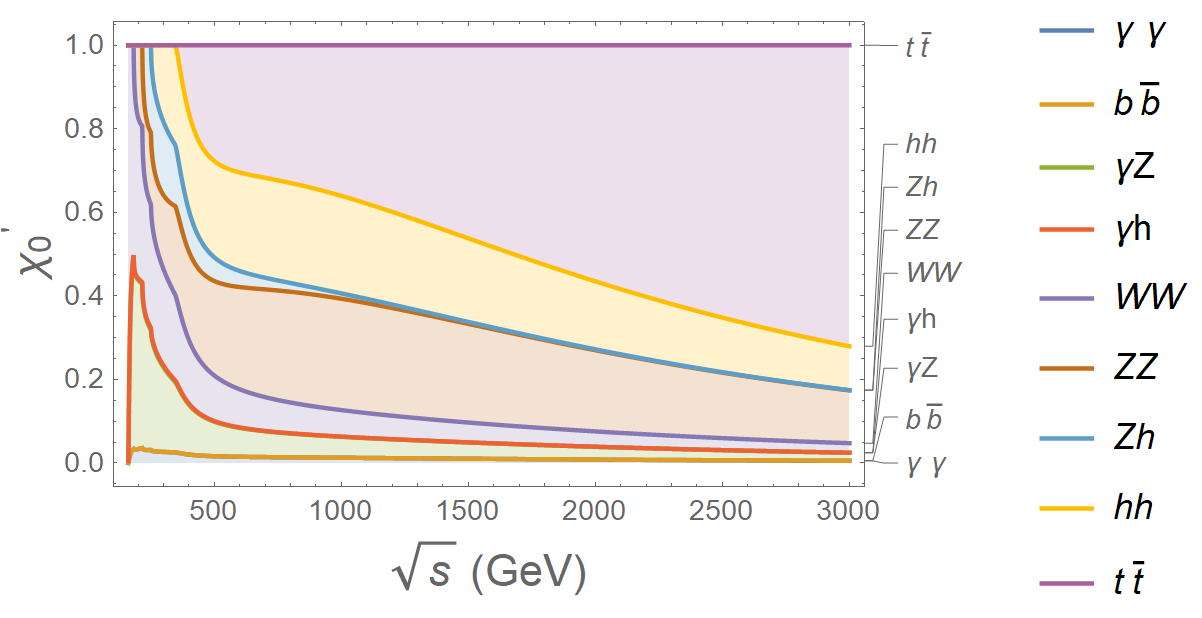}  
  \caption{$J=0$ p-PWA: contribution of each channel for $c_1=0.90$ and $a=b=d_3=1$ }
  \label{fig:r0ratio_prime_c109}
\end{subfigure}
\caption{}
\label{fig:ratioj0partial}
\end{figure}

\begin{figure}[!t]    
\centering
\begin{subfigure}{.4 \columnwidth}
  \centering
  \includegraphics[width=1\linewidth]{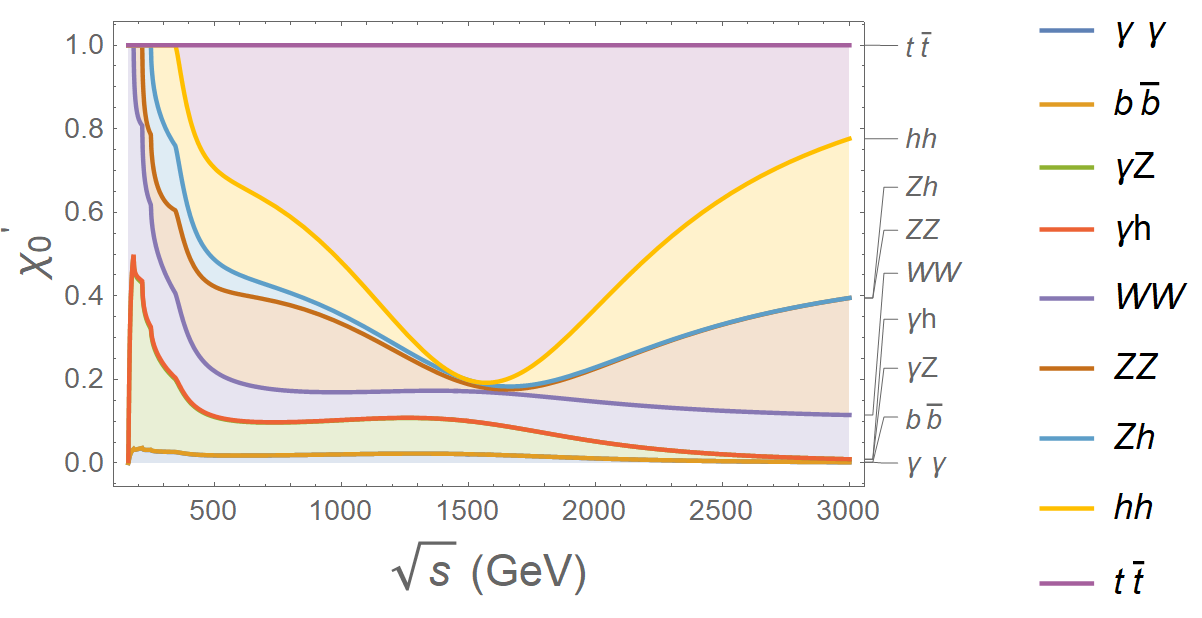}  
  \caption{$J=0$ p-PWA: largest fermion-loop contribution of 80\% at 1.5~TeV  for $a=1.011$, $b=1.045$, $c_1=0.900$ and  $d_3=1.094$. }
  \label{fig:ratio_r0_prime_highest15tev}
\end{subfigure}
\hspace*{0.75cm}
\begin{subfigure}{.4 \columnwidth}
  \centering
  \includegraphics[width=1 \linewidth]{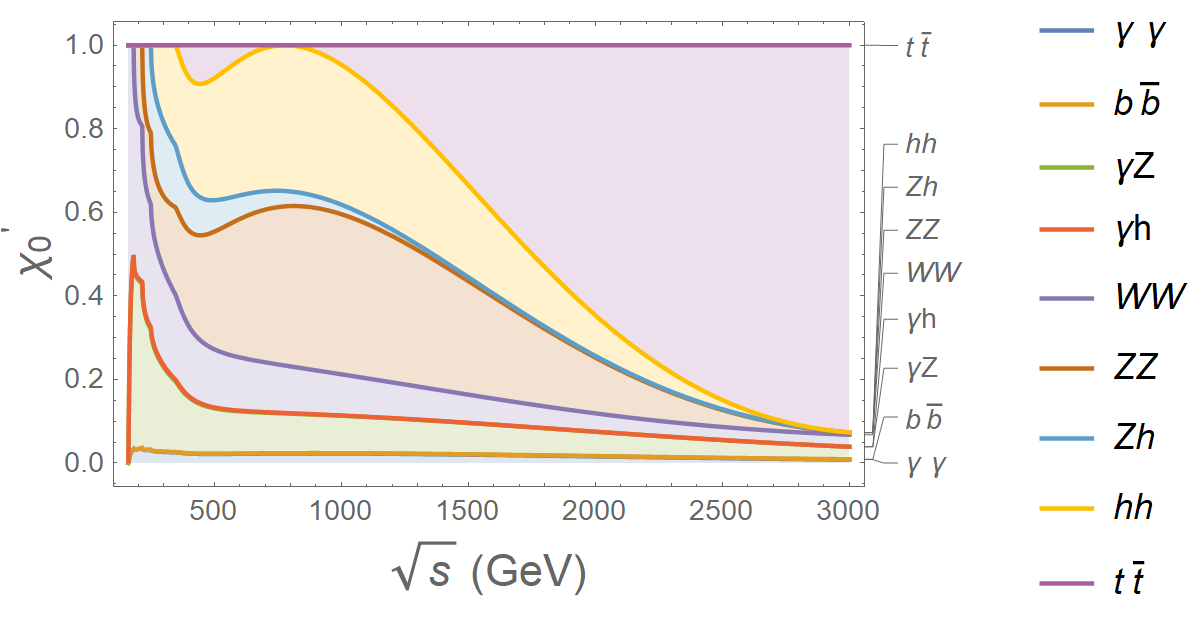}   \caption{$J=0$ p-PWA: largest fermion-loop contribution of 93\% at 3~TeV happens for$a=1.003$, $b=1.011$, $c_1=1.100$ and  $d_3=1.100$. }
  \label{fig:ratio_r0_prime_highest3tevv}
\end{subfigure}
\caption{}
\label{fig:ratio_r0_prime_highest}
\end{figure}

We provide the Figs. \ref{fig:r0ratio_prime_sm}, \ref{fig:ratioj0partial} and \ref{fig:ratio_r0_prime_highest} corresponding to the ratios $\chi'_0$ for a $\abs{\cos \theta}\leq $0.9  angular integration for all absorptive cuts  explained in explained in Sec.~\ref{section:r0prime}.

\subsection{$J=1$  pseudo-PWA: $\chi'_1$}

\begin{figure}[!t]    
\includegraphics[width=10cm]{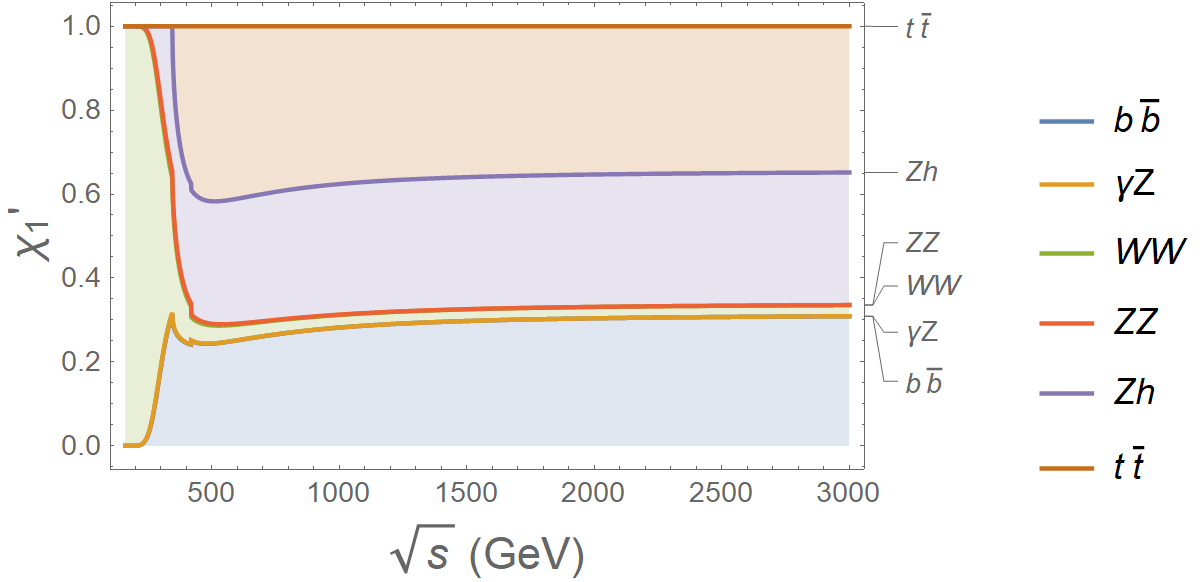}
\centering
\caption{\small Ratio for the $R'_1$ p-PWA at the SM}
\label{fig:r1ratio_prime_allcuts_sm}
\end{figure}

\begin{figure}[!t]    
\centering
\begin{subfigure}{.4 \columnwidth}
  \centering
  \includegraphics[width=1\linewidth]{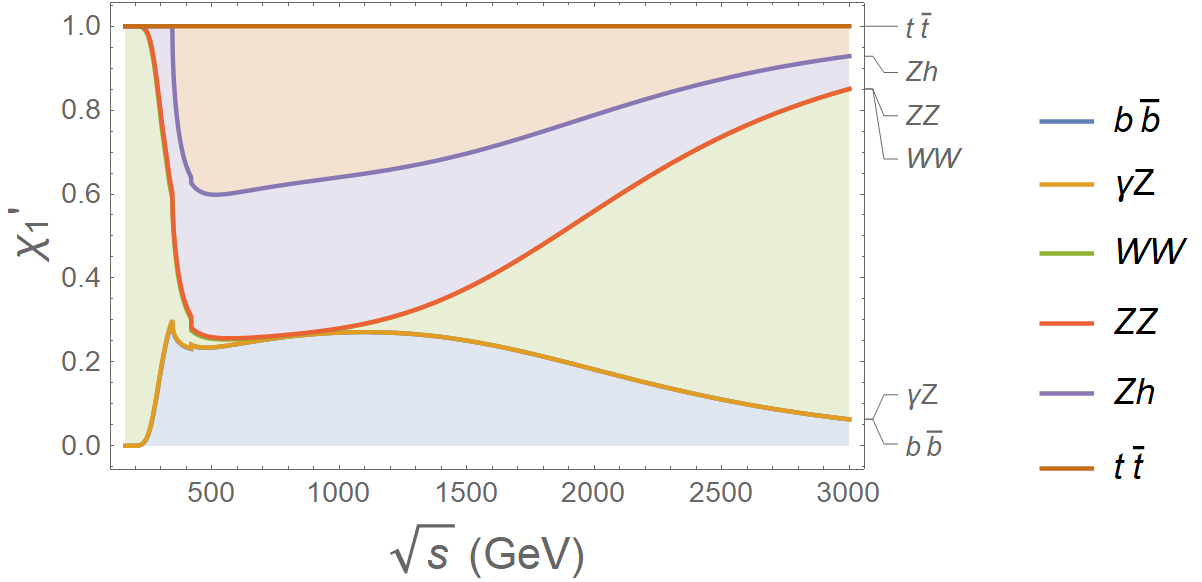}  \caption{$J=1$ p-PWA: contribution of each channel for $a=1.10$.}
  \label{fig:ratior1a11_prime}
\end{subfigure}
\hspace*{0.75cm}
\begin{subfigure}{.4 \columnwidth}
  \centering
  \includegraphics[width=1 \linewidth]{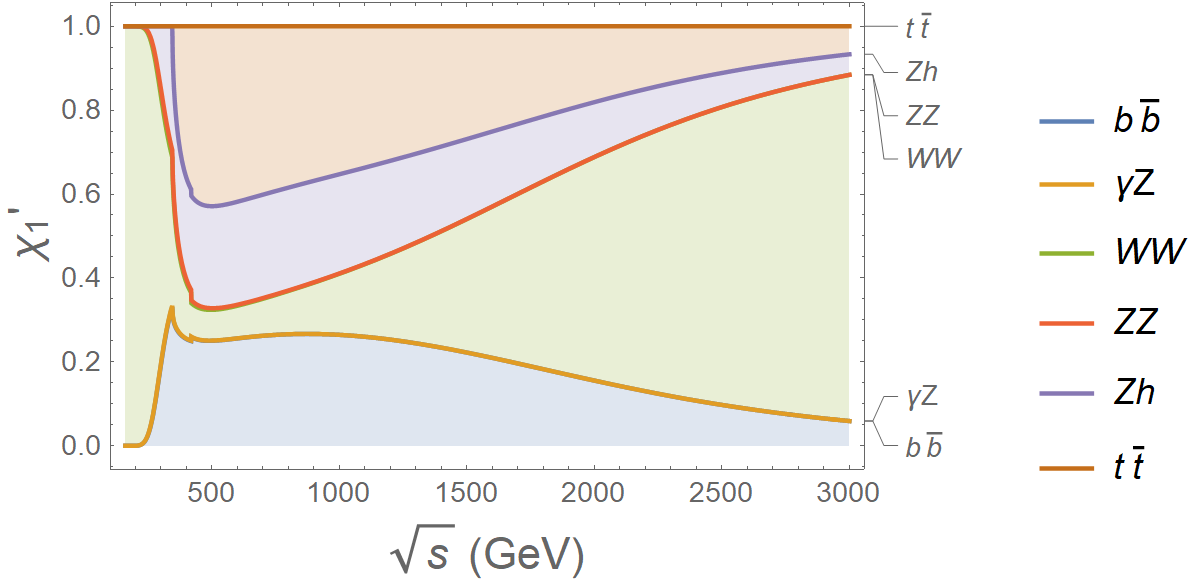}  
  \caption{$J=1$ p-PWA: contribution of each channel for $a=0.90$.}
  \label{fig:ratior1a09_all_prime}
\end{subfigure}
\caption{}
    \label{fig:ratior1_all_partial}
\end{figure}

\begin{figure}[!t]    
\centering
\begin{subfigure}{.4 \columnwidth}
  \centering
  \includegraphics[width=1\linewidth]{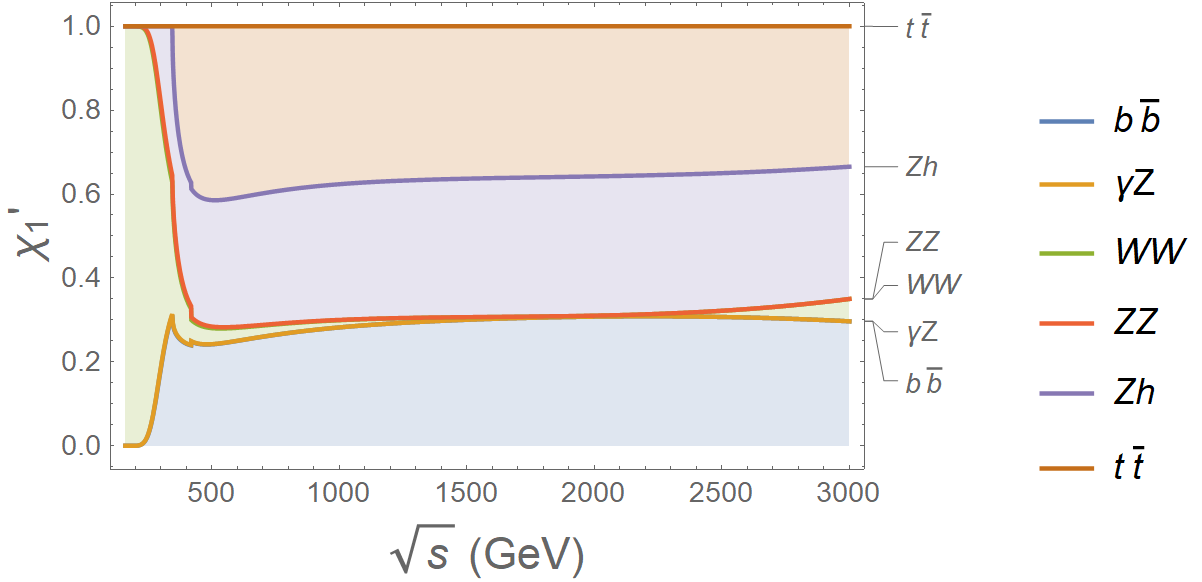}  \caption{$J=1$ p-PWA: largest fermion-loop contribution of 75\% at 1.5 TeV for $a=1.019$.}
  \label{fig:ratior1a11_prime_all}
\end{subfigure}
\hspace*{0.75cm}
\begin{subfigure}{.4 \columnwidth}
  \centering
  \includegraphics[width=1 \linewidth]{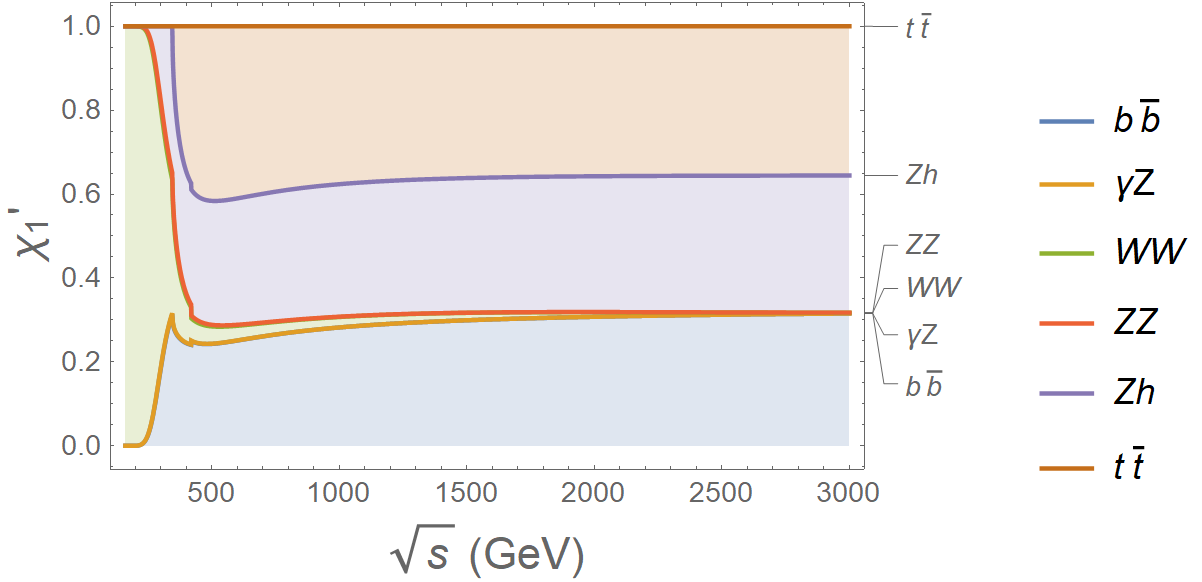}  
  \caption{$J=1$ p-PWA: largest fermion-loop contribution of 76\% at 3 TeV for $a=1.007$. }
  \label{fig:ratior1a09_prime_all}
\end{subfigure}
\caption{}
\label{fig:ratior1partialbest}
\end{figure}

We provide the Figs.~\ref{fig:r1ratio_prime_allcuts_sm}, \ref{fig:ratior1_all_partial} and~\ref{fig:ratior1partialbest} corresponding to the ratios $\chi'_1$ for a $\abs{\cos \theta}$=0.9  angular integration for all absorptive cuts  explained in Sec.~\ref{section:r1prime}.

\subsection{$J=0$  pseudo-PWA: sensitivity of $R'_0$ to the optimal points}
\label{app:sensitivity-r0pr}

\begin{figure}[!t]  
\begin{subfigure}{.5\textwidth}
  \centering
  \includegraphics[width=.9\linewidth]{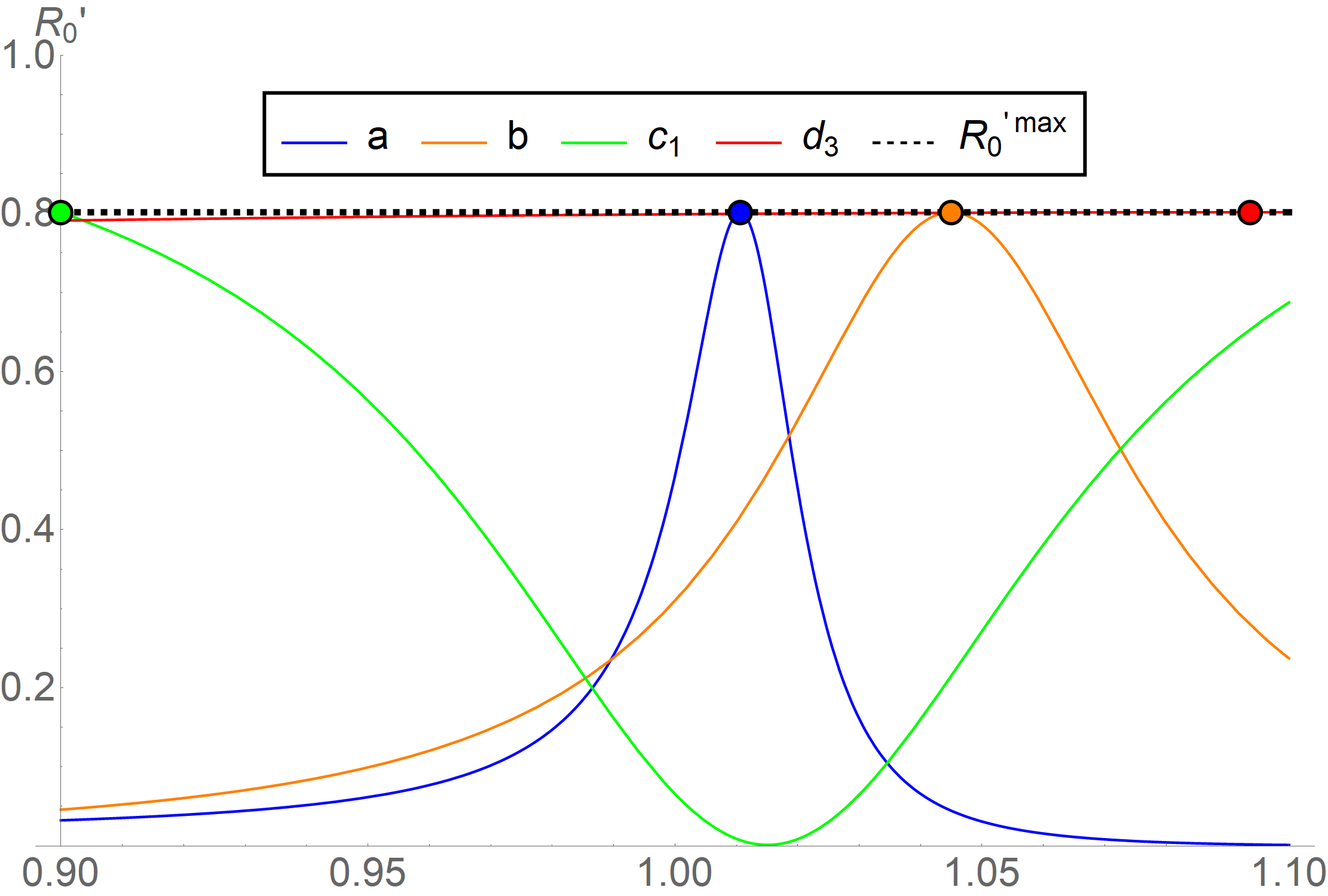}  
  \caption{}
  \label{fig:r0_prime__best1500_sensitivity}
\end{subfigure}
\begin{subfigure}{.5\textwidth}
  \centering
  \includegraphics[width=.9\linewidth]{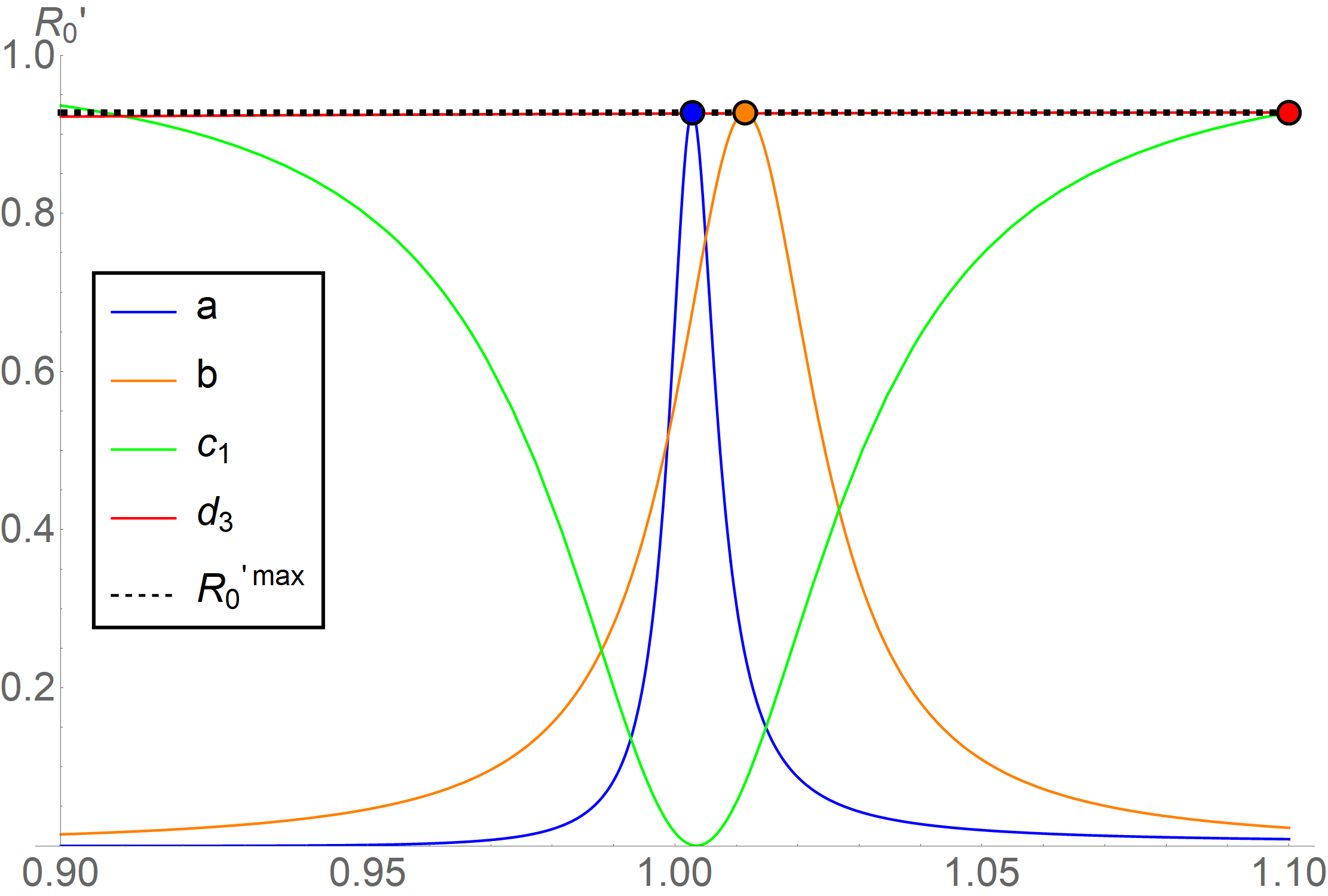}  
  \caption{}
  \label{fig:r0_prime_best3000_sensitivity}
\end{subfigure}
\caption{Sensitivity of $R'_0$ to each parameter when the rest are set to the highest correction value at $\sqrt{s}$=1.5 TeV (left) and $\sqrt{3}$ (right) TeV.}
\label{fig:figr0_prime_sensitivity}
\end{figure}

For the case of the $J=0$ p-PWA $R_0'$, we have plotted the sensitivity to the optimal parameters $(a,b,c_1,d_3)$ in Fig.~\ref{fig:figr0_prime_sensitivity}. All notations are analogous to those for $R_0$  in Fig.~\ref{fig:figr0sensitivity}.

\subsection{$J=1$  pseudo-PWA: sensitivity of $R'_1$ to the optimal points}
\label{app:sensitivity-r1pr}

\begin{figure}[!t]    
\includegraphics[width=10cm]{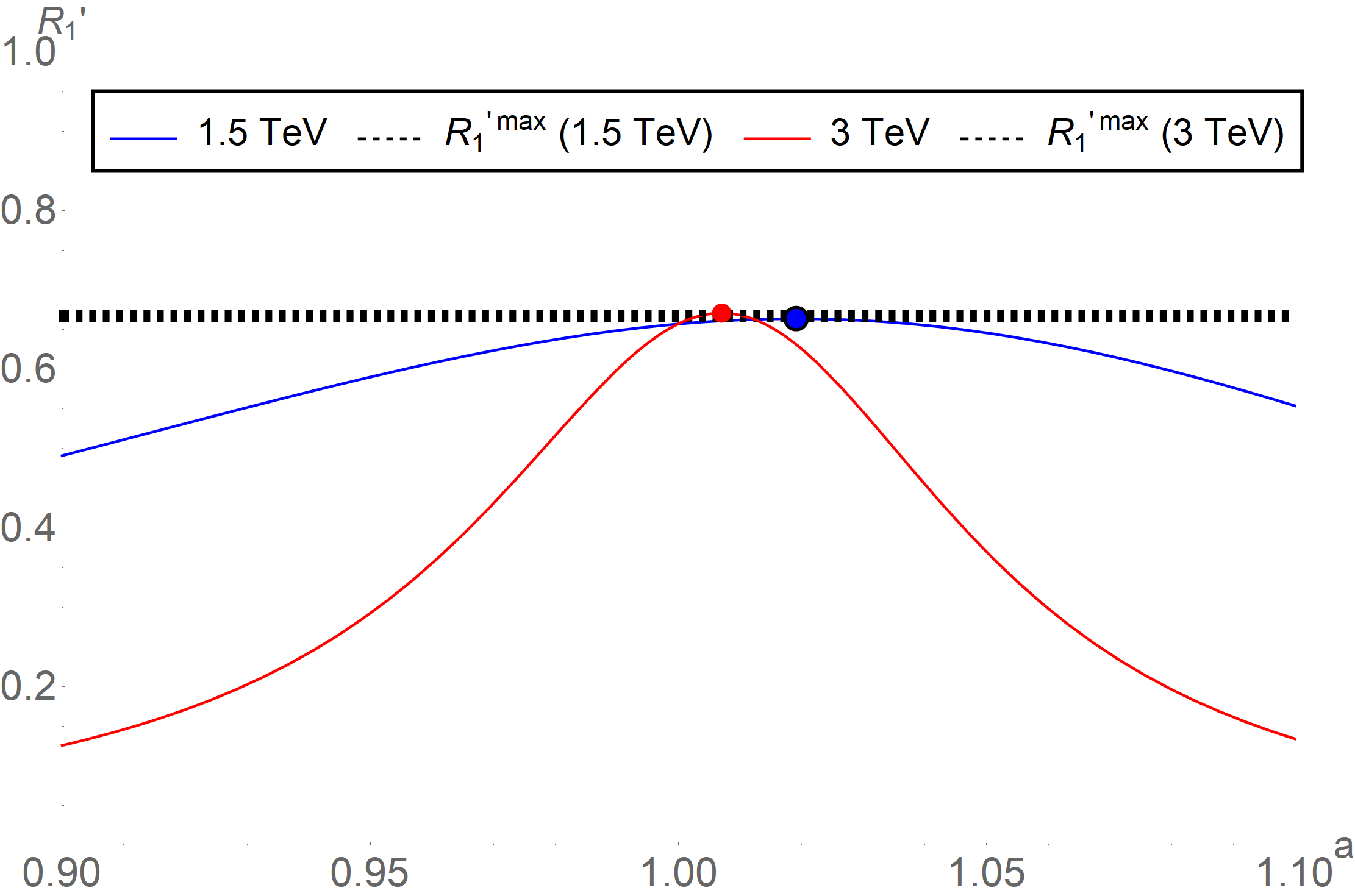}
\centering
\caption{\small Sensitivity of $R'_1$ to the $a$ parameter for the highest contribution at $\sqrt{s}$= 1.5 TeV and $\sqrt{s}$=3 TeV}
\label{fig:r1_prime_sensitivity}
\end{figure}

For the case of the $J=1$ p-PWA $R_1'$, we have plotted the sensitivity to the $a$ parameter in Fig.~\ref{fig:r1_prime_sensitivity}. All notations are analogous to those for $R_1$ in Fig.~\ref{fig:r1_sensitivity}.

\end{document}